\documentclass[a4paper,11pt]{article}
\usepackage{amsmath,amsthm,amssymb,mathrsfs,mathtools,graphicx,color,yhmath,comment}
\usepackage[titletoc,title]{appendix}
\usepackage[numbers ,sort&compress]{natbib}
\newcommand{\nc}{\newcommand}
\nc{\rnc}{\renewcommand}
\nc{\nn}{\nonumber}

\nc{\del}{{\partial}}
\rnc{\Im}{{\textrm{Im}\,}}
\rnc{\Re}{{\textrm{Re}\,}}

\nc{\bra}{\langle}
\nc{\ket}{\rangle}
\nc{\wt}{\widetilde}
\nc{\ms}{\mathsf}

\nc{\A}{\textrm{A}}
\nc{\B}{\textrm{B}}
\nc{\C}{\textrm{C}}
\nc{\D}{\textrm{D}}
\nc{\ul}{\underline}
\nc{\dr}{\textrm{dr}}
\nc{\eff}{\textrm{eff}}

\nc{\tcr}{\textcolor{red}}
\nc{\tcb}{\textcolor{blue}}
\nc{\bs}{\boldsymbol}

\DeclareMathOperator{\sh}{sh}
\DeclareMathOperator{\ch}{ch}
\DeclareMathOperator{\tnh}{th}
\DeclareMathOperator{\Tr}{Tr}

\DeclareMathOperator{\End}{End}

\DeclareMathOperator{\sgn}{sgn}

\newtheorem{theorem}{Theorem}[section]
\newtheorem{lemma}[theorem]{Lemma}

\theoremstyle{definition}

\numberwithin{equation}{section}

\begin{document}

\title{Spin Drude weight for the 
integrable XXZ chain\\ with arbitrary spin}

\author{
Shinya Ae\thanks{E-mail: 1221701@ed.tus.ac.jp}\ \ \ and \ 
Kazumitsu Sakai\thanks{E-mail: k.sakai@rs.tus.ac.jp}
\\\\
\textit{Department of Physics,
Tokyo University of Science,}\\
 \textit{Kagurazaka 1-3, Shinjuku-ku, Tokyo 162-8601, Japan} \\
\\\\
\\
}

\date{}

\maketitle
\begin{abstract}
Using generalized hydrodynamics (GHD), we exactly evaluate the 
finite-temperature spin Drude weight at zero magnetic field 
for the integrable XXZ chain with arbitrary spin and easy-plane anisotropy. 
First, we construct the fusion hierarchy of the quantum transfer 
matrices ($T$-functions) and derive functional relations ($T$- and
 $Y$-systems) satisfied by the $T$-functions and certain combinations 
of them ($Y$-functions). Through analytical arguments, the $Y$-system is 
reduced to a set of non-linear integral equations, equivalent to the
thermodynamic Bethe ansatz (TBA) equations. Then, employing 
GHD, we calculate the spin Drude weight at arbitrary finite temperatures. 
As a result, a characteristic fractal-like structure of the Drude weight 
is observed at arbitrary spin, similar to the spin-1/2 case. 
In our approach, the solutions to the TBA equations (i.e., the 
$Y$-functions) can be explicitly written in terms of the $T$-functions, 
thus allowing for a systematic calculation of the high-temperature limit of 
the Drude weight.
\end{abstract}
\section{Introduction}\label{introduction}
Quantum integrable systems are characterized by an infinite 
number of local or quasi-local conserved charges, which have led 
to extensive investigations into their unique physical properties. 
While conventional systems are understood through Gibbs ensembles, 
integrable systems, due to their multitude of extensive conserved charges, 
are described by Generalized Gibbs Ensembles (GGE) 
\cite{rigol2006hard,rigol2007relaxation} (refer to 
\cite{vidmar2016generalized} for a review). This distinction
 becomes particularly evident in thermalization processes, 
where integrable systems often deviate from the predictions of 
frameworks like the eigenstate thermalization hypothesis 
\cite{rigol2012alternatives}.

On the other hand, our understanding of non-equilibrium properties 
in integrable systems has seen rapid advancements, largely attributed 
to the theoretical framework of Generalized Hydrodynamics (GHD)
 \cite{castro2016emergent,bertini2016transport} (see also \cite{doyon2020lecture} 
for lecture notes). GHD effectively merges the concepts of fluid 
mechanics and GGE, providing a novel perspective in studying the
 non-equilibrium properties of quantum integrable systems 
(recent reviews are available in \cite{bulchandani2021superdiffusion, 
de2022correlation,essler2022short}). Moreover, the GHD framework has been extended
 to weakly broken integrable systems (refer to \cite{bastianello2021hydrodynamics, 
gopalakrishnan2023anomalous} for a review).

Among these advancements, the finite-temperature spin transport properties
of the spin-1/2 XXZ chain, 
which are deeply relevant to this work, have been extensively 
studied for a considerable period \cite{bertini2021finite}.
The spin-1/2 XXZ chain is given by
\begin{equation}
\mathcal{H}_0=\frac{J}{4}\sum_{j=1}^{L}\left[
\sigma_j^x\sigma_{j+1}^x+\sigma_j^y\sigma_{j+1}^y+
\Delta\left(\sigma_j^z\sigma^z_{j+1}+1\right)\right],
\label{XXZ0}
\end{equation}
where $\sigma_j^{x}$, $\sigma_j^{y}$, and $\sigma_j^z$ are the 
spin-1/2 spin operators (Pauli matrices), and $\Delta$ 
stands for the anisotropy parameter.
In this model, the spin current is not a conserved quantity except for 
$\Delta = 0$, which corresponds to a free fermion system. 
For $|\Delta| < 1$,  the system exhibits ballistic 
spin transport, characterized by a finite spin Drude weight 
\cite{zotos1999finite,prosen2013families,urichuk2019spin,klumper2019spin} and also by 
a dynamical exponent $z = 1$  that signifies 
a scaling relation $x \sim t^{1/z}=t$ between the magnetization displacement 
$x$ and time $t$.  In contrast, when $ \Delta > 1 $, the spin transport is
believed to be diffusive with a dynamical exponent $z=2$ 
\cite{vznidarivc2011spin,ilievski2018superdiffusion}.
Interestingly, at the isotropic point $\Delta = 1$, i.e., at the boundary between the diffusive and ballistic regimes, the spin transport at infinite temperature is considered to exhibit superdiffusive behavior with a dynamical exponent $z = 3/2$ \cite{vznidarivc2011spin,ljubotina2017spin,gopalakrishnan2019kinetic}.
Further investigations \cite{ljubotina2019kardar} suggest that, particularly 
at this isotropic point, the transport properties are likely to belong to the
 Kardar-Parisi-Zhang (KPZ) universality class \cite{kardar1986dynamic}. On the
other hand, a recent investigation \cite{krajnik2022absence} indicates that 
the transport at $\Delta=1$ cannot be explained by the KPZ type, which 
has also been supported by recent quantum simulations 
\cite{rosenberg2023dynamics}. From this perspective, the topic continues to
 be a subject of ongoing debate. More rigorous and quantitative analysis of 
spin transport at the isotropic point is desired, especially a more accurate 
evaluation of the exponent.  

For $|\Delta=\cos(\pi/p_0)|<1$, the spin Drude weight at finite temperatures had 
long been a topic of debate since Zotos derived it in \cite{zotos1999finite}
by combining Kohn's 
formula \cite{kohn1964theory} with the Thermodynamic Bethe Ansatz (TBA) 
\cite{takahashi1972one}. Taking a notably different approach, Prosen and 
Ilievski evaluated the Drude weight at infinite temperature \cite{prosen2013families}.
Specifically, they constructed quasi-local conserved charges and 
optimized the Mazur bound of the Drude weight using them.
Remarkably, 
this optimized lower bound, 
referred to as the Prosen-Ilievski bound, exhibits a fractal-like 
structure, more aptly described as popcorn-like 
\cite{ilievski2023popcorn}\footnote{
The term popcorn refers to the characteristic dependence on $p_0$. For 
$p_0$ close to,  but not exactly on, the irrational value, 
the Drude weight appears to behave as a 
continuous function, but for rational values of $p_0$, it exhibits 
discontinuities everywhere. This contrasts with a fractal, which is 
continuous everywhere yet not differentiable at any point.
Note that the Drude weight for the case that $p_0$ is exactly on 
irrational numbers remains unclear.}.
Moreover, with the emergence of GHD, the spin Drude weight 
has been formulated in a universal manner in \cite{ilievski2017ballistic, 
doyon2017drude}. The result from GHD reproduces  Zotos's formula. 
Intriguingly, the high-temperature limit of the Drude weight exactly 
coincides with the Prosen-Ilievski bound 
\cite{urichuk2019spin}.  This popcorn structure is 
intrinsically linked to the structures of strings \cite{takahashi1972one} 
that define the thermodynamics of the XXZ model. In fact, recent studies 
suggest that similar popcorn structures also appear in  
higher-spin integrable XXZ models and the sine-Gordon 
model \cite{ilievski2023popcorn, nagy2023thermodynamics,nagy2023thermodynamic2}, 
where these
thermodynamics are governed by the same strings as in the XXZ model.

This paper focuses on the finite-temperature spin Drude weight at zero
  magnetic field for an integrable XXZ chain with arbitrary spin $S$. We 
investigate the model in the critical regime, characterized by the anisotropy 
parameter  $\Delta = \cos(\pi/p_0)$ where 
$p_0 \in \mathbb{Q}_{\ge 2}$. This model was introduced by Kirillov and 
Reshetikhin \cite{kirillov1987exact1} through a ``fusion'' of the XXZ model. 
For $S=1$, the model corresponds to the Zamolodchikov-Fateev model 
\cite{Zamolodchikov1980}. Regarding the spin Drude weight, a suboptimal 
Mazur bound was derived for the high-temperature limit in the $S=1$ case 
\cite{piroli2016quasi}. More recently, the exact high-temperature behavior
 of this quantity has been further clarified for the specific case where 
$S=1$ and $p_0 \in \mathbb{N}_{\ge 2}$ \cite{ilievski2023popcorn}. 

As reported in \cite{kirillov1987exact1,frahm1990integrable}, for 
$S\ge 1$, the anisotropy $\Delta$ must be constrained by the value of 
$S$ to ensure that the Hamiltonian derived from the fusion procedure 
possesses real eigenvalues. Specifically, for an integer spin $S$ greater than $2$, 
the permissible region of $\Delta$ is composed of several separate 
open intervals. This also holds for half odd integer spin $S$ 
greater than $7/2$. (See Table~\ref{interval} in the main text.) 
Importantly, the low-energy properties in each region, for a given $S$, 
are governed by a distinct conformal field theory \cite{frahm1990integrable,
frahm1990finite}. From this perspective, 
it is crucial to delve into how this rich phase structure affects 
the transport properties. 

We investigate the spin Drude weight of this model using 
GHD across a broad range of temperatures and anisotropies. To systematically 
derive the exact expressions at infinite temperature, we derive the TBA equations 
using the quantum transfer matrices (QTMs)
and their functional relations 
referred to as the $T$ and $Y$-systems \cite{kuniba1998continued}. Notably, 
the resulting Drude weight, at both finite and infinite temperatures, 
prominently displays a popcorn structure that depends on $\Delta$ and 
exhibits discontinuities throughout. As expected from the different excitation 
properties in each region explained earlier, the behavior of the Drude weight 
also varies distinctly across regions.

The paper is organized as follows. In the subsequent section, we provide a 
comprehensive overview of the integrable XXZ chain with arbitrary spin, 
including its derivation using the fusion procedure. In section~3, we construct 
the QTMs and derive their functional relations, called the $T$ and $Y$-systems. 
By analyzing their analytic structure, we derive the TBA equations. 
In section~4, using the solutions to the TBA equations, we calculate the
finite-temperature Drude 
weight at zero magnetic field over a wide range of 
anisotropies and spins. Section~5 presents the exact expression for the 
Drude weight in the high-temperature limit. Section~6 is devoted to a 
summary and discussion. Technical details and supplementary information 
necessary for the main text are provided in the Appendices.

\section{Integrable XXZ chain with arbitrary spin}
In this section, we construct the integrable XXZ chain with
arbitrary spin $S=\sigma/2$ ($\sigma\in\mathbb{N}_{+}$),
applying the so-called fusion procedure to the six-vertex model. 
The energy spectrum of the model is derived by means of the 
Bethe ansatz as in \cite{kirillov1987exact1,frahm1990integrable}.
%

\subsection{Higher spin generalization of the six-vertex model}
%
The six-vertex model is a classical statistical model defined on
a two-dimensional square lattice, where the configuration variables 
(classical spins) $\{1,2\}$ are assigned on the edges between each 
vertex of the lattice. Admissible configurations of the model are described by 
the following local states with the Boltzmann weights
\begin{equation}
\includegraphics[width=0.91\textwidth]{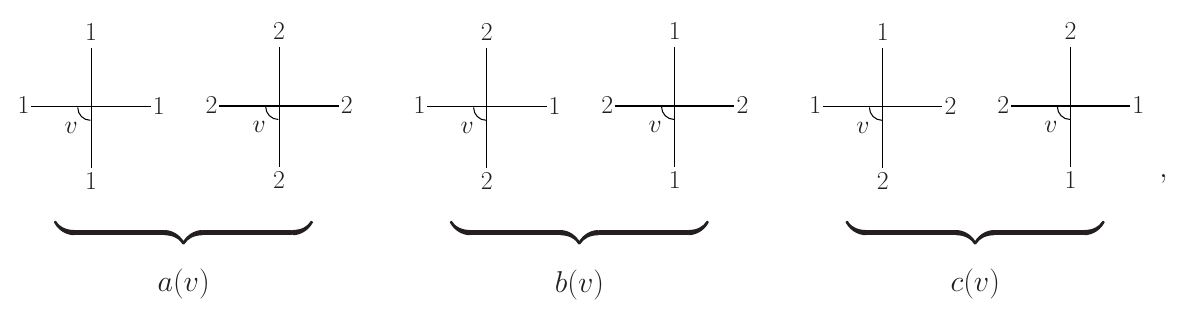}
\label{6-vertex}
\end{equation}
where
\begin{equation}
a(v):=\frac{[v+2]}{[2]},\quad b(v):=\frac{[v]}{[2]}, \quad c(v):=1,\quad
[v]:=\frac{\sin\frac{\theta}{2}v}{\sin{\frac{\theta}{2}}}.
\end{equation}
The weights for all other configurations are set to zero. 
Let $V$ be a two-dimensional vector space spanned 
by an orthonormal basis $\{|1\ket, |2\ket\}$. Then the $R$-matrix
$R(v)\in \End\left(V\otimes V\right)$, whose elements are defined
by
\begin{equation}
\includegraphics[width=0.75\textwidth]{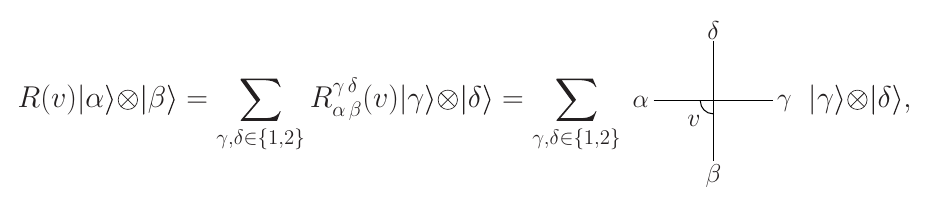}
\end{equation}
satisfies the Yang-Baxter equation (YBE):
\begin{equation}
R_{23}(v-w)R_{13}(u-w)R_{12}(u-v)=
R_{12}(u-v)R_{13}(u-w)R_{23}(v-w).
\label{ybe}
\end{equation}
Here $R_{jk}(v)$ stands for the operator $R(v)$ acting
non-trivially on $V_j\otimes V_k$ as
\begin{equation}
\includegraphics[width=0.27\textwidth]{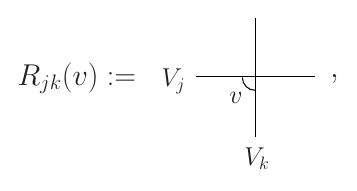}
\end{equation}
where $V_j$ is an indexed copy of $V$.
Graphically, the YBE \eqref{ybe} is depicted as
\begin{equation}
\includegraphics[width=0.65\textwidth]{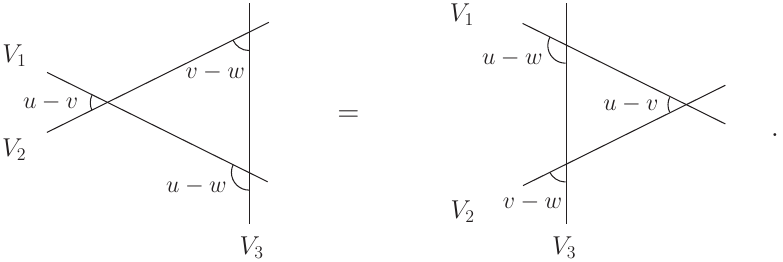}
\label{ybe-graph}
\end{equation}
This model is related to the spin-1/2 XXZ chain as seen later 
(\eqref{Hamiltonian} and \eqref{XXZ}). 

Starting from the six-vertex model, we can construct 
a higher spin version for the six-vertex model by the fusion procedure
\cite{kulish1981yang} (see also \cite{reshetikhin2010lectures} and \cite{fonseca2015higher}).
Let $V^{(m)}\subset V^{\otimes m}=V_1\otimes\dots\otimes V_m$ 
be the $(m+1)$-dimensional irreducible $\mathfrak{sl}_2$-module
spanned by $\{|m;1\ket,\dots,|m;m+1\ket\}$;
\begin{equation}
|m;n\ket:=\frac{1}{\sqrt{\binom{m}{n-1}}}\sum_{1\le j_1<\dots<j_{n-1}\le m}
\sigma^-_{j_1}\cdots \sigma^-_{j_{n-1}} |1\ket_{1}\otimes\cdots\otimes |1\ket_{m} 
\quad (1\le n\le m+1),
\label{state}
\end{equation}
where $\sigma^{-}_j$ is the lowering operator
acting on the $j$th space $V_j$ as
$\sigma_j^{-}|1\ket_j=|2\ket_j$.
Since 
$R_{j j+1}(-2)|m;n\ket=0$
$(1\le j\le m-1)$, a vector $|v\ket\in V^{\otimes m}$ ($m\ge 2$) is an element of
$V^{(m)}$ if and only if
$R_{j j+1}(-2)|v\ket=0$ for all $1\le j\le m-1$.
With this fact in mind, we define the operator $R^{(m,n)}(v)\in 
\End\left(V^{\otimes m}\otimes V^{\otimes n}\right)$ as
\begin{equation}
\includegraphics[width=0.54\textwidth]{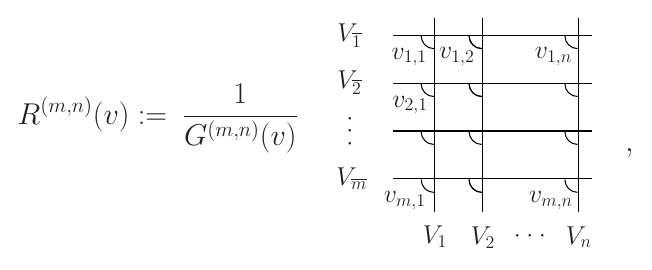}
\end{equation}
where $v_{j,k}:=v-m+n+2(j-k)$ and
\begin{equation}
G^{(m,n)}(v):=\prod_{j=1}^{m}\prod_{k=1}^{n-1} b\left(v-m+n+2(j-k)\right).
\end{equation}
One finds that $V^{(m)}\otimes V^{(n)}$
is an invariant subspace of  $R^{(m,n)}(v)$. In fact, for the case
that
$|w\ket\otimes |x\ket\in V^{(m)}\otimes V^{(n)}$ and $R_{j j+1}$
acts on $|x\ket$,  we find
\begin{equation}
R_{jj+1}(-2)\left(R^{(m,n)}(v)|w\ket\otimes |x\ket\right)
=P_{jj+1}\left.R^{(m,n)}(v)\right|_{\substack{v_{k,j}\leftrightarrow 
v_{k,j+1}\\
(1\le k \le m)}}P_{jj+1} \left(|w\ket\otimes R_{jj+1}(-2)|x\ket\right)=0
\label{invariant}
\end{equation}
for all $1\le j\le n-1$,
where $P_{jj+1}$ is the permutation operator on $V_j\otimes V_{j+1}$.
Note that the first equality  in the above follows from the YBE \eqref{ybe-graph},
and the second equality is the consequence of $|x\ket\in V^{(n)}$.
Eq.~\eqref{invariant} indicates that  $V^{(n)}$ is preserved under the 
action of $R^{(m,n)}(v)$. Similarly, considering the action
of $R_{\overline{j} \overline{j+1}}$ on $|w\ket$, one 
concludes that $R^{(m,n)}(v)|w\ket\otimes|x\ket\in
V^{(m)}\otimes V^{(n)}$. Thus, we define the fusion $R$-matrix 
as  $R^{(m,n)}(v)$ restricted to the subspace
$V^{(m)}\otimes V^{(n)}$, and denote it by
\begin{equation}
\includegraphics[width=0.62\textwidth]{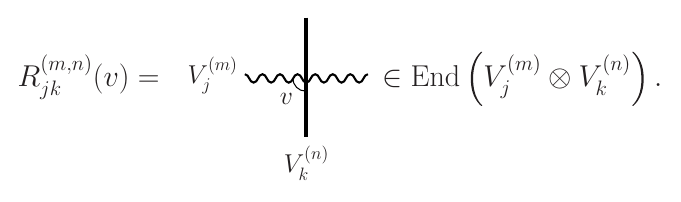}
\end{equation}
The elements of $R^{(m,n)}_{jk}(v)$ are given by
\begin{equation}
\includegraphics[width=1\textwidth]{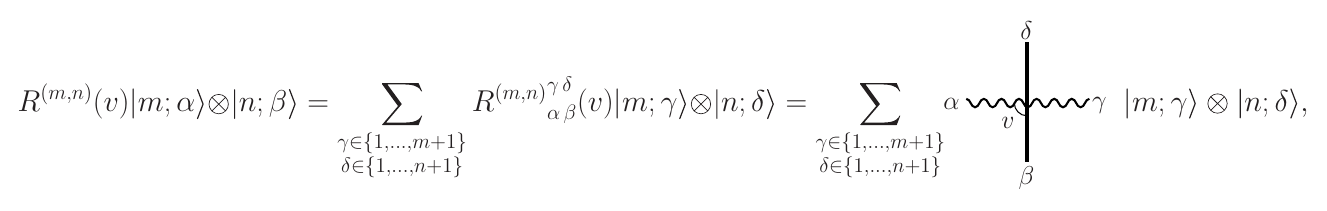}
\end{equation}
where $\alpha,\gamma \in\{1,2,\dots,m+1\}$ and $\beta,\delta \in\{1,2,\dots,n+1\}$.
Explicitly, they read
\begin{equation}
\includegraphics[width=0.95\textwidth]{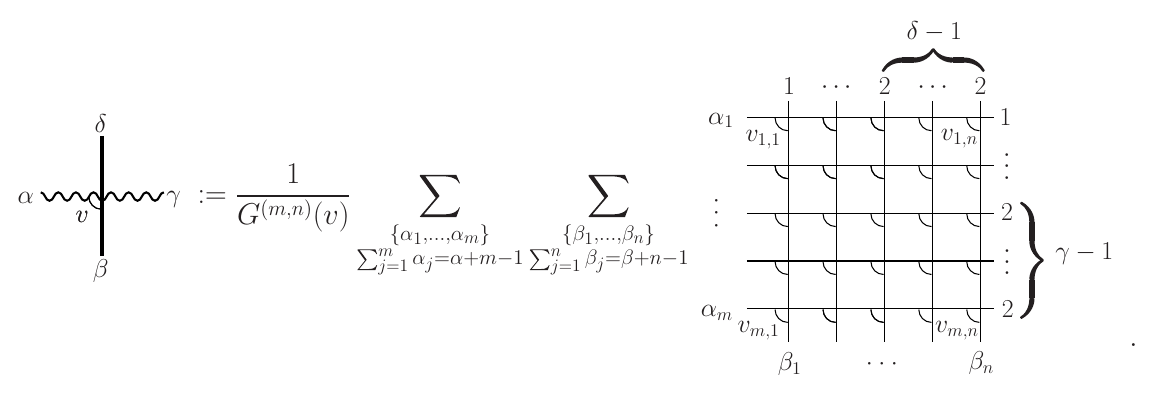}
\label{fusion-Boltzmann}
\end{equation}
In the above, the position of the $\delta-1$ (resp. $\gamma-1$) 
2's on  uppermost (resp. rightmost) edges can be taken arbitrarily,
since  $V^{(m)}$  and $V^{(n)}$ are preserved under the 
action of $R^{(m,n)}(v)$ as explained before.
By repeatedly applying the YBE \eqref{ybe-graph}, one can show that
the fusion $R$-matrix also
satisfies the YBE:
\begin{equation}
R_{23}^{(m,n)}(v-w)R_{13}^{(l,n)}(u-w)R_{12}^{(l,m)}(u-v)=
R_{12}^{(l,m)}(u-v)R_{13}^{(l,n)}(u-w)R_{23}^{(m,n)}(v-w).
\label{ybe-gen}
\end{equation}

Here, let us provide some remarks about the fusion $R$-matrix $R^{(m,n)}(v)$
constructed by \eqref{fusion-Boltzmann}. 
The $R^{(m,n)}(v)$ possesses 
the following symmetries ($C$ and $P$ invariances):
\begin{equation}
R^{(m,n)\,\gamma\delta}_{\quad \, \ \ \ \alpha\beta}(v)=
R^{(m,n)\, m+2-\gamma \ n+2-\delta}_{\quad \,\ \ \  m+2-\alpha \ n+2-\beta}(v),
\quad
R^{(m,m)\,\gamma\delta}_{\quad \, \, \ \ \ \alpha\beta}(v)=
R^{(m,m)\, \delta \gamma}_{\quad \, \, \ \ \  \beta \alpha}(v).
\end{equation}
The charge conservation also holds:
\begin{equation}
R^{(m,n)\,\gamma\delta}_{\quad\, \ \ \ \alpha\beta}(v)=0
\ \
\text{
unless $\alpha+\beta=\gamma+\delta$}.
\label{conserved}
\end{equation}
In general, our construction does not guarantee $T$ invariance. Namely,
\begin{equation}
R^{(m,n)\,\gamma\delta}_{\quad\, \ \ \ \alpha\beta}(v)
\ne R^{(m,n)\,\alpha\beta}_{\quad\, \ \ \ \gamma\delta}(v)
\label{broken-T}
\end{equation}
for certain combinations of $\alpha$, $\beta$, $\gamma$, and $\delta$. 
This is in contrast to the $R$-matrix  used in \cite{kirillov1987exact1, 
frahm1990integrable} (see also \cite{kulish1981quantum,sogo1983new}), where the fusion procedure is designed 
to ensure the $T$ invariance of the $R$-matrix. However, a similarity 
transformation
\begin{equation}
\ms{R}^{(m,n)}_{jk}(v):={F_k^{(n)}}^{-1}{F_j^{(m)}}^{-1} R^{(m,n)}_{jk}(v) F^{(m)}_jF^{(n)}_k
\label{T-invariantR}
\end{equation}
renders a $T$-invariant matrix \cite{kirillov1987exact1,nepomechie2002solving,
frappat2007complete}, which 
coincides with the $R$-matrix in \cite{kirillov1987exact1, frahm1990integrable}. 
Here, $F_j^{(m)}$ acting non-trivially on $V_j^{(m)}$ is a
$v$-independent diagonal matrix. Though the explicit form of
$F_j^{(m)}$ is not necessary in the following arguments, we
list them for a few small values of $m$:
\begin{align}
&F_j^{(1)}=\textrm{diag}\left(1,1\right), \nn \\
&F_j^{(2)}=\textrm{diag}\left(1,\sqrt{2\cos\theta},1\right),\nn \\
&F_j^{(3)}=\textrm{diag}\left(1,\sqrt{1+2\cos2\theta},\sqrt{1+2\cos2\theta},
1\right),\nn \\
&F_j^{(4)}=\textrm{diag}\left(1,\sqrt{\sin 4\theta/\sin\theta},
\sqrt{2 (1+\cos2\theta+\cos4\theta)},\sqrt{\sin 4\theta/\sin\theta},1
\right).
\label{S}
\end{align}

Since the similarity transformation does not change the eigenvalues of the 
transfer matrices, in this paper, we will develop 
the argument based on the $R$-matrix \eqref{fusion-Boltzmann} instead of 
\eqref{T-invariantR}, whose elements 
can be intuitively evaluated graphically.
%
\subsection{Integrable XXZ chain with arbitrary spin}
%
The integrable XXZ model with arbitrary spin $S=\sigma/2$ 
($\sigma\in\mathbb{N}_+$) can be derived by first taking the 
logarithmic derivative of the row transfer matrix of the fusion 
six-vertex model with respect to the spectral parameter 
$v \in \mathbb{C}$ and then applying a suitable transformation to 
the resulting expression to make it a Hermitian operator.
Let $T^{\textrm{R}}_n(v)\in
\End\left({V^{(\sigma)}}^{\otimes L}\right)$
be the row transfer matrix:
\begin{equation}
\includegraphics[width=0.7\textwidth]{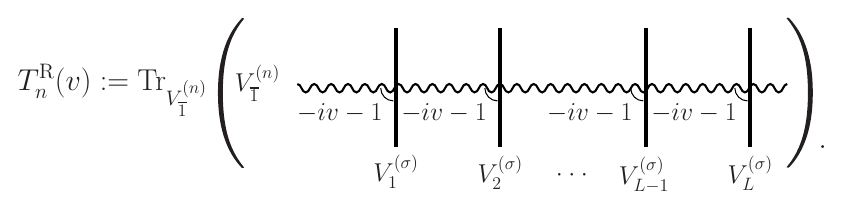}
\label{fusion-T}
\end{equation}
As a consequence of the YBE \eqref{ybe-gen}, the transfer matrices
commute for different values of the spectral parameter and
different fusion levels
\begin{equation}
[T^{\textrm{R}}_n(v),T^{\textrm{R}}_{n'}(v')]=0.
\label{commutativity}
\end{equation}

Let us define the Hamiltonian $\mathcal{H}^{(\sigma)} \in 
\End\left({V^{(\sigma)}}^{\otimes L}\right)$ for the spin-$\sigma$/2 
integrable XXZ model, defined on a 1D periodic lattice with $L$ sites 
(i.e., $V_{L+1}^{(\sigma)} = V_1^{(\sigma)}$). First, take the 
logarithmic derivative of $T^{\textrm{R}}_\sigma(v)$ with respect to 
$v$ to define
\begin{align}
\wt{\mathcal{H}}^{(\sigma)} &:= \wt{\mathcal{H}}^{(\sigma)}_0 - 
h\sum_{j=1}^L\sigma_j^z, \nn \\
\wt{\mathcal{H}}_0^{(\sigma)} &:= \frac{J\sin\theta}{\theta}\left.\del_v
\log T^{\textrm{R}}_{\sigma}(iv+i)\right|_{v=0}
=\frac{J\sin\theta}{\theta}
\sum_{j=1}^L
\frac{P_{jj+1}^{(\sigma)}
{R_{jj+1}^{(\sigma,\sigma)}}'(0)}{\prod_{k=1}^{\sigma}b(2k)},
\label{Hamiltonian}
\end{align}
where $P_{jj+1}^{(\sigma)}$ is the permutation operator on 
$V_{j}^{(\sigma)}\otimes V_{j+1}^{(\sigma)}$, $\sigma_j^z$ is the 
$z$-component of the spin-$\sigma/2$ operator acting on the state 
$|\sigma;n\ket_j \in V_j^{(\sigma)}$ ($1\le n\le \sigma+1$) on the 
$j$th site (see ~\eqref{state}) as
\begin{equation}
\sigma^z_j|\sigma;n\ket_j = (\sigma-2(n-1))|\sigma;n\ket_j,
\label{sigma-z}
\end{equation}
$J\lessgtr 0$ is a coupling constant, and $h$ is the external magnetic 
field. The second equality in \eqref{Hamiltonian} comes from 
\begin{equation}
R^{(\sigma,\sigma)}_{jk}(0)=\left[\prod_{l=1}^{\sigma}b(2l)\right]
P_{jk}^{(\sigma)},
\label{permutation}
\end{equation}
which is a consequence of $R_{jk}(0)=P_{jk}$ and $R_{jk}(v)R_{kj}(-v)=
a(v)a(-v)P_{jk}$. The $z$-component of the total spin
commutes with  $\wt{\mathcal{H}}_0^{(\sigma)}$, i.e.,
\begin{equation}
\left[\wt{\mathcal{H}}_0^{(\sigma)},\sum_{j=1}^L \sigma_j^z\right]=0,
\end{equation}
due to the charge conservation \eqref{conserved}.
It is crucial to point out that the operator $\wt{\mathcal{H}}^{(\sigma)}$ is, in 
general, \textit{non}-Hermitian (more precisely, a real asymmetric matrix) 
except for $\sigma=1$ due to the breaking of 
the $T$ invariance for the $R$-matrix (see \eqref{broken-T}). 
However, a similarity transformation corresponding to \eqref{T-invariantR}
reproduces the Hamiltonian $\mathcal{H}^{(\sigma)}$ given by 
\cite{kirillov1987exact1,frahm1990integrable}:
\begin{equation}
\mathcal{H}^{(\sigma)} := \mathcal{H}^{(\sigma)}_0 - h\sum_{j=1}^L\sigma_j^z, \quad 
\mathcal{H}_0^{(\sigma)} := {F^{(\sigma)}}^{-1}\wt{\mathcal{H}}_0^{(\sigma)}
F^{(\sigma)},
\label{Hamiltonian1}
\end{equation}
where 
\begin{equation}
F^{(\sigma)}:=\prod_{j=1}^L F^{(\sigma)}_{j}\in \End\left({V^{(\sigma)}}^{\otimes L}\right)
\end{equation}
is a diagonal matrix.
In particular, for $S=1/2$ ($\sigma=1$), 
$\wt{\mathcal{H}}^{(1)}=\mathcal{H}^{(1)}$, which is 
Hermitian by definition
of the six-vertex model, 
is the well-known spin-1/2 XXZ model:
\begin{equation}
\mathcal{H}^{(1)}=\wt{\mathcal{H}}^{(1)}
=\frac{J}{4}\sum_{j=1}^L\left[
\sigma_j^x\sigma_{j+1}^x+\sigma_j^y\sigma_{j+1}^y+
\Delta\left(\sigma_j^z\sigma^z_{j+1}+1\right)\right]-h\sum_{j=1}^L\sigma_j^z.
\label{XXZ}
\end{equation}
The Hamiltonian for the higher-spin {\it integrable} XXZ chain 
is not simply obtained by replacing the spin operators 
$\sigma_j^{x,y,z}$ in \eqref{XXZ} with those of a higher spin 
representation. (Note that $[\sigma_j^a,\sigma_j^b]=
2i\epsilon_{abc}\sigma_j^c$, where $a, b, c \in \{x, y, z\}$, 
and $\epsilon_{xyz}=1$ is a completely antisymmetric tensor.)
For instance, for $S=1$ ($\sigma=2$), we can derive
$\wt{\mathcal{H}}^{(2)}_0=\sum_{j=1}^L 
\left(\wt{\mathcal{H}}^{(2)}_0\right)_{jj+1}$
from \eqref{Hamiltonian}:
\begin{align}
\left(\wt{\mathcal{H}}^{(2)}_0\right)_{jj+1}&=
\frac{J}{64\cos\theta}
\biggl\{
4\bs{\sigma}_j\cdot\bs{\sigma}_{j+1}-(\bs{\sigma}_j\cdot\bs{\sigma}_{j+1})^2
\nn \\
&\ \ +
c_1\left[4\sigma_j^z\sigma_{j+1}^z+4(\sigma_j^z)^2+4(\sigma_{j+1}^z)^2-
(\sigma_j^z\sigma_{j+1}^z)^2\right] \nn \\
&\ \
+c_2\left[(\sigma_j^x\sigma_{j+1}^x+\sigma_j^y\sigma_{j+1}^y)\sigma_j^z\sigma_{j+1}^z
\right] 
+c_3\left[\sigma_j^z\sigma_{j+1}^z
(\sigma_j^x\sigma_{j+1}^x+\sigma_j^y\sigma_{j+1}^y)\right] 
+c_4
\biggr\},
\label{non-hermitian}
\end{align}
where
$\sigma_j^{x,y,z}$ are the spin-1
operators,  $\bs{\sigma}_j:=(\sigma_j^x,\sigma_j^y,\sigma_j^z)$
and
\begin{equation}
c_1=-2\sin^2\theta,\quad
c_2=-2\cos2\theta,\quad
c_3=1, \quad c_4=48.
\end{equation}
The operator $\wt{\mathcal{H}}_0^{(2)}$ defined above is obviously non-Hermitian, 
however it can be transformed to the Hermitian operator 
$\mathcal{H}_0^{(2)}$ by
the similarity transformation 
\eqref{Hamiltonian1}, where 
$F^{(2)}_{j}$ is explicitly given by \eqref{S}.
The resultant $\mathcal{H}_0^{(2)}$  is obtained by
replacing
\begin{equation}
c_2,\ c_3\to 2(1-\cos\theta)
\end{equation}
 in \eqref{non-hermitian}, which 
exactly coincides
with the Zamolodchikov-Fateev model \cite{Zamolodchikov1980}.  
Up to authors' knowledge,
the explicit form of the Hamiltonian 
has been found only for the cases of
$\sigma=1$, $\sigma=2$ \cite{Zamolodchikov1980} and $\sigma=3$ 
\cite{bytsko2003integrable}.

The eigenvalues of $T^{\textrm{R}}_n(v)$ \eqref{fusion-T} and
$\wt{\mathcal{H}}^{(\sigma)}$ are preserved under the 
similarity transformation \eqref{T-invariantR} and \eqref{Hamiltonian1}.
Also, the expectation values and the matrix elements of any operator $o$ with respect 
to the eigenstates of $\mathcal{H}^{(\sigma)}$ coincide with  those for the 
corresponding operator 
\begin{equation}
\wt{o}={F^{(\sigma)}}^{-1}o F^{(\sigma)}
\label{ot}
\end{equation}
with respect 
to the eigenstates of $\wt{\mathcal{H}}^{(\sigma)}$. For instance, 
the equilibrium
thermal averages
\begin{equation}
\bra o \ket:=\lim_{L\to\infty}\frac{\Tr(o e^{-\beta\mathcal{H}^{(\sigma)}})}
{\Tr( e^{-\beta\mathcal{H}^{(\sigma)}})},
\qquad
\bra \wt{o} \ket:=\lim_{L\to\infty}\frac{\Tr(\wt{o} e^{-\beta\wt{\mathcal{H}}^{(\sigma)}})}
{\Tr( e^{-\beta\wt{\mathcal{H}}^{(\sigma)}})},
\label{thermal0}
\end{equation}
are exactly the same, i.e.,
\begin{equation}
\bra o \ket=\bra \wt{o} \ket.
\label{thermal1}
\end{equation}
Physically, the spin Drude weight 
discussed in this paper is the higher-spin integrable XXZ chain 
\eqref{Hamiltonian1}, which can be evaluated by using the model
\eqref{Hamiltonian} through the correspondence as in \eqref{thermal1}.

The eigenvalues of the row transfer matrix $T^{\textrm{R}}_n(v)$
can be obtained by the analytic Bethe ansatz \cite{reshetikhin1983functional},
which is a shortcut procedure that allows 
one to find the eigenvalues of a transfer matrix without knowing the details of 
the eigenstates. The resultant eigenvalues (denoted by the same notation 
 $T^{\textrm{R}}_n(v)$)  is written in the 
dressed vacuum form (DVF) \cite{kirillov1987exact1}:
\begin{align}
T_n^{\textrm{R}}(v)&=\sum_{j=1}^{n+1}
\left[\prod_{k=1}^{j-1}a\left(-iv-1-\sigma+n-2(j-k)\right)
\prod_{k=j}^n a\left(-iv-1+\sigma-n-2(j-k)\right)\right]^L\nn \\
&\quad \times
\frac{q\left(v+i(n+1)\right)q\left(v-i(n+1)\right)}
{q\left(v+i(n+1-2j)\right)q\left(v+i(n+3-2j)\right)},
\label{eigen-transfer}
\end{align}
where 
\begin{equation}
q(v):=\prod_{j=1}^M\sh\frac{\theta}{2}(v-\lambda_j).
\end{equation}
Note that $M=0$ (i.e., $q(v)=1$) corresponds to the vacuum state 
\begin{equation}
|0\ket_{\textrm{q}}:=\bigotimes_{j=1}^L|\sigma;1\ket_j,
\label{vac-RTR}
\end{equation}
which is an eigenstate of $T^{\textrm{R}}_n(v)$. The vacuum eigenvalue
is actually calculated by the substitution of
\begin{equation}
{R^{(n,\sigma)}}^{j1}_{j 1}(v)=
\prod_{k=1}^{j-1}a\left(v-\sigma+n-2(j-k)\right)
\prod_{k=j}^n a\left(v+\sigma-n-2(j-k)\right),
\end{equation} 
which easily follows from \eqref{fusion-Boltzmann}, into
${}_\textrm{q}\bra 0|T^{\textrm{R}}_{n}(v)|0\ket_{\textrm{q}}$.
The $M$-particle states are spanned by the basis set
\begin{equation}
\left.\left\{\bigotimes_{j=1}^{L}
|\sigma;M_j\ket_{j}\right|\sum_{j=1}^{L} (M_j-1)=M
\right\}.
\end{equation}
The unknown $\{\lambda_j\}$ is determined by the Bethe ansatz equation (BAE):
\begin{equation}
\left[\frac{\sh\frac{\theta}{2}(\lambda_j+i\sigma)}
{\sh\frac{\theta}{2}(\lambda_j-i\sigma)}\right]^L
=-\frac{q(\lambda_j+2i)}{q(\lambda_j-2i)}.
\label{BAE}
\end{equation}
The BAE is a consequence of the fact that
$\left[G^{(n,\sigma)}(-iv-1)\right]^LT^{\textrm{R}}_n(v)$
(see \eqref{fusion-Boltzmann} and \eqref{fusion-T}) is an entire function on the complex $v$-plane.
The eigenvalues $E^{(\sigma)}$ of the Hamiltonian \eqref{Hamiltonian1} or the operator
\eqref{Hamiltonian} are then 
obtained from the DVF \eqref{eigen-transfer} as
\begin{equation}
E^{(\sigma)}=
-J \sin\theta \sum_{j=1}^M\frac{\sin(\theta\sigma)}{\ch(\theta \lambda_j)-\cos(\theta\sigma)}+
\frac{LJ\sin\theta}{2}\sum_{j=1}^\sigma \cot(\theta j)-h(L\sigma-2M).
\label{energy-spectrum}
\end{equation}
Up to the overall factor and the constant term (the second term), 
the energy spectra agree with those obtained in 
\cite{kirillov1987exact1,frahm1990integrable}.

\begin{table}[tb]
\centering
\begin{tabular}{|c| l l |}
\hline 
$\sigma=2S$ & \multicolumn{2}{c|}{$p_0$} \\
\hline
1 &  $[2,\infty)$ &  \\
3 & $(3,\infty)$& \\
5 & $(5,\infty)$& \\
7 & $(\frac{5}{2},3)$ & $(7,\infty)$\\
9 & $(3,\frac{7}{2})$ & $(9,\infty)$\\
\hline 
\end{tabular}
\hspace*{1cm}
\begin{tabular}{|c| l l l l l |}
\hline
$\sigma=2S$ & \multicolumn{5}{c|}{$p_0$}\\
\hline
2 &  $(2,\infty)$ & & & &  \\
4 & $(2,3)$ & $(4,\infty)$ & & &  \\
6 & $(2,\frac{5}{2})$ & $(3,4)$ & $(6,\infty)$ & & \\
8 & $(2,\frac{7}{3})$ & $(4,5)$ & $(8,\infty)$ & & \\
10 & $(2,\frac{9}{4})$ & $(\frac{8}{3},3)$ & $(\frac{7}{2},4)$ 
& $(5,6)$ & $(10,\infty)$\\
\hline
\end{tabular}
\caption{Intervals of $p_0=\pi/\theta$ \eqref{critical} fulfilling \eqref{restriction1} 
\cite{frahm1990integrable}.}
\label{interval}
\end{table}

In this paper, we consider the case where the model 
is in the critical regime, i.e., 
\begin{align}
&0\le\Delta:=\cos\theta< 1 \quad (0< \theta \le \pi/2),\nn \\
&p_0:=\frac{\pi}{\theta} \quad (p_0\ge 2).
\label{critical}
\end{align}
In this case, it should be noted that for a given $\sigma$, except for $\sigma=1$, 
the anisotropy parameter 
$\Delta$ cannot necessarily take any value in the range $0\le\Delta<1$.
Namely, for a given $\sigma$, only those values of the parameter  $\theta$ 
that satisfy the condition 
\begin{align}
&v(\sigma+1)\sin(\theta j)\sin\left(\theta(\sigma+1-j)\right)>0\ \
\text{for all $j\in\{1,2,\dots,\sigma\}$}.
\label{restriction1}
\end{align}
Here $v(\sigma+1)$ is the spin parity, where $v(n)$ 
is defined as
\begin{equation}
v(n):=\exp\left(i\pi
\left\lfloor \frac{\theta}{\pi}(n-1)\right\rfloor\right)
=\exp\left(i\pi
\left\lfloor \frac{n-1}{p_0}\right\rfloor\right),
\label{spin-parity}
\end{equation}
and 
$\lfloor x\rfloor$ denotes the largest integer less than or equal 
to $x$. See Table~\ref{interval} for the admissible region
of the parameter $p_0$ for several given $\sigma$.

Originally, the relationship \eqref{restriction1} 
between the spin $\sigma/2$ and the anisotropy parameter
 $\Delta=\cos\theta$
was found by Kirillov and Reshetikhin in 
\cite{kirillov1985,kirillov1988classification}.
They discovered this by considering the conditions for 
the existence of string-type solutions to the BAE 
\eqref{BAE}, which are given by the form
\begin{align}
\lambda_{\mu,k}^{n_j}&\equiv \lambda_{\mu}^{n_j}
+i(n_j+1-2k)+i\frac{1- v_j v(\sigma+1)}{2}p_0+O(e^{-\delta L})
\ \ \textrm{mod} \ 2p_0 i 
\nn \\
&\equiv  \lambda_{\mu}^{n_j}
+i(n_j+1-2k)+i(w_j+\wt{w}_{j_\sigma+1})p_0+O(e^{-\delta L})
\ \ \textrm{mod} \ 2p_0 i.
\label{string}
\end{align}
Here, $ \delta > 0 $, $ n_j \in \mathbb{N}_+ $ is the admissible string 
length known as the Takahashi-Suzuki (TS) number defined by 
\eqref{TS-numbers}, and $ \lambda_{\mu}^k \in \mathbb{R} $ ($ \mu \in \mathbb{N}_+ $, 
$ k \in \{1,2,\dots,n_j\} $) is the center of the string. The integers 
$ v_j\in \{1,-1\} $ and $ v(\sigma+1)\in \{1,-1\} $ are, respectively, 
the string parity defined as 
\eqref{string-parity} and the spin parity \eqref{spin-parity}. These 
can also be rewritten using the sequences $ w_j $ and $ \wt{w}_j $ 
defined as \eqref{def-w} as $ v_j=(-1)^{w_j} $ and $ v(\sigma+1) 
= (-1)^{\wt{w}_{j_\sigma+1}} $, where $ j_\sigma $ is later defined 
as in \eqref{jsigma}. The condition \eqref{restriction1} is equivalent to 
the requirement that the admissible string lengths $\{n_j\}$ must contain $\sigma+1$:
\begin{equation}
\sigma+1\in \{n_j\}.
\label{S-restriction}
\end{equation}
(To be more precise, this condition is valid when $p_0$ is an irrational 
number. On the other hand, when $p_0$ is a rational number, the 
condition becomes \eqref{restriction2}. See also Appendix~\ref{TS2} for details.)
Subsequently, Frahm, Yu, and Fowler found that the conditions in \eqref{restriction1} are
 simply those necessary for the Hamiltonian to possess real 
eigenvalues \cite{frahm1990integrable}.
Clearly, for $\sigma=1$, the arbitrary $\theta\in(0,\pi/2]$ 
satisfies \eqref{restriction1}. 
On the other hand, for a given $\sigma \ge 2$,
\eqref{restriction1} is fulfilled by some open intervals for $\theta$
(see Table~\ref{interval} for an example and  \cite{frahm1990integrable} for details).

The ground state, low-lying excitations, and finite-temperature properties 
of the model for arbitrary $(\sigma,\theta)$ satisfying \eqref{restriction1} 
have been investigated in \cite{kirillov1987exact1,kirillov1987exact2}. 
This was done by analyzing the thermodynamic Bethe ansatz (TBA) equations 
derived in \cite{kirillov1987exact1}. Specifically, $f$, representing the 
free energy per site, is given by
\begin{equation}
f=-\frac{T}{2\pi} 
\sum_j \int_{-\infty}^{\infty} dv\, \xi_jp'_{j}(v)\log\left(1+\eta^{-1}_j(v)\right),
\label{free-xxz}
\end{equation}
where $p_j(v)$ is the bare momentum, explicitly defined in 
\cite{kirillov1987exact1,frahm1990integrable}. Additionally, 
\begin{equation}
\xi_j:=\sgn(q_j)
\label{sgn}
\end{equation}
 is a sign factor, with $q_j$ defined in \eqref{def-q}.
The function $ \eta_j(v) $ is determined by the TBA equations
\begin{equation}
\log \eta_j(v) = \beta \varepsilon_j(v)+2\beta h n_j + 
\sum_{k} \xi_k K_{jk} * \log\left(1 + \eta^{-1}_k\right)(v),
\label{TBA-xxz}
\end{equation}
where $f*g(v)$ represents the convolution defined by 
$f*g(v) = \int_{-\infty}^{\infty} f(v-x) g(x) \, dx$, and $ \varepsilon_j(v) $ 
is the bare energy defined as 
\begin{equation}
\varepsilon_j(v) := \frac{1}{\epsilon} p'_j(v), \quad \epsilon =
-\frac{\theta}{J\sin\theta}.
\end{equation}
For the explicit form of the integration kernel $ K_{jk}(v) $ in \eqref{TBA-xxz}, 
see \cite{kirillov1987exact1,frahm1990integrable}.

The conformal properties have also 
been studied in \cite{johannesson1988central,johannesson1988universality,
frahm1990integrable,frahm1990finite}. 
Characteristically, for some intervals $(\sigma, \theta)$ satisfying
\eqref{restriction1}, the ground state is described by  multiple types
of string solutions. Thus, there are multiple branches of low-lying excitations 
with different Fermi velocities within these intervals, and as a result, the free 
energy per site $f$ at low temperatures $T$ behaves like
\begin{equation}
f=e_0-\frac{\pi T^2}{6}\sum_{j}\frac{c_j}{v^{\textrm{f}}_j}+o(T^2),
\end{equation}
where $e_0$ is the ground state energy per 
site, $v^{\textrm{f}}_j$ is the Fermi velocity
for the $n_j$-string describing the ground state,
and $c_j$ is the central charge, which is, in general, 
given by the form of $c_j=3k_j/(k_j+2)$ ($k_j\in\mathbb{N}_+$).
The integer $k_j$ is determined through a rather complicated 
procedure in \cite{kirillov1987exact2}. 
For example, for $p_0$ in one of the intervals 
$k<p_0<k+k/(\sigma-k)$ ($k\in\{1,2,\cdots,\sigma\}$),
the ground state in this interval is described by
a single type of string, and then the central charge is 
given by $c=3k/(k+2)$ \cite{frahm1990integrable}.
\section{QTM and TBA}
Let us formulate the model at finite temperatures. The bulk thermodynamic 
quantities are described by the TBA equations, a set of coupled non-linear 
integral equations. Concerning transport properties, the Drude weights are 
universally given by using the solutions to the TBA equations 
\cite{ilievski2017ballistic,doyon2017drude}. The TBA equations 
for the present model have already been established in \cite{kirillov1987exact1},
whose explicit form is given by \eqref{TBA-xxz}. 
In this section, however, we revisit the equations using the quantum transfer 
matrix (QTM) approach \cite{suzuki1976relationship, suzuki1985transfer, suzuki1987st,
koma1987thermal, suzuki1990new, takahashi1991correlation,
destri1992new, klumper1993thermodynamics,klumper1992free} (see also books
\cite{takahashi1999thermodynamics,essler2005one,
vsamaj2013introduction}), enabling us to develop a systematic analytical 
formulation for the Drude weight at high temperatures.
%
\subsection{QTM}
\begin{figure}
\centering
\includegraphics[width=0.81\textwidth]{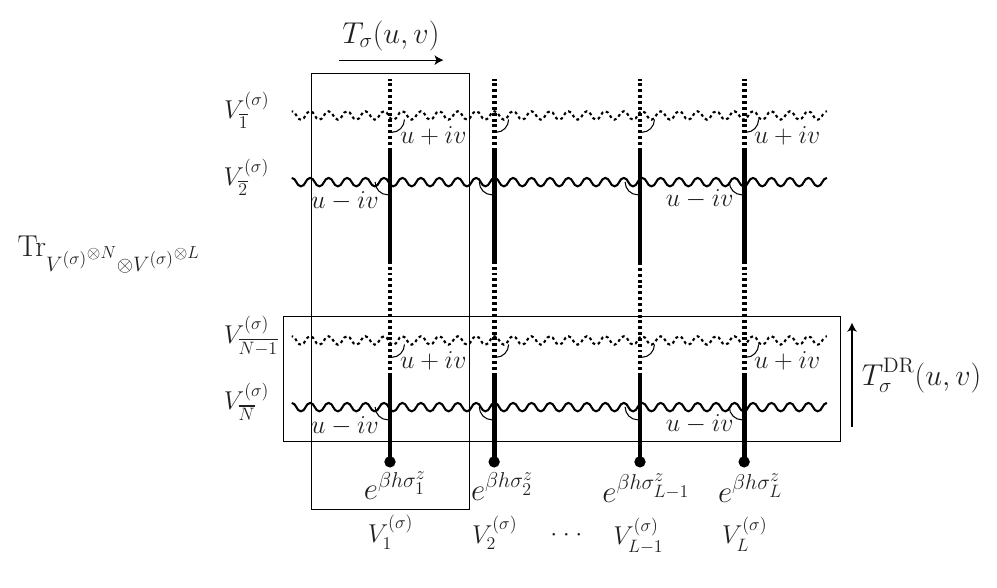}
\caption{The QTM $T_{\sigma}(u,v)$ 
and the double-row transfer matrix $T^{\textrm{DR}}_{\sigma}(u,v)$
defined on the $N\times L$ square lattice, where $N\in 2\mathbb{N}_+$ is 
the Trotter number.}
\label{2D}
\end{figure}

To consider the thermodynamics of the original quantum system, let us go back to
the two-dimensional classical system, i.e., the higher-spin
generalization of the six-vertex model whose local Boltzmann weights are defined as
\eqref{fusion-Boltzmann}. On the $N\times L$ square lattice, we define the following 
double-row transfer matrix $T^{\textrm{DR}}_\sigma(u,v)\in 
\End\left({V^{(\sigma)}}^{\otimes L}\right)$:
\begin{equation}
\includegraphics[width=0.86\textwidth]{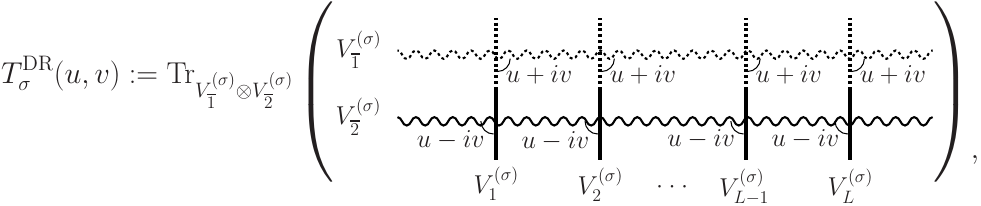} 
\label{doubule-T}
\end{equation}
where the matrix depicted by the dashed line (we denote it by $\overline{R}^{(m,n)}(v)$)
\begin{equation}
\includegraphics[width=0.64\textwidth]{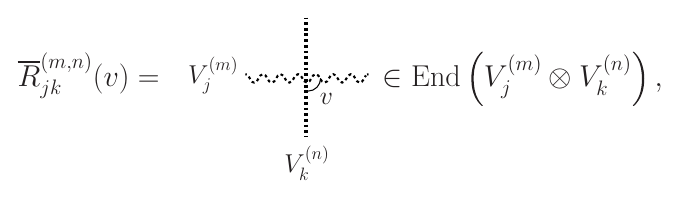}
\label{fusion-R2}
\end{equation}
is defined by the elements
\begin{equation}
\includegraphics[width=0.88\textwidth]{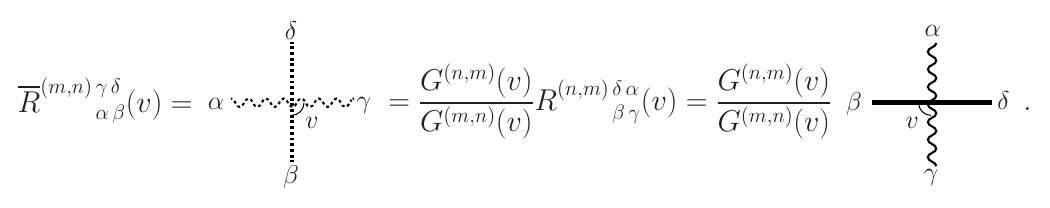}
\label{fusion-Boltzmann2}
\end{equation}
Using \eqref{Hamiltonian}, \eqref{permutation}, \eqref{fusion-R2} and \eqref{fusion-Boltzmann2},
$T^{\textrm{DR}}_{\sigma}(u,0)$ can be expanded with respect to $u$ as
\begin{equation}
T^{\textrm{DR}}_{\sigma}(u,0)=\left[\prod_{j=1}^{\sigma}b(2j)\right]^{2L}
\left(1+\frac{2\theta u}{J\sin\theta}\wt{\mathcal{H}}_0^{(\sigma)}\right)+O(u^2),
\label{T-expand}
\end{equation}
which gives the partition function $Z$. That is, by setting
\begin{equation}
u_N:=-\frac{\beta J\sin\theta}{\theta N}, 
\label{un}
\end{equation}
the free energy per site $f$ can be
formally expressed as
\begin{align}
f&=- \lim_{L\to\infty} \frac{1}{L\beta}\log Z, \nn \\
Z&=\Tr_{{V^{(\sigma)}}^{\otimes L}}\exp\left[-\beta\mathcal{H}^{(\sigma)}\right]\nn \\
&=\Tr_{{V^{(\sigma)}}^{\otimes L}}\exp\left[-\beta\wt{\mathcal{H}}^{(\sigma)}\right]\nn \\
&=\lim_{N\to\infty}\Tr_{{V^{(\sigma)}}^{\otimes L}}
\frac{T^{\textrm{DR}}_{\sigma}(u_N,0)^{N/2}}{\prod_{j=1}^{\sigma}b(2j)^{LN}}
e^{\beta h\sum_{j=1}^L\sigma_j^z}.
\label{free}
\end{align}
Here $N$ is the Trotter number $N\in2\mathbb{N}_+$, and $\beta$ is the reciprocal 
temperature: $\beta:=1/T$. To actually evaluate \eqref{free}, we now introduce 
the QTM $T_n(u,v)\in \End \left({V^{(\sigma)}}^{\otimes N}\right)$:
\begin{equation}
\includegraphics[width=0.73\textwidth]{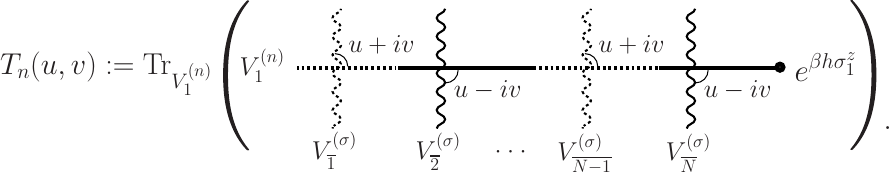}
\label{QTM}
\end{equation}
See also Fig.~\ref{2D}. Note that the operator $\sigma_1^z$ 
in \eqref{QTM} is the $z$-component of the
spin-$n/2$ operator acting on $V_1^{(n)}$ (see \eqref{sigma-z} for the definition
of $\sigma_j^z$).
The free energy \eqref{free} is then given by the largest
eigenvalue of $T_\sigma(u_N,0)$, denoted by $\Lambda(u_N,0)$, as
\begin{equation}
f=-\lim_{L\to\infty}\lim_{N\to\infty}\frac{1}{L\beta}
\log \Tr_{{V^{(\sigma)}}^{\otimes N}}
\left[\frac{T_{\sigma}(u_N,0)}{\prod_{j=1}^{\sigma}b(2j)^N}\right]^L=
-\lim_{N\to\infty}\frac{1}{\beta}\log\left[
\frac{\Lambda(u_N,0)}{\prod_{j=1}^\sigma b(2j)^N}
\right].
\label{free-largest}
\end{equation}
Here we have used the exchangeability of the two 
limits in the first equality 
as proved in 
\cite{suzuki1985transfer,suzuki1987st} and the fact that 
there exists a finite gap between the largest eigenvalue 
$\Lambda(u_N,0)$ and the subleading eigenvalues.

Applying the YBE \eqref{ybe-gen}, we can show that the QTMs commute with each other for 
different values of $v$ and different levels of $n$, as long as the parameter 
$u_N$ is the same:
\begin{equation}
[T_n(u_N,v),T_{n'}(u_N,v')]=0.
\label{commutation}
\end{equation}
The eigenvalues of $T_n(u_N,v)$, also denoted by the same symbol $T_n(u_N,v)$, 
are written in the DVF by the analytic Bethe ansatz.
From \eqref{commutation} and the fact that the algebraic structure of  $V^{(n)}$  in 
\eqref{QTM} is identical to that of the fusion QTM for $\sigma=1$ defined in 
\cite{kuniba1998continued}, the DVF is the same as for $\sigma=1$ except for 
the vacuum part. The vacuum state $|0\ket_{\textrm{a}}$ of $T_n(u_N,v)$ is given by
\begin{equation}
|0\ket_{\textrm{a}}:=\bigotimes_{j=1}^{N/2}\left[
|\sigma;1\ket_{\overline{2j-1}}
\otimes |\sigma;\sigma+1\ket_{\overline{2j}}
\right]
\end{equation}
(cf. \eqref{vac-RTR} for the row transfer matrix), where the state 
$|m;n\ket$ is defined by \eqref{state}. 
On the other hand, the $M$-particle eigenstates of the QTM are spanned by the basis set
\begin{equation}
\left.\left\{\bigotimes_{j=1}^{N/2}
|\sigma;M_{\overline{j}}\ket_{\overline{j}}\right|\sum_{j=1}^{N/2} (M_{\overline{2j-1}}-1)+
\sum_{j=1}^{N/2} (\sigma+1-M_{\overline{2j}})=M
\right\}.
\end{equation}
Noticing \eqref{sigma-z}, and substituting 
the following elements of fusion $R$-matrix
\begin{equation}
{R^{(\sigma,n)}}^{\sigma+1\,j}_{\sigma+1\, j}(v)=\prod_{k=1}^\sigma 
b(v+\sigma-n-2(k-j)),
\
{\overline{R}^{(\sigma,n)}}^{1j}_{1 j}(v)=\prod_{k=1}^\sigma 
a(v-\sigma+n+2(k-j)),
\end{equation}
which can easily be derived from \eqref{fusion-Boltzmann},
one evaluates the vacuum eigenvalue 
${}_\textrm{a}\bra 0|T_{n}(u,v)|0\ket_{\textrm{a}}$.
The resulting DVF (henceforth referred to as the $T$-function) 
is given by
\begin{align}
&T_n(u,v)=\sum_{j=1}^{n+1}\left[\prod_{k=1}^\sigma
\phi\left(v+i(u-n+\sigma-2(k-j))\right)
\phi\left(v-i(u+n-\sigma+2+2(k-j))\right)\right]\nn \\
&\qquad \qquad \quad \times
\frac{Q(v+i(n+1))Q(v-i(n+1))}{Q(v+i(2j-n-1))Q(v+i(2j-n-3))}e^{\beta h(n-2(j-1))},\nn \\
&\phi(v):=\left(\frac{\sh\frac{\theta}{2}v}{\sin\theta}\right)^{N/2},
\quad Q(v):=\prod_{j=1}^M\sh\frac{\theta}{2}(v-\omega_j).
\label{DVF}
\end{align}
Note that the dress part in the above formula is also  equivalent to 
that for the row transfer matrix \eqref{eigen-transfer} after changing 
the variables as
$v\to-v$, and $v_j\to-\omega_j$.
The $M$ unknown numbers $\omega_j$ $(1\le j\le M)$ are determined
by the BAE:
\begin{equation}
\frac{\phi\left(\omega_j+i(u+\sigma+1)\right)\phi\left(\omega_j-i(u-\sigma+1)\right)}
{\phi\left(\omega_j-i(u+\sigma+1)\right)\phi\left(\omega_j+i(u-\sigma+1)\right)}
=
-\frac{Q(\omega_j+2i)}{Q(\omega_j-2i)}e^{2\beta h}.
\label{BAE-QTM}
\end{equation}
%

In the critical regime \eqref{critical} restricted by the condition 
\eqref{restriction1}, the largest eigenvalue
$\Lambda(u_N,0)$ that gives the free energy of the model
(see \eqref{free-largest}) lies in the sector $M=N\sigma/2$. More explicitly, 
the BAE roots $\{\omega_j\}_{j=1}^{M}$ 
associated with the largest eigenvalue $\Lambda(u_N,0)$
of the QTM $T_\sigma(u_N,0)$ are given in the
following form: $\{\omega_j\}=\{\omega_l^k\}$,
\begin{equation}
\omega_l^{k}\equiv \epsilon_l^k+i\delta_l^k+i(\sigma+1-2k)\ \
\textrm{mod} \ 2p_0 i \quad
(1\le k \le \sigma,
\  1\le l \le N/2).
\label{roots}
\end{equation}
Here, $\epsilon_l^k \in \mathbb{R}$ $(1 \leq l \leq N/2)$ are 
symmetrically distributed about the imaginary axis for a given $k$, 
and $\delta_l^k \in \mathbb{R}$. In particular, 
in the high-temperature limit $\beta\to 0$ (i.e.,
$u_N\to 0$), we observe that the BAE roots \eqref{roots} are reduced to 
$\omega_l^{k}=i(\sigma+1-2k)$. In fact, substituting them into \eqref{DVF}
yields 
\begin{equation}
\log\Lambda(u_N=0,v)=\log T_\sigma(u_N=0,v)=
\log\left[\sum_{j=1}^{\sigma+1}\prod_{k=1}^\sigma\phi(v+2i k)\phi(v-2ik)\right],
\label{u=0}
\end{equation}
and hence
\begin{equation}
\frac{\Lambda(u_N=0,0)}{\prod_{j=1}^\sigma b(2j)^N}=\sigma+1.
\label{T-high}
\end{equation} 
Thus,
from  \eqref{free-largest}, $-\lim_{\beta\to0}\beta f$ (i.e, the 
entropy per site at $\beta=0$)
reproduces the physically relevant  result $-\lim_{\beta\to 0}\beta f=\log (\sigma+1)$.

%
\subsection{TBA equation via $T$ and $Y$-functions}
Our first goal is to relate the solutions to the TBA equations
in terms of the $T$-functions \eqref{DVF}.  As shown in \eqref{free-largest},
the thermodynamic quantities of the spin-$\sigma/2$ XXZ chain
are described by the largest eigenvalue of the QTM $T_\sigma(u_N,v)$,
which is given by \eqref{DVF} via the solutions to the BAE \eqref{BAE-QTM}.
However, in general, the Trotter limit $N\to\infty$ in \eqref{free-largest} 
cannot be taken analytically by directly solving the BAE,
except for the high-temperature limit $T=\infty$
(see \eqref{T-high}). To overcome this, we embed
 $T_\sigma(u_N,v)$ into the functional relations ($T$ and $Y$-systems)
that are fulfilled by the $T$-functions and their suitable combinations
called $Y$-functions. By analyzing analytic properties, the $Y$-systems 
are mapped to  the form of nonlinear integral equations, 
in which the Trotter limit can be taken analytically.
These nonlinear integral equations are the TBA equations.

In what follows, the common variable $u$ in the $T$-functions is often 
omitted to simplify the notation. As previously described, the algebraic 
structure of $V^{(n)}$ of the QTM \eqref{QTM} is identical to that of 
the QTM for the spin-1/2 model. Consequently, 
the $T$-systems defined by \eqref{DVF} are the same as those obtained for 
the spin-1/2 model in \cite{kuniba1998continued}:
\begin{equation}
T_{n-1}(v+iy)T_{n-1}(v-iy)=T_{n+y-1}(v)T_{n-y-1}(v)+T_{y-1}(v+in)T_{y-1}(v-in),
\label{T-system1}
\end{equation}
which holds for any $v\in\mathbb{C}$, for any $\omega_j\in\mathbb{C}$ (referred to 
as the off-shell roots, implying that there is no requirement for $\omega_j$ 
to satisfy the BAE \eqref{BAE-QTM}), and  for all  
integers $n,y\in\mathbb{N}_+$ 
such that $n\ge y\ge 1$.
Note here that 
\begin{align}
T_{-1}(v)=0, \quad T_0(v)=\prod_{k=1}^\sigma\phi
\left(v+i\left(u+\sigma-2\left(k-1\right)\right)\right)
\phi
\left(v-i\left(u-\sigma+2k
\right)\right),
\end{align}
and the $T$-functions have the periodicity:
\begin{equation}
T_n(v+2p_0i)=T_n(v).
\label{periodicity}
\end{equation}

Now let us restrict ourselves to the case that $p_0$ defined in \eqref{critical}
 (see also \eqref{restriction1}) is a rational number and express it as the
continued fraction form:
\begin{equation}
p_0=\left[\nu_1,\nu_2,\dots,\nu_\alpha\right]=
\nu_1 + \cfrac{1}{\nu_2 + \cfrac{1}{\cfrac{\ddots}{ \nu_{\alpha-1}+\cfrac{1}{\nu_\alpha}}}},
\label{cfrac}
\end{equation}
where $\alpha\ge 1$, $\nu_2,\dots,\nu_{\alpha-1}\in\mathbb{N}_+$ and $\nu_\alpha\in
\mathbb{N}_{\ge 2}$. We also set $\nu_1\in\mathbb{N}_{\ge 2}$ 
which comes from the condition $p_0\ge 2$ (see \eqref{critical}). Note that
irrational $p_0$ corresponds to $\alpha=\infty$.
In Appendix~\ref{TS}, we list
the sequences required in this section:
$\{m_r \}_{r=1}^\alpha$, $\{p_r\}_{r=1}^\alpha$, $\{y_r\}_{r=-1}^{\alpha}$,
$\{z_r\}_{r=-1}^\alpha$, the TS-numbers $\{n_j\}_{j\ge 1}$ \cite{takahashi1972one}
defining the admissible lengths of the strings \eqref{string}, and 
its slight modification $\{\wt{n}_j\}_{j\ge 1}$ \cite{kuniba1998continued}.
We also introduce the sequence $\{w_j\}_{j\ge 1}$ and $\{\wt{w}_j\}_{j\ge 1}$
in \eqref{def-w}. 
related to the string parity \eqref{string-parity}. All of these sequences are uniquely 
determined for a given value of $p_0$.
In Appendix~\ref{TS2}, we prove that the 
restriction 
between the spin and the anisotropy, given by  \eqref{restriction1},
is equivalent to the condition
\begin{equation}
\sigma+1\in\{\wt{n}_j\}_{j=1}^{m_{\alpha}}.
\label{restriction2}
\end{equation}
Hereafter, we consider the \textit{physical} case where $\sigma$ and $p_0$ 
satisfies \eqref{restriction2}. Correspondingly, let us define the index 
$j_\sigma\in\{1,2,\dots,m_{\alpha}-1\}$ and $r_\sigma\in\{1,2,\dots,\alpha\}$
such that
\begin{equation}
\wt{n}_{j_\sigma+1}=\sigma+1, \quad m_{r_\sigma-1}<j_\sigma+1\le m_{r_\sigma}
\ (m_{r_\sigma-1}\le j_\sigma< m_{r_\sigma}).
\label{jsigma}
\end{equation}
Note that $j_{\sigma=1}=1$ and $r_{\sigma=1}=1$ for any $p_0$ satisfying
\eqref{cfrac}.

With the definitions provided above, we introduce a functional 
relation involving $ T_n(v) $, as described in \cite{kuniba1998continued}:
\begin{equation}
T_{y_\alpha+y_{\alpha-1}-1}(v) = T_{y_{\alpha}-y_{\alpha-1}-1}(v) 
+ 2(-1)^{M z_\alpha}\ch(y_\alpha\beta h)T_{y_{\alpha-1}-1}(v+i y_\alpha),
\label{T-system2}
\end{equation}
where $ M $  denotes the number of BAE roots for \eqref{BAE-QTM}. 
Note that this relation holds true for any off-shell roots, just as in 
equation \eqref{T-system2}. The proof can be straightforwardly completed 
using equation \eqref{p0-yz} and the following periodicities:
\begin{equation}
Q(v+2iy_\alpha)=(-1)^{Mz_\alpha}Q(v), \quad 
\phi(v+2iy_\alpha)=(-1)^{N z_\alpha/2}\phi(v).
\label{periodicity-Q}
\end{equation}
It is important to note that the relation \eqref{T-system2} is valid only 
when $ p_0 $ is an arbitrary rational number as given by \eqref{cfrac}.

The $Y$-functions $\{Y_j(v)\}_{j=1}^{m_\alpha}$ 
are defined by the following combinations
of the $T$-functions. For $m_r\le j<m_{r+1}$ $(0\le r\le \alpha-1$)
and $j\ne m_{\alpha}-1$,
\begin{equation}
Y_j(v):=\frac{T_{\wt{n}_{j+1}+y_r-1}\left(v+i(w_j+\wt{w}_{j_\sigma+1})p_0\right)
              T_{\wt{n}_{j+1}-y_r-1}\left(v+i(w_j+\wt{w}_{j_\sigma+1})p_0\right)}
{T_{y_r-1}\left(v+i\wt{n}_{j+1}+i(w_j+\wt{w}_{j_\sigma+1})p_0\right)
 T_{y_r-1}\left(v-i\wt{n}_{j+1}+i(w_j+\wt{w}_{j_\sigma+1})p_0\right)},
\label{y-function1}
\end{equation}
while for $j=m_{\alpha}-1$ and $j=m_\alpha$,
\begin{align}
&Y_{m_{\alpha}-1}(v):=e^{y_\alpha \beta h}(-1)^{Mz_\alpha} 
\frac{T_{y_\alpha-y_{\alpha-1}-1}
\left(v+i(w_{m_\alpha-1}+\wt{w}_{j_\sigma+1})p_0\right)}
{T_{y_{\alpha-1}-1}
\left(v+iy_\alpha+i(w_{m_\alpha-1}+\wt{w}_{j_\sigma+1})p_0\right)},
\nn \\
&Y_{m_{\alpha}}(v)^{-1}:=e^{-y_\alpha \beta h}(-1)^{Mz_\alpha} 
\frac{T_{y_\alpha-y_{\alpha-1}-1}
\left(v+i(w_{m_\alpha-1}+\wt{w}_{j_\sigma+1})p_0\right)}
{T_{y_{\alpha-1}-1}
\left(v+iy_\alpha+i(w_{m_\alpha-1}+\wt{w}_{j_\sigma+1})p_0\right)}.
\label{y-function2}
\end{align}
We also set $Y_0(v):=0$ and $Y_{-1}(v)=\infty$.
From the $T$-systems \eqref{T-system1}
and \eqref{T-system2}, we have
\begin{equation}
1+Y_j(v)=\frac{T_{\wt{n}_{j+1}-1}\left(v+iy_r+i(w_j+\wt{w}_{j_\sigma+1})p_0\right)
                T_{\wt{n}_{j+1}-1}\left(v-iy_r+i(w_j+\wt{w}_{j_\sigma+1})p_0\right)}
               {T_{y_r-1}\left(v+i\wt{n}_{j+1}+i(w_j+\wt{w}_{j_\sigma+1})p_0\right)
                T_{y_r-1}\left(v-i\wt{n}_{j+1}+i(w_j+\wt{w}_{j_\sigma+1})p_0\right)}
\label{y-function3}
\end{equation}
for  $m_r\le j<m_{r+1}$ $(0\le r\le \alpha-1$)
and $j\ne m_{\alpha}-1$, and 
\begin{align}
&\left(1+Y_{m_{\alpha}-1}(v)\right)\left(1+Y_{m_{\alpha}}(v)^{-1}\right)\nn \\
&\quad =
\frac{T_{y_\alpha-1}\left(v+iy_{\alpha-1}+i(w_{m_{\alpha}-1}+\wt{w}_{j_\sigma+1})p_0\right)
     T_{y_\alpha-1}\left(v-iy_{\alpha-1}+i(w_{m_{\alpha}-1}+\wt{w}_{j_\sigma+1})p_0\right)}
     {T_{y_{\alpha-1}-1}\left(v+iy_\alpha+i(w_{m_\alpha-1}+\wt{w}_{j_\sigma+1})p_0\right)
      T_{y_{\alpha-1}-1}\left(v-iy_\alpha+i(w_{m_\alpha-1}+\wt{w}_{j_\sigma+1})p_0\right)}.
\end{align}
One finds that the above definitions of the $Y$-functions essentially 
the same with those introduced in \cite{kuniba1998continued} for the 
case where $h=0$ and $\sigma=1$. The sole distinction arises from a 
uniform shift in the argument that depends on $\sigma$, namely, 
$i\wt{w}_{j_\sigma+1} p_0$, which reflects the spin parity for the strings 
\eqref{string}. Consequently, the $Y$-functions satisfy the same 
$Y$-systems as given by \cite{kuniba1998continued}:
\begin{align}
&\textrm{For } m_{r-1} \le j \le m_r - 2\,\, ( 1 \le r \le \alpha), \nn \\
&\quad Y_j(v+ip_r)Y_j(v-ip_r) = \left( 1+Y_{j-1}(v)\right)^{1-2\delta_{j,m_{r-1}}}
\left(1+Y_{j+1}(v)\right)
 \left( 1+Y_{j+2}(v)^{-1}\right)^{\delta_{j,m_{\alpha}-2}}
,\nn \\
&\textrm{for } j = m_r -1\,\, (1 \le r \le \alpha -1 ),  \nn \\
&\quad Y_j(v + ip_r + ip_{r+1})Y_j(v + ip_r - ip_{r+1})
Y_j(v - ip_r + ip_{r+1})Y_j(v - ip_r - ip_{r+1}) \nn \\
&\qquad = \left[ \left(1+Y_{j-1}(v+ip_{r+1})\right)
\left(1+Y_{j-1}(v-ip_{r+1})\right) \right]^{1-2\delta_{j,m_{r-1}}}
\left(1+Y_{j+1}(v+ip_r)\right)\nn \\
&\qquad \quad\times \left(1+Y_{j+1}(v-ip_r)\right)
\left(1+Y_j(v + ip_r - ip_{r+1})\right)
\left(1+Y_j(v - ip_r + ip_{r+1})\right),
\nn \\
&\quad Y_{m_\alpha-1}(v+ip_\alpha)Y_{m_\alpha-1}(v-ip_\alpha)
=e^{2 y_\alpha \beta h}(1+Y_{m_\alpha-2}(v)), \nn \\
&\quad \left(Y_{m_\alpha}(v+ip_\alpha)Y_{m_\alpha}(v-ip_\alpha)\right)^{-1}
=e^{-2 y_\alpha \beta h}(1+Y_{m_\alpha-2}(v)),
\label{y-system}
\end{align}
which are valid for any off-shell roots.
One can easily prove them by using 
\eqref{y-function1}, \eqref{y-function2} and \eqref{y-function3}
 in conjunction with property \eqref{periodicity} and 
the definitions of the sequences $\{m_r\}$, $\{y_r\}$, $\{z_r\}$, $\{\wt{n}_j\}$ 
and $\{w_j\}$ given in Appendix~\ref{TS}.
By replacing $Y_{m_{\alpha-1}}(v)$ and $Y_{m_{\alpha}}(v)$ in  \eqref{y-system}
as \begin{equation}
K(v):=e^{-y_\alpha\beta h}Y_{m_{\alpha-1}}(v) = 
e^{y_\alpha \beta h}Y_{m_{\alpha}}(v)^{-1},
\end{equation}
the last two equalities in \eqref{y-system}
reduce to
\begin{equation}
K(v+ip_\alpha)K(v-ip_\alpha)=1+Y_{m_{\alpha}-2}(v).
\end{equation}
Then, by newly introducing
\begin{equation}
1+Y_{m_{\alpha-1}}(v):=(1+e^{y_{\alpha}\beta h}K(v))(1+e^{-y_{\alpha}\beta h}K(v))=K(v)^2+
2\ch(y_{\alpha}\beta h)K(v)+1
\end{equation}
in \eqref{y-system},
we reproduce the $Y$-system originally
introduced in \cite{kuniba1998continued}. 
In this paper, however,
we use \eqref{y-system}, because the physical structure
of the quasi-particles, which is essential in the GHD formalism,
is more clear.

By substituting  a set of the BAE roots (on-shell roots) that gives the 
largest eigenvalue $\Lambda(u_N,0)$ of the QTM  $T_\sigma(u_N,0)$, and 
examining the analytical properties of the $Y$-functions, the $Y$-systems
 are transformed into the following non-linear 
integral equations through the Fourier transform as explained in
Appendix~\ref{NLIE-derivation}.

In the non-linear
integral equations (NLIEs), the Trotter limit $N\to\infty$ can be taken 
analytically. The resultant NLIEs read
\begin{align}
\text{For}\ &m_{r-1}\le j\le m_r-2\ (1\le r\le \alpha),\nn \\
& \log\eta_j(v)=\frac{2\pi \beta J \sin\theta}{\theta}(-1)^{r_\sigma}
    s_{r_\sigma}(v)\delta_{j,j_\sigma}+(1-2\delta_{j,m_{r-1}})s_r*\log(1+\eta_{j-1})(v)
    \nn \\
&\qquad \qquad \quad + s_r*\log(1+\eta_{j+1})
(1+\eta_{j+2}^{-1})^{\delta_{j,m_{\alpha}-2}}(v), \nn \\
\text{for}\ & j=m_r-1 \ (r_{\sigma}\le r\le \alpha-1) ,  \nn \\
& \log \eta_j(v)=\frac{2\pi \beta J \sin\theta}{\theta}(-1)^{r_\sigma}
    s_{r_\sigma}(v)\delta_{j,j_\sigma}+(1-2\delta_{j,m_{r-1}})s_r*\log(1+\eta_{j-1})(v)
    \nn \\ 
&\qquad \qquad \quad +d_r*\log(1+\eta_j)(v) +s_{r+1}*\log(1+\eta_{j+1})(v), \nn\\
\text{for}\ & j=m_r-1 \ (1\le r \le r_{\sigma}-1), \nn \\ 
& \log \eta_j(v)=\frac{2\pi \beta J \sin\theta}{\theta}(-1)^r d_{r,\sigma}(v) 
+(1-2\delta_{j,m_{r-1}})s_r*\log(1+\eta_{j-1})(v)    \nn \\ 
&\qquad \qquad \quad +d_r*\log(1+\eta_j)(v) +s_{r+1}*\log(1+\eta_{j+1})(v), \nn\\
& \log \eta_{m_{\alpha}-1}(v)=y_{\alpha}\beta h+ 
\frac{2\pi \beta J \sin\theta}{\theta}
(-1)^\alpha s_\alpha(v)\delta_{j_\sigma, m_\alpha-1}
+s_\alpha*\log(1+\eta_{m_\alpha-2})(v), \nn \\
& \log \eta_{m_{\alpha}}(v)^{-1}=-y_{\alpha}\beta h+ 
\frac{2\pi \beta J \sin\theta}{\theta}
(-1)^\alpha s_\alpha(v)\delta_{j_\sigma, m_\alpha-1}
+s_\alpha*\log(1+\eta_{m_\alpha-2})(v),
\label{TBA-QTM}
\end{align}
where 
\begin{equation}
\eta_j(v):=\lim_{N\to\infty} Y_j(v) \quad (1\le j\le m_\alpha),
\label{eta-Y}
\end{equation}
and
\begin{align}
&s_r(v):=\frac{1}{4p_r\ch\left(\frac{\pi v}{2p_r}\right)},\quad
d_r(v):=\int_{-\infty}^{\infty}e^{ik v}
\frac{\ch(p_r-p_{r+1})k}{4\pi\ch(p_r k)\ch(p_{r+1}k)}dk,\nn \\
&
d_{r,\sigma}(v):=\int_{-\infty}^{\infty}e^{ik v}
\frac{\ch(\wt{q}_{j_{\sigma}+1}k)}{4\pi\ch(p_r k)\ch(p_{r+1}k)}dk
=\int_{-\infty}^{\infty}e^{ik v}
\frac{\ch\left(q_{j_\sigma}+(-1)^{r_\sigma}p_{r_\sigma}\right)k}{4\pi\ch(p_r k)\ch(p_{r+1}k)}dk,
\end{align}
with $\{q_j\}$ and $\{\wt{q}_j\}$ defined by \eqref{def-q}.

We have numerically confirmed for wide ranges of $\sigma$ and $p_0$
that the solutions $\eta_j(v)$ to the above TBA
equations agree with those obtained by the string hypothesis \cite{kirillov1987exact1}
as given in \eqref{TBA-xxz}.
The free energy \eqref{free-largest} can be expressed in terms of
$\eta_j(v)$ as \eqref{free-TBA}. See Appendix~\ref{free-energy} for detailed derivation
of the free energy.

\section{Drude weight}\label{drude weight}
Now we evaluate the finite-temperature spin Drude weight at zero magnetic field
(i.e., $h=0$).
A general formula describing the Drude weight in terms of quantities determined by 
the generalized TBA equations \eqref{TBA-general} 
is explained in Appendix~\ref{derivation-Drude},
which is based on the arguments in \cite{doyon2017drude, doyon2020lecture}.

In general, the spin Drude weight is defined by the Kubo formula using 
the dynamical correlation functions in the equilibrium state:
\begin{equation}
D_\textrm{s}(\beta) :=
\beta \lim_{t\to\infty}\frac{1}{t}\int_0^t ds
\int dx \bra j_\textrm{s}(s,x)j_\textrm{s}(0,0)\ket^\textrm{c}
= \beta \lim_{t\to\infty}
\int dx \bra j_\textrm{s}(t,x)j_\textrm{s}(0,0)\ket^\textrm{c},
\label{Drude-text}
\end{equation}
where $ j_\textrm{s} $ represents the spin current density, and
$ \bra \cdot \ket^\textrm{c} $ denotes the thermal average 
of the connected correlation function: $\bra o_1 o_2 \ket^\textrm{c}:=
\bra o_1o_2\ket-\bra o_1\ket\bra o_2\ket$ with $\bra \cdot \ket$
defined as \eqref{thermal0}.
Note that the quantity $D_\textrm{s}(\beta)/\beta$ 
is sometimes used as an alternative definition of the Drude weight. 
In such cases, the factor $\beta$ in front of the integral in 
\eqref{Drude-text} becomes unnecessary. However, in this paper, we adhere 
to the definition \eqref{Drude-text} as used in \cite{zotos1999finite, 
urichuk2019spin}.
For a generic expression of the Drude weight, see \eqref{def-Drude}. 
The Drude weight $\mathsf{D}_{00}(\beta)$ in Appendix~\ref{derivation-Drude} 
is denoted as $D_\textrm{s}(\beta)$ in this section.
By projecting the currents onto the space spanned by the 
local or quasi-local conserved charges, the dynamical current-current 
correlation function, defining the Drude weight, can be expressed in terms of 
equal-time current-charge and charge-charge correlation functions 
\cite{doyon2017drude,doyon2020lecture} as given by \eqref{drude-matrix}.

For the present system \eqref{Hamiltonian1}, the local spin current
$(j_\textrm{s})_{kk+1}$ ($1\le k \le L$) is determined by 
\begin{equation}
(j_\textrm{s})_{kk+1}=i\left[(\mathcal{H}^{(\sigma)})_{kk+1},(q_\textrm{s})_{k+1}\right],
\quad 
(q_\textrm{s})_k:=\frac{\sigma -\sigma_k^z}{2}
\label{cc}
\end{equation}
which follows from 
the discrete form of the continuity equation (cf. \eqref{continuity}):
\begin{equation}
\del_t (q_\textrm{s})_k+\left[(j_\textrm{s})_{kk+1}-(j_\textrm{s})_{kk+1}\right]=0,
\end{equation}
where $\del_t(q_\textrm{s})_k=i\left[\mathcal{H}^{(\sigma)},(q_\textrm{s})_k\right]$.
The corresponding quantities $(\wt{q}_\textrm{s})_k$ and $(\wt{j}_\textrm{s})_k$ 
(see \eqref{ot}) are also given by
\begin{align}
&(\wt{q}_\textrm{s})_k=F^{(\sigma)} \left[(q_\textrm{s})_k\right] {F^{(\sigma)}}^{-1}=
(q_\textrm{s})_k,\nn \\
&(\wt{j}_\textrm{s})_{kk+1}=F^{(\sigma)} \left[(j_\textrm{s})_{kk+1}\right]
 {F^{(\sigma)}}^{-1}=i\left[(\wt{\mathcal{H}}^{(\sigma)})_{kk+1},(\wt{q}_\textrm{s})_{k+1}\right]
=
i\left[(\wt{\mathcal{H}}^{(\sigma)})_{kk+1},(q_\textrm{s})_{k+1}\right],
\end{align}
where the second equality in the first equation is due to the fact 
that $F^{(\sigma)}$ and 
$\sigma_k^z$ are diagonal matrices.

The averages of \eqref{cc} over the GGE (see \eqref{cc-GGE}) can 
be expressed in terms of a 
mode decomposition using the $n_j$-string ($1 \le j \le m_\alpha$) as shown 
in \eqref{charge}, \eqref{q-dress}, and \eqref{j-dress}. They reduce to
\begin{align}
&\left\bra 
(q_\textrm{s})_k \right \ket= 
\left\bra
(\wt{q}_\textrm{s})_k \right\ket
 = \sum_{j=1}^{m_{\alpha}} \int_{-\infty}^{\infty} dv \, 
\rho_j(v) h_{\textrm{s};j}(v),\nn \\
&\left\bra 
(j_\textrm{s})_{kk+1} \right\ket= 
\left\bra (\wt{j}_\textrm{s})_{kk+1}
\right\ket= \sum_{j=1}^{m_{\alpha}} \int_{-\infty}^{\infty} dv \, 
\rho_j(v) v_j^{\eff}(v) h_{\textrm{s};j}(v),
\label{current}
\end{align}
where 
\begin{equation}
\left\bra 
(j_\textrm{s})_{kk+1} \right\ket= 
\left\bra (\wt{j}_\textrm{s})_{kk+1}
\right\ket=0
\label{zero-c}
\end{equation} 
must hold, since $\bra \cdot \ket$ and $\bra \wt{\cdot} \ket$
defined as \eqref{thermal0} denote the equilibrium thermal average (cf.
\eqref{cc-GGE}).
In \eqref{current}, the first equalities are derived from 
\eqref{thermal1}, $\rho_j(v)$ is the distribution function associated with the 
$n_j$-string, and
$h_{\textrm{s};j}(v)$ (corresponding to $h_{0;j}(v)$ in 
Appendix~\ref{derivation-Drude}) denotes the bare spin, which is the number 
of magnons composing the $n_j$-string, i.e., $h_{\textrm{s};j}(v)=
h_{0;j}(v)=n_j$. 
Accordingly, by \eqref{TBA-xxz} and \eqref{TBA-general}, 
we identify 
\begin{equation}
\beta^0=2\beta h
\label{beta0}
\end{equation} 
in the present case. 
The effective velocity, $v_j^\eff(v)$, is generally described by 
\eqref{velocity} via the generalized TBA equations \eqref{TBA-general}. 
In the context of our present case, it is defined as
\begin{equation}
v_j^\eff(v) =\frac{\beta^{-1} \del_v \log \eta_j(v)}
{\epsilon \del_\beta \log \eta_j(v)}, \quad \epsilon =-\frac{\theta}{J \sin \theta},
\label{velocity-eta}
\end{equation}
where $\eta_j(v)$ is determined by the standard TBA equations 
\eqref{TBA-QTM}, which are equivalent to \eqref{TBA-xxz}.
As $v_j^{\text{eff}}(v)$ in \eqref{velocity-eta} is an 
odd function while $\rho_j(v)$ and $h_{\textrm{s};j}(v)$ are even functions,
we can easily confirm \eqref{zero-c}.
Then, using the procedure described in \cite{doyon2017drude,doyon2020lecture}, 
which is also summarized in Appendix~\ref{derivation-Drude}, we can express 
the spin Drude weight as
\begin{equation}
D_{\textrm{s}}(\beta)=
\beta \sum_{j=1}^{m_\alpha} \int dv\, \rho_j(v)\left(1-\vartheta_j(v)\right)
\left[v^\eff_j(v)  h_{0;j}^\dr(v)\right]^2.
\label{Drude-spin}
\end{equation}
Here, $h_{0;j}^{\dr}(v)$ represents the dressed spin, which is defined as
\begin{equation}
h_{0;j}^{\dr}(v) = \del_{\beta^0} \log \eta_j(v)=\del_{2\beta h} \log \eta_j(v),
\end{equation}
as indicated in \eqref{beta0} and \eqref{dress}. Moreover, according to the TBA 
equations \eqref{TBA-QTM}, for zero magnetic field ($h=0$), the value of 
$h_{0;j}^\dr(v)$ is
\begin{equation}
h_{0;j}^\dr(v) = \begin{dcases}
\frac{y_\alpha}{2} & \text{if $j = m_\alpha, m_{\alpha-1}$} \\
0 & \text{otherwise}
\end{dcases}.
\label{dress-spin}
\end{equation}
Inserting $ \eta_j(v) :=\rho^\textrm{h}_j(v)/
\rho_j(v) $, 
where $ \rho^\textrm{h}_j(v) $ is the hole density of 
the $ n_j $-string, 
into the fermionic distribution function $ \vartheta_j(v) $ 
(eq.~\eqref{stat-factor})
with the density of states 
$ \rho^\textrm{tot}_j(v) := \rho_j(v) + \rho^\textrm{h}_j(v) $, 
we obtain
\begin{equation}
\rho_j(v)(1-\vartheta_j(v)) = \frac{\rho^{\textrm{tot}}_j(v)}
{(1+\eta_j(v))(1+\eta^{-1}_j(v))}=
 \frac{\epsilon\xi_j \del_\beta \log \eta_j(v)}
{2\pi  (1+\eta_j(v))(1+\eta^{-1}_j(v))}.
\label{stat-factor2}
\end{equation}
Here $\xi_j$ is a sign factor defined in \eqref{sgn}.
To derive the last equality, we have used \eqref{rho-dress}
with $\varepsilon_j^\dr(v)=\del_{\beta^1}\log\eta_j(v)=
\del_{\beta}\log\eta_j(v)$
which follows from \eqref{dress}, \eqref{bare-energy}
and \eqref{beta1}.
By substituting \eqref{velocity-eta}, 
\eqref{dress-spin} and \eqref{stat-factor2} into \eqref{Drude-spin},
and taking into account
\begin{equation}
\xi_{m_{\alpha}}=\sgn(q_{m_{\alpha}})=-\xi_{m_\alpha-1}=
(-1)^{\alpha}
\end{equation}
along with the identity $\eta_{m_\alpha}(v)^{-1}=\eta_{m_{\alpha}-1}(v)$
which holds for $h=0$, one sees that only the last two
$\eta$-functions $\eta_{\alpha-1}$ and $\eta_\alpha$ contribute
to the Drude weight at $h=0$. Explicitly it reads
\begin{equation}
D_\textrm{s}(\beta)=(-1)^\alpha
\frac{\beta y_\alpha^2 J \sin\theta}{4\pi \theta} 
\int_{-\infty}^{\infty}\frac{\left(
\beta^{-1}\del_v\log\eta_{m_\alpha-1}(v)\right)^2}
{\left(1+\eta_{m_\alpha-1}(v)\right)
\left(1+\eta_{m_\alpha-1}^{-1}(v)\right)
\del_\beta\log\eta_{m_\alpha-1}(v)}
dv.
\label{Drude}
\end{equation}
The formula presented above \eqref{Drude} is formally the same as that for 
the spin-1/2 model 
\cite{zotos1999finite, benz2005finite, urichuk2019spin, klumper2019spin}. 
The spin dependency is given through the $\eta$-functions \eqref{TBA-QTM}.
\begin{figure}
\centering
\includegraphics[width=0.52\textwidth]{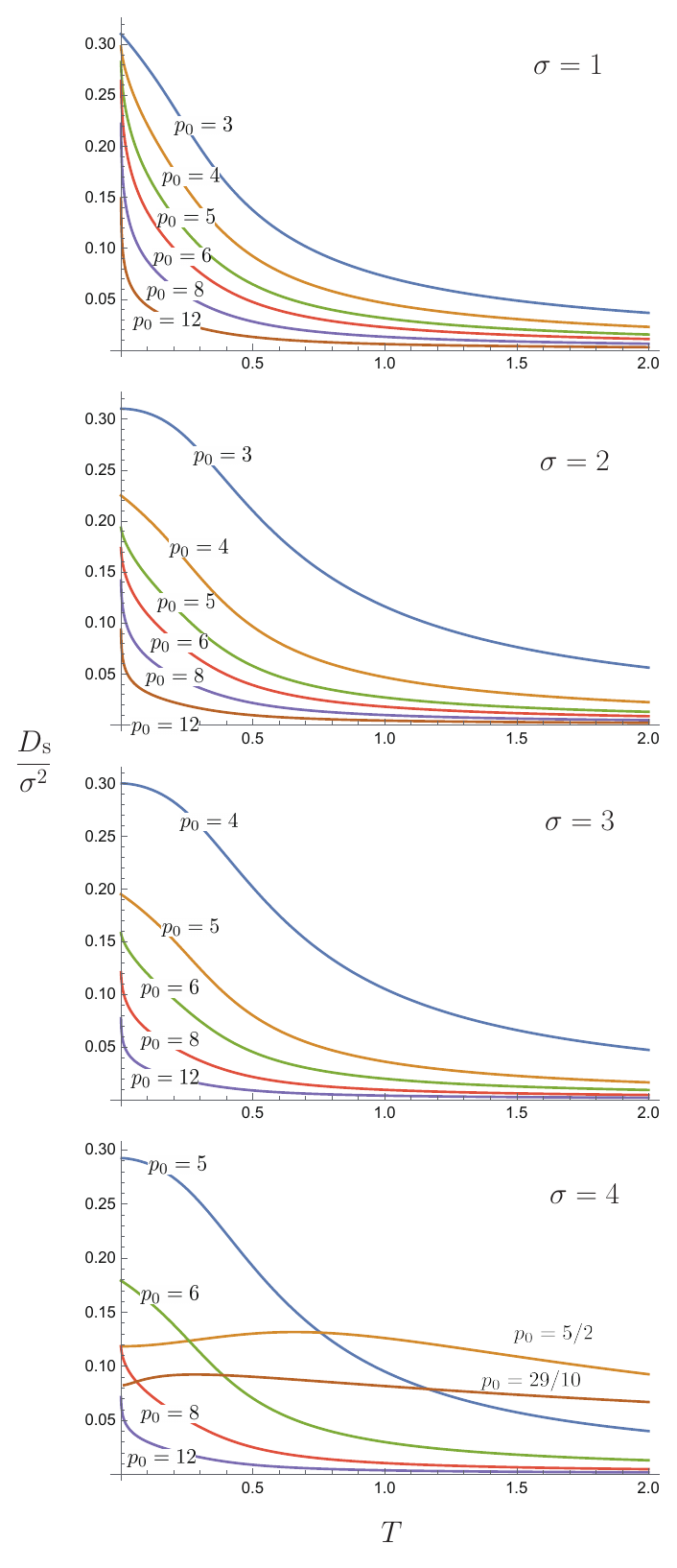}
\caption{
Temperature-dependence of the spin Drude weight (divided by $\sigma^2$)
 $D_\textrm{s}/\sigma^2$ for $J=1$, 
$S = \sigma / 2$ ($\sigma = 1, 2, 3, 4$) and 
various anisotropies $\Delta=\cos(\pi/p_0)$. $D_\textrm{s}/\sigma^2$ 
smoothly varies with respect to $T$ and approaches zero at 
the isotropic limit $p_0 \to \infty$ ($\Delta\to 1$). For $\sigma=4$, 
the admissible region of $p_0$ is separated into two intervals:
 $p_0\in(2,3)$ and $p_0\in(4,\infty)$ as in Table~\ref{interval}. 
The behavior of $D_\textrm{s}/\sigma^2$ in each region distinctly differs, 
reflecting the fact that the structure of the energy spectrum characterized 
by the strings is essentially different in each interval.}
\label{finite-T1}
\end{figure}

\begin{figure}
\centering
\includegraphics[width=0.57\textwidth]{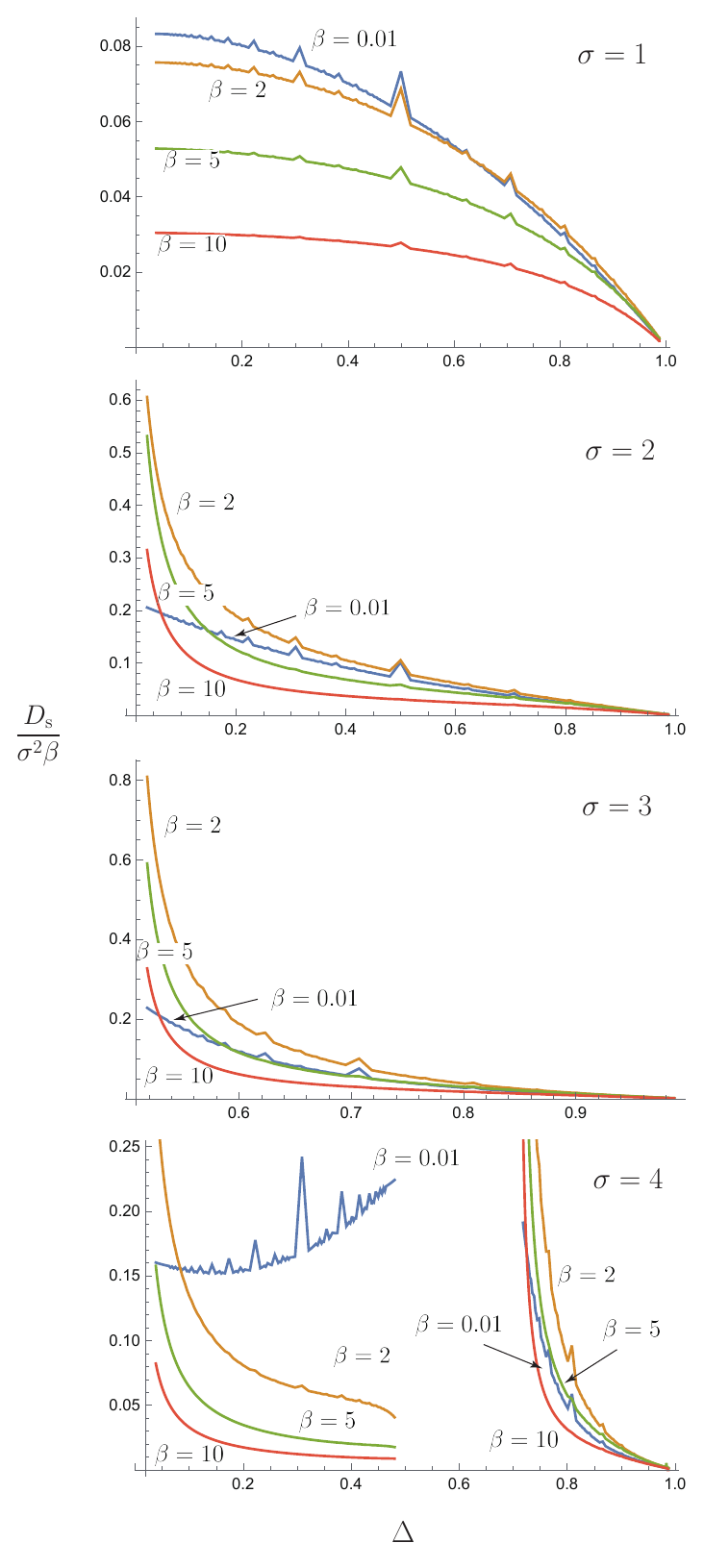}
\caption{Popcorn structure of the Drude weight (divided by temperature
 $\beta=1/T$ and $\sigma^2$) $D_\textrm{s}/(\sigma^2\beta)$ for $J=1$,
$S = \sigma / 2$
 ($\sigma = 1, 2, 3, 4$). $D_\textrm{s}/(\sigma^2\beta)$ exhibits 
discontinuities throughout the anisotropy $\Delta=\cos(\pi/p_0)$.
 The physically admissible region for $\Delta$ depends on $S$. 
Refer to \eqref{restriction1} and Table~\ref{interval} for details.
}
\label{finite-T2}
\end{figure}

We solved the TBA equations numerically to compute the spin Drude weight 
\eqref{Drude}. In Fig.~\ref{finite-T1}, 
the temperature-dependence
of the Drude weight (divided  by $\sigma^2$)  $D_\textrm{s}(\beta)/\sigma^2$
is shown for the cases where the model with 
$\sigma=1,2,3,4$ ($S=\sigma/2$) in the antiferromagnetic regime $J=1$.
 Although the dependencies 
for $\sigma=1$ have already been known in \cite{zotos1999finite, urichuk2019spin}, 
we included them for the sake of comparison. As mentioned earlier, the 
physically admissible region of $\Delta$ is restricted as 
in \eqref{restriction1} (refer to Table~\ref{interval} for concrete examples).
For a given $\Delta$ and $\sigma$, the Drude weight varies smoothly with
the temperature, and approaches zero at the isotropic
limit $p_0 \to \infty$ ($\Delta\to 1$). Notably, when the admissible region of $\Delta$ 
consists of separate intervals (as with $\sigma=4$ in the figure,
where the admissible region of $p_0$ is separated into two intervals:
 $p_0\in(2,3)$ and $p_0\in(4,\infty)$), the behavior 
of the Drude weight differs distinctly in each region. 
As pointed out in \cite{kirillov1987exact1,kirillov1987exact2,frahm1990integrable,
frahm1990finite}, the ground state and the low-lying excitations in each region 
are characterized by different types of strings. Accordingly, the
low-energy properties in each region are described by a different 
conformal field theory \cite{frahm1990finite}. The different behavior
of the Drude weight can be interpreted by these differences in the 
structure of the energy spectrum.

In  Fig.~\ref{finite-T2}, we 
also depict the anisotropy dependence of the Drude weight for $J=1$ divided 
by $\beta=1/T$ and $\sigma^2$, i.e., $D_{\rm s}/(\beta^2\sigma^2)$. Again, 
for comparison, we include the known results \cite{urichuk2019spin}
for $\sigma=1$.
For a given $T$ and $\sigma$, 
the anisotropy dependence of the Drude weight exhibits the characteristic 
popcorn structure. 
The popcorn structure is suppressed with the decrease 
in temperature and will eventually exhibit a smooth curve depending on 
$\Delta$ at the zero-temperature limit.

%
\section{High-temperature limit}
We aim to derive an analytical expression for the high-temperature behavior of 
the Drude weight at $h=0$ \eqref{Drude}, 
namely $\lim_{\beta\to0}D_\textrm{s}(\beta)/\beta$. 
Given the complexity of the TBA equations, we take a different approach by 
adopting and extending the method developed in \cite{urichuk2019spin} for 
the spin-1/2 case. Namely, we evaluate this limit by directly examining 
the high-temperature behavior of the $T$ and $Y$-functions, 
as defined in equations \eqref{DVF}, \eqref{y-function1} and \eqref{y-function2}.

In fact, according to the definition of free energy given by equations \eqref{free} 
through \eqref{T-expand} and \eqref{un}, considering only the case $N=2$ would 
suffice for evaluating contributions of $O(\beta)$. However, in this work, 
we extend the discussion to arbitrary even values of $N$ to make the argument 
more generally applicable. The behavior of $T_n(u_N, v)$ at  $O(\beta)$ 
can be evaluated by inserting the BAE roots, specifically those given by 
\eqref{roots}, up to terms of $O(\beta)$. Since the real parts of 
the roots are symmetrically distributed with respect to the imaginary axis, 
we have $\sum_{l=1}^{N/2}\epsilon_l^k=0$ for $1\le k\le \sigma$. Therefore, 
the $O(\beta)$ contributions from $\epsilon_l^k$ to $T(u_N, v)$ cancel each 
other out. Conversely, the contributions from  $\delta_l^k$ are retained
in the form of the total sum $\delta^k := \sum_{l=1}^{N/2} \delta_l^k$.

The quantity $\delta^k$ is obtained as follows. First,
substituting $u=u_N$ and \eqref{roots} into \eqref{BAE-QTM}, and using the fact 
that $z_l^k:=\epsilon_l^k+i\delta_l^k$ and $u_N$ are both $O(\beta)$, 
we obtain
\begin{equation}
P_k(z_l^k) = 0 \quad (1 \le k \le \sigma, \ 1 \le l \le N/2),
\label{sol}
\end{equation}
where
\begin{equation}
P_k(z):=\prod_{l=1}^{N/2}(z-z_l^{k-1})+\prod_{l=1}^{N/2}(z-z_l^{k+1}),
\label{poly}
\end{equation}
and $z_l^0 := -i u_N$ and $z_l^{\sigma+1}:=i u_N$ ($1 \le l \le N/2$).
Second, due to \eqref{sol}, we can rewrite $P_k(z)$ as
$P_k(z)=2\prod_{l=1}^{N/2}(z-z_l^k)$, 
where the overall constant has been determined by the 
leading coefficient of $z^{N/2}$ in \eqref{poly}. Then, 
by comparing the coefficients of $z^{N/2-1}$, we obtain
$2\sum_lz_l^k=\sum_l(z_l^{k-1}+z_l^{k+1})$,
which leads to $\delta^k=(\delta^{k-1}+\delta^{k+1})/2$.
Here we have used the equality $\sum_{l=1}^{N/2}\epsilon^k_l=0$.
Finally, imposing the condition $\delta^1=-\delta^\sigma$, which comes from the 
symmetrical argument, we arrive at
\begin{equation}
\delta^k:=\sum_{l=1}^{N/2}\delta_l^k=-\frac{Nu_N}{2}\left(1-\frac{2k}{\sigma+1}\right)=
\frac{J\sin\theta}{2\theta}\left(1-\frac{2k}{\sigma+1}\right)\beta.
\label{delta}
\end{equation}
By inserting this equation into \eqref{DVF}, we can explicitly calculate 
the $O(\beta)$ behavior of $\eta_j(v)$ ($1\le j \le m_{\alpha}$) via
the formula \eqref{y-function1}, \eqref{y-function2} and \eqref{eta-Y}, 
and consequently, also determine the $O(\beta)$ behavior of the Drude
weight \eqref{Drude}.

According to \eqref{Drude}, only the function $ Y_{m_{\alpha-1}}(v) $ needs to be 
considered for the Drude weight at $h = 0$. Combining \eqref{y-function2} 
and \eqref{T-system2} at 
$ h=0 $, we obtain
\begin{equation}
1 + Y_{m_{\alpha}-1}(v) = \frac{(-1)^{Mz_\alpha} T_{y_\alpha + y_{\alpha-1} - 1}(v') - T_{y_{\alpha-1} - 1}(v' + i y_\alpha)}{T_{y_{\alpha-1} - 1}(v' + i y_\alpha)},
\end{equation}
where we introduce
\begin{equation}
v' := v + i(w_{m_{\alpha}-1} + \widetilde{w}_{j_\sigma + 1}) p_0,
\label{vp}
\end{equation}
to simplify subsequent notations. Inserting equation \eqref{DVF}, 
decomposing the summation appearing in $ T_{y_\alpha + y_{\alpha-1} - 1}(v') $ 
into two parts as $ \sum_{j=1}^{y_\alpha + y_{\alpha-1}} = \sum_{j=1}^{y_{\alpha-1}} 
+ \sum_{j=y_{\alpha-1}+1}^{y_\alpha+y_{\alpha-1}} $, 
and using the periodicities \eqref{periodicity-Q}, 
one obtains
\begin{align}
&1+Y_{m_{\alpha}-1}(v) = \frac{f(v)}{g(v)},\nn \\ 
&f(v):=\frac{T_{y_{\alpha}-1}(v'+i y_{\alpha-1})}
{Q\left(v'+i(y_\alpha+y_{\alpha-1})\right)
Q\left(v'-i(y_\alpha-y_{\alpha-1})\right)},\nn \\
&g(v):=\frac{T_{y_{\alpha-1}-1}(v'+i y_{\alpha})}
{Q\left(v'+i(y_\alpha+y_{\alpha-1})\right)
Q\left(v'+i(y_\alpha-y_{\alpha-1})\right)}.
\end{align}
Due to the BAE \eqref{BAE-QTM} and periodicities given in 
\eqref{periodicity-Q}, $f(v)$ is an entire function and is 
bounded. Therefore, by Liouville's theorem, 
$f(v)$ must be a constant which is calculated by taking the limit
$v\to\infty$. Using $\sum_{j=1}^M \omega_j=0$, it explicitly reads
\begin{equation}
f(v)=\frac{y_{\alpha}}{(\sin\theta)^{2M}}.
\end{equation}
Taking into account \eqref{un} and \eqref{delta}, and 
expanding $g(v)$ up to $O(\beta)$, we obtain
\begin{align}
&1+\eta_{m_\alpha-1}(v)=\lim_{N\to\infty}(1+ Y_{m_\alpha-1}(v))\nn \\
&\qquad=\frac{y_\alpha}{y_{\alpha-1}}
\Bigg\{1-i\frac{\beta J\sin\theta}{2y_{\alpha-1}}
\sum_{k=1}^{\sigma}k\left(1-\frac{k}{\sigma+1}\right) 
 \left[
\coth\frac{\theta}{2}\left(v'-i(y_\alpha+y_{\alpha-1}+\sigma+1-2k)\right)\right.
\nn \\
&\qquad \qquad \qquad \quad \left.-
\coth\frac{\theta}{2}\left(v'-i(y_\alpha-y_{\alpha-1}+\sigma+1-2k)\right)
\right]\Biggr\}+O(\beta^2).
\end{align}
Substituting \eqref{vp} in the above and
using 
\begin{equation}
y_{\alpha}\pm y_{\alpha-1}+w_{m_{\alpha-1}}p_0 
\equiv \pm(-1)^{\alpha-1}p_\alpha+p_0
\ \
\textrm{mod} \ 2p_0
\end{equation}
lead to
\begin{align}
&1+\eta_{m_\alpha-1}(v)\nn \\
&\  =\frac{y_\alpha}{y_{\alpha-1}}
\left[1-\frac{2\pi\beta J\sin\theta}{\theta y_{\alpha-1}}
\sum_{k=1}^\sigma k\left(1-\frac{k}{\sigma+1}\right)
a_{m_{\alpha}-1}\left(v-i(\sigma+1-2k+\wt{w}_{j_\sigma+1}p_0)
\right)
\right]+O(\beta^2),
\label{y-expansion}
\end{align}
where the function $a_j(v)$ is defined as
\begin{equation}
a_j(v):=\frac{\theta}{2\pi}\frac{\sin(\theta q_j)}
{\ch(\theta v)+\cos(\theta q_j)}.
\end{equation}
Finally, the insertion of \eqref{y-expansion} into \eqref{Drude} yields
\begin{align}
\lim_{\beta\to 0}\beta^{-1}D_\textrm{s}(\beta)&=
\frac{(-1)^{\alpha+1} y_\alpha}{2}\left(\frac{J\sin\theta}{\theta}\right)^2 \nn \\
&\ \times
\int_{-\infty}^{\infty}
\frac{\left[\sum_{k=1}^\sigma k\left(1-\frac{k}{\sigma+1}\right) 
a'_{m_\alpha-1}\left(v-i(\sigma+1-2k+\wt{w}_{j_\sigma+1}p_0)\right)\right]^2}
{\sum_{k=1}^\sigma k\left(1-\frac{k}{\sigma+1}\right) 
a_{m_\alpha-1}\left(v-i(\sigma+1-2k+\wt{w}_{j_\sigma+1}p_0)\right)} dv.
\label{Drude-high-T}
\end{align}

For $S=1/2$  ($ \sigma = 1 $), \eqref{Drude-high-T} 
coincides with the result obtained in \cite{urichuk2019spin} up to an overall 
factor, which is due to the difference in the definition of the coupling 
constant $ J $. In this case, the integral can be explicitly computed 
in \cite{urichuk2019spin}, yielding:
\begin{equation}
\left. \lim_{{\beta \to 0}} D_{\textrm{s}}(\beta) \right|_{\sigma=1} = 
\frac{(J \sin \theta)^2}{8 \sin^2 \frac{\pi}{y_\alpha}} 
\left( 1 - \frac{y_\alpha}{2 \pi} \sin \frac{2 \pi}{y_\alpha} \right).
\end{equation}
This result is consistent with the Prosen-Ilievski bound 
\cite{prosen2013families} derived
via the construction of quasi-local conserved charges 
\cite{prosen2014quasilocal,pereira2014exactly}. 
On the other hand, for spins higher than $1/2$, the integrand 
becomes more complicated. It appears difficult to calculate it explicitly
 when $ p_0 $ is a generic rational number. For $S=1$
 ($ \sigma = 2 $), the integral can be explicitly calculated 
for $ \alpha = 1 $, i.e., $ p_0 = \pi/\nu_1 $. The result is
\begin{equation}
\left. \lim_{{\beta \to 0}} D_{\textrm{s}}(\beta) \right|_{\sigma=2,\alpha=1} =
 \frac{J^2}{12 \cos^2 \frac{\pi}{\nu_1}} \left( 1 - \frac{\nu_1}{4 \pi}
 \sin \frac{4 \pi}{\nu_1} \right),
\end{equation}
which exactly coincides with the result recently derived in \cite{ilievski2023popcorn}.

\begin{figure}
\centering
\includegraphics[width=1\textwidth]{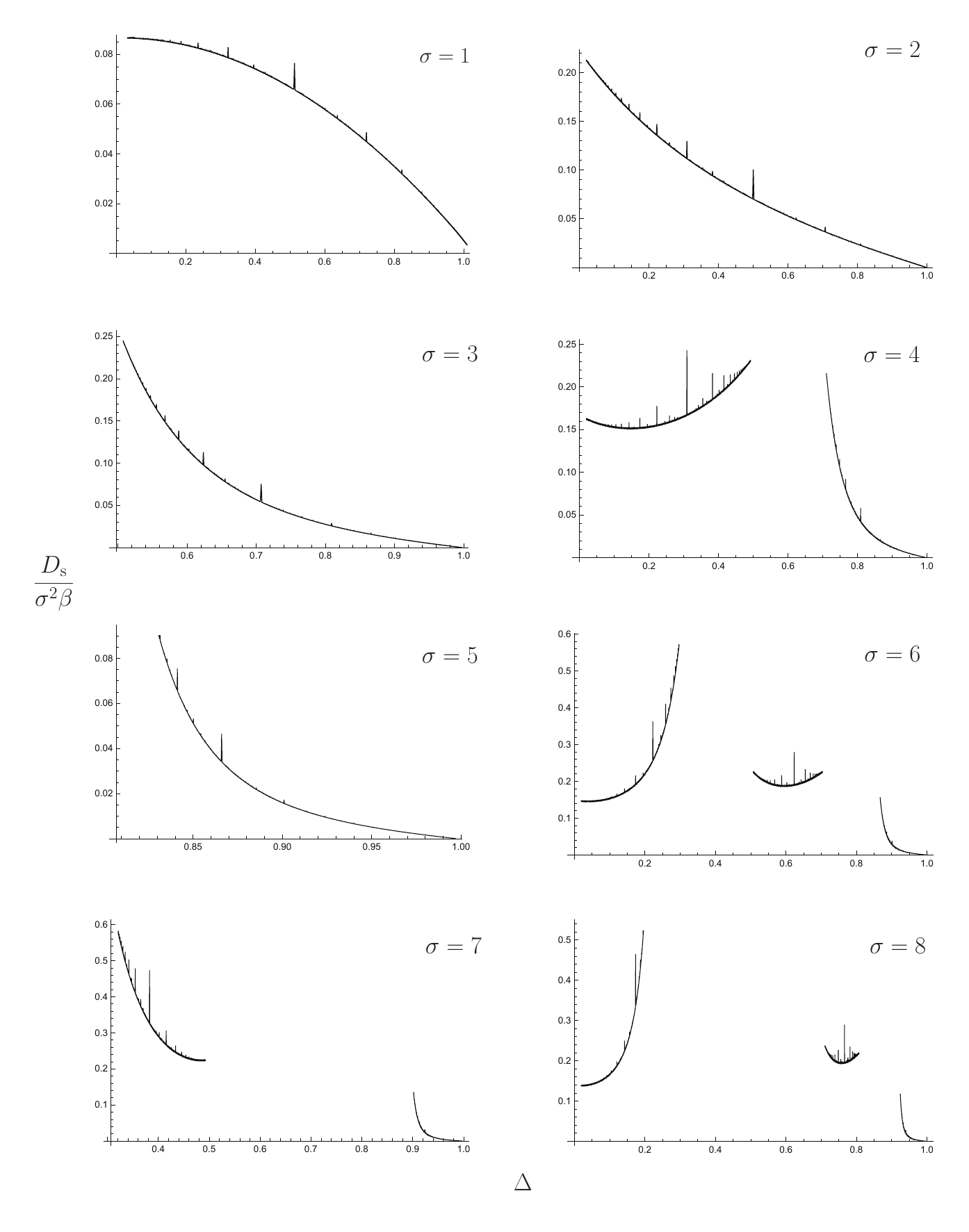}
\caption{High-temperature limit of the spin Drude weight (divided by $ \beta = 1/T $) 
with $J=\pm 1$ and  $ S = \sigma / 2 $ ($1 \le \sigma \le 8$) for the anisotropy dependence. 
The graph displays popcorn structures, indicating everywhere discontinuity 
with respect to $ \Delta $. For the physically admissible region of $ \Delta $, 
refer to Eq.~\eqref{restriction1} and Table~\ref{interval}.
}
\label{fractal}
\end{figure}

In Fig.~\ref{fractal}, we depict the high-temperature limit 
$\beta\to 0$ of $D_\textrm{s}(\beta)/(\sigma^2 \beta)$. The result for $\sigma=1$ is 
referred to as the Prosen-Ilievski bound \cite{prosen2013families}.
For $\sigma\ge 2$, the spin Drude weights exhibit a 
prominent popcorn structure similar to the $\sigma=1$ case. 
For  $p_0$ close to,  but not exactly on, the irrational value, 
which can be obtained with the desired precision by taking the limit 
$\alpha\to\infty$, the Drude 
weights are approximately represented by the points on the curve. 
Namely, the Drude weight behaves like a continuous function 
for $p_0$ close to the irrational value, 
but is discontinuous everywhere for rational $p_0$. 
The Drude weight cannot be obtained in the present approach when $p_0$ 
is exactly an irrational number.
\section{Summary and Discussion}
We have examined the finite-temperature spin Drude weight for the integrable XXZ 
chain with arbitrary spin $S$ at zero magnetic field. In the critical 
regime, the admissible string lengths are described by the TS-numbers, 
akin to the spin-1/2 case. In relation to this, the Drude weight 
exhibits a behavior similar to that observed in the spin-1/2 XXZ, displaying 
a pronounced popcorn structure that is discontinuous everywhere with 
respect to the anisotropy parameter $\Delta$. For a given $S$, the admissible 
region of $\Delta$, in general, comprises several separate intervals,
which is a fundamental distinction from the spin-1/2 case.
Importantly, the structure of the TBA equations in each region is 
distinct, resulting in the Drude weight behaving drastically 
differently in each region. 
In this work, we have constructed the TBA equations, using the QTMs and their 
functional relations. As a significant advantage, this approach 
enables us to perform a systematic exact calculation of the spin Drude
 weight in the high-temperature limit.

Let us briefly discuss the open problems that remain unresolved 
in this work. First, the low-temperature behavior, including the 
zero-temperature limit of the Drude weight, has not yet been 
calculated. In contrast, for the spin-1/2 case, the low-temperature 
behavior of the Drude weight has been recently addressed and 
calculated \cite{urichuk2021analytical}. While the low-temperature properties of the 
present model are more intricate than those of the spin-1/2 case, 
the procedure used there might be useful to our current model.

Second, the investigation of spin diffusion at the isotropic point
 $\Delta=1$ is crucial. At this point, along with the massive region 
$\Delta>1$, the spin Drude weight vanishes, suggesting diffusive spin 
transport \cite{ilievski2018superdiffusion, de2019diffusion, ilievski2021superuniversality}. 
In fact, behavior similar to the divergence observed in the spin-1/2 case 
of the diffusion constant has also been noted in higher spin cases 
\cite{ilievski2018superdiffusion}, indicating superdiffusive behavior. 
There is considerable interest in examining this property through more 
quantitative approaches, such as deriving a rigorous dynamical exponent.

Third, the Drude weight cannot be derived in our current procedure
 when $p_0$ is 
exactly on irrational numbers. Of course, as described before, taking the limit 
$\alpha \to \infty$, we may obtain the Drude weight when $p_0$ is in arbitrary 
vicinity of an irrational number. Indeed, the expression at infinite temperature 
(see \eqref{Drude-high-T}) indicates that the Drude weight converges to a certain 
finite value in this limit. However, it remains unclear whether 
this convergence value actually corresponds to the real Drude weight 
on $p_0$ when it is irrational.

Fourth, the spin Drude weight may also be evaluated using Kohn's formula
\cite{kohn1964theory}. 
Namely, by imposing a twisted boundary condition on the row transfer
 matrix \eqref{fusion-T}, achieved through the multiplication of the
 factor $\exp\left[-i\phi(n-\sigma_{\overline{1}}^z)/2 \right]$ inside
 the trace, where $\sigma_{\overline{1}}^z$ is the $z$-component of
 the spin-$n/2$ operator acting on $V_{\overline{1}}^{(n)}$, the 
Drude weight is determined by the expectation value of the curvature 
of the energy spectrum with respect to the twist angle $\phi$:
\begin{equation}
D_\textrm{s}(\beta)=\lim_{L\to\infty}\frac{1}{L}\sum_n 
\left.\frac{d^2 E_n(\phi)}{d\phi^2}\right|_{\phi=0}e^{-\beta E_n(0)}.
\label{Kohn}
\end{equation}
In the above, $E_n(\phi)$ denotes the $n$th eigenvalue of the operator 
obtained by taking the logarithmic derivative of the twisted 
row transfer matrix. Note that the integrable structure 
\eqref{commutativity} is preserved even under the twisted 
boundary condition.
Note that the integrable structure 
\eqref{commutativity} is preserved even under the twisted
 boundary condition. Furthermore, the Hamiltonian is 
obtained by the same similarity transformation as in 
\eqref{Hamiltonian1}, with eigenvalues that are formally
in the same form as \eqref{energy-spectrum}. The effect of the 
twisted boundary condition is incorporated into the BAE 
\eqref{BAE} by multiplying the RHS with the factor
 $e^{-i\phi}$, leading to a shift in the string center in
\eqref{string}
from $\lambda_{\mu}^{n_j}$ to $\lambda_{\mu}^{n_j}(\phi)$. 
By expanding $E_n(\phi)$ up to second order in $\phi$ and 
explicitly evaluating \eqref{Kohn}, in a manner similar to Zotos's approach 
for the $\sigma=1$ case,  it might be possible to reproduce 
the formula given by \eqref{Drude}.

Finally, utilizing the QTMs developed in this study, we can systematically 
calculate the infinite temperature Onsager matrix element 
associated with spin transport for the $S=1/2$ chain in 
a zero magnetic field. This becomes feasible 
because the dressed scattering kernel
for the $S=1/2$ model  at infinite temperature 
can be determined using the TBA equations 
for an $S>1/2$ model. Our detailed analysis uncovers an intriguing
popcorn structure of the Onsager matrix element. Further 
details will be reported in an upcoming publication
(see \cite{ae2023spin}).

\section*{Acknowledgment}
%
The present work was partially supported by Grant-in-Aid for Scientific
Research (C) No. 20K03793 from the Japan Society for the 
Promotion of Science.

\begin{appendix}

\section{Takahashi-Suzuki (TS) numbers}\label{TS}

Given anisotropy $\theta$, it is known that the admissible lengths of string solutions
of the Bethe ansatz equation \eqref{BAE} are given by the so-called 
Takahashi-Suzuki (TS) numbers generated by the continued 
fraction expansion of $p_0:=\pi/\theta$ (eq.~\eqref{cfrac}).
Here, we summarize the TS-numbers \cite{takahashi1972one,kuniba1998continued}
and related sequences required for the formulation.

Given $\{\nu_r\}_{r=1}^\alpha$ in \eqref{cfrac},
we define the sequences $\{m_r\}_{r=0}^{\alpha+1}$ and $\{p_r\}_{r=0}^{\alpha+1}$ by
\begin{equation}
m_r=\nu_1+\nu_2+\cdots+\nu_r \quad (0 \le r \le \alpha),\quad
m_{\alpha+1}=\infty,
\label{def-m}
\end{equation}
and 
\begin{equation}
p_0=\frac{\pi}{\theta}=[\nu_1,\dots,\nu_{\alpha}], \quad p_1=1,\quad
p_r=p_{r-2}-\nu_{r-1}p_{r-1} \quad (2\le r \le \alpha+1),
\label{def-p}
\end{equation}
respectively.  More explicitly, $p_r$ ($r\ge 1$) reads
\begin{equation}
p_r=\prod_{s=2}^r\frac{1}{[\nu_s,\dots,\nu_{\alpha}]} \ \ \text{for $1\le r\le \alpha$}.
\label{def-p2}
\end{equation}
From this expression and \eqref{def-p}, one easily finds that
\begin{equation}
p_{\alpha+1}=0,\quad
[\nu_r,\dots,\nu_{\alpha}]=\frac{p_{r-1}}{p_r},\quad 
 \nu_r=\left\lfloor 
\frac{p_{r-1}}{p_{r}} \right\rfloor.
\end{equation}
Define the sequences $\{y_r\}_{r=-1}^\alpha$ and $\{z_r\}_{r=-1}^\alpha$
by
\begin{alignat}{5}
&y_{-1}=0,&&\quad y_{0}=1,&&\quad y_r=y_{r-2}+\nu_r y_{r-1}\ \ &&(1\le r \le \alpha), \nn \\
&z_{-1}=1,&&\quad z_{0}=0,&&\quad z_r=z_{r-2}+\nu_r z_{r-1}\ \ &&(1\le r \le \alpha).
\label{def-yz}
\end{alignat}
Then by induction, one sees that
\begin{equation}
[\nu_1,\dots,\nu_r]=\frac{y_r}{z_r} \quad (1\le r\le \alpha),\quad
y_r=p_0 z_r+(-1)^r p_{r+1} \quad (-1\le r\le \alpha),
\label{yz-ratio}
\end{equation}
where we extend integers $\nu_r$ $(1\le r \le \alpha)$ formally to real numbers and use 
\begin{equation}
[\nu_1,\dots,\nu_{r},\nu_{r+1}]=\left[\nu_1,\dots,\nu_{r-1},\nu_r+\frac{1}{\nu_{r+1}}\right]
\end{equation}
to prove the first equality.
In particular, 
\begin{equation}
p_0=\frac{y_\alpha}{z_\alpha}.
\label{p0-yz}
\end{equation}

Let $y_r=:f_r(\nu_1,\dots,\nu_r)$. From \eqref{def-yz}, the equality
$z_{r+1}=f_{r}(\nu_2,\dots,\nu_{r+1})$ holds. Also, one finds that
\begin{equation}
f_r(\nu_1,\dots,\nu_r)=f_r(\nu_r,\dots,\nu_1),
\label{inverse}
\end{equation}
 which 
can be shown by induction. Indeed, one directly sees that
it holds for $1\le r \le 4$. Assume that the equality also holds for 
$r=s-k$ ($s\ge 5$; $1\le k\le 4$). Then, using \eqref{def-yz} and the assumptions, one obtains
\begin{align}
f_s(\nu_1,\dots,\nu_s)&=\nu_s f_{s-1}(\nu_1,\dots,\nu_{s-1})+f_{s-2}(\nu_1,\dots,\nu_{s-2})\nn \\
&=\nu_s f_{s-1}(\nu_{s-1},\dots,\nu_1)+f_{s-2}(\nu_{s-2},\dots,\nu_1)\nn \\
&=\nu_s\left[\nu_1f_{s-2}(\nu_{s-1},\dots,\nu_2)+f_{s-3}(\nu_{s-1},\dots,\nu_3)\right]\nn \\
&\qquad+
\nu_1f_{s-3}(\nu_{s-2},\dots,\nu_2)+f_{s-4}(\nu_{s-2},\dots,\nu_3)\nn \\
&=\nu_1f_{s-1}(\nu_{s},\dots,\nu_2)+f_{s-2}(\nu_s,\dots,\nu_3)\nn \\
&=f_s(\nu_s,\dots,\nu_1).
\end{align}
Thus, \eqref{inverse} holds. \eqref{def-yz} yields
\begin{align}
&f_r(\nu_r,\dots,\nu_1)=\nu_1 f_{r-1}(\nu_r,\dots,\nu_2)+f_{r-2}(\nu_r,\dots,\nu_3), \nn \\
&f_{r-1}(\nu_r,\dots,\nu_2)=\nu_2 f_{r-2}(\nu_r,\dots,\nu_3)+f_{r-3}(\nu_r,\dots,\nu_4), \nn \\
&\cdots \nn \\
&f_2(\nu_r,\nu_{r-1})=\nu_{r-1}f_1(\nu_r)+1.
\label{euclid}
\end{align}
By applying the Euclidean algorithm and noting $y_r=f_r(\nu_r,\dots,\nu_1)$ and
$z_r=f_{r-1}(\nu_r,\dots,\nu_2)$,  \eqref{euclid} implies that 
$y_r$ and $z_r$ are coprime.  Hence, from the first equality in \eqref{yz-ratio}, 
$z_r$ is given by the numerator of the irreducible
fraction of $[\nu_2,\dots,\nu_r]$, which leads to
\begin{align}
&y_r=\prod_{s=1}^r[\nu_s,\dots,\nu_r]=\prod_{s=1}^r[\nu_r,\dots,\nu_s] \ \ (1\le r\le\alpha),\nn \\
&z_r=\prod_{s=1}^{r-1}[\nu_{s+1},\dots,\nu_r]=\prod_{s=1}^{r-1}[\nu_r,\dots,\nu_{s+1}] 
\ \ (1\le r\le \alpha).
\end{align}
The above equalities, \eqref{def-p2} and  \eqref{p0-yz} 
lead to
\begin{equation}
p_{\alpha}=\frac{1}{z_{\alpha}}=\frac{p_0}{y_\alpha}.
\end{equation}
This identity can also be proved by induction as in \cite{urichuk2019spin}.

Now we define the TS numbers $\{n_j\}_{j\ge 1}$ \cite{takahashi1972one}
and their modifications $\{\wt{n}_j\}_{j\ge 1}$ \cite{kuniba1998continued}
by the following sequences:
\begin{alignat}{3}
&n_j=y_{r-1}+(j-m_r)y_r \quad &&(m_r\le j < m_{r+1}), \nn \\
&\wt{n}_j=y_{r-1}+(j-m_r)y_r \quad &&(m_r< j \le m_{r+1}).
\label{TS-numbers}
\end{alignat}
In this paper, we treat the first $m_{\alpha}$ sequences, i.e., 
$\{n_j\}_{j=1}^{m_\alpha}$ and $\{\wt{n}_j\}_{j=1}^{m_\alpha}$.
One finds $\wt{n}_j=n_j$ except for $j=m_r$ ($1\le r\le \alpha)$
($n_{m_r}=y_{r-1}$, while $\wt{n}_{m_r}=y_{r}$). In particular, 
$n_j$ duplicates at $j=1$ and $j=m_1$ as 
$n_1=n_{m_1}=1$, whereas 
$\wt{n}_j$ strictly increases with respect to $j$.
Replacing $\{y_r\}$ with $\{z_r\}$ in the above
sequences, we introduce $\{w_j\}_{j\ge 1}$ and $\{\wt{w}\}_{j\ge 1}$
by
\begin{alignat}{3}
&w_j=z_{r-1}+(j-m_r)z_r-1 \quad &&(m_r\le j < m_{r+1}), \nn \\
&\wt{w}_j=z_{r-1}+(j-m_r)z_r-1 \quad &&(m_r< j \le m_{r+1}).
\label{def-w}
\end{alignat}
%
One finds that the sequences $ \{w_j\}_{j \ge 1} $ and $ \{\wt{w}\}_{j \ge 1} $ 
are related to the parity of the system:
\begin{equation}
v(\wt{n}_j) = (-1)^{\wt{w}_j}, \quad v_j = (-1)^{w_j}.
\label{v-w}
\end{equation}
Here, $v(n)$ is given by \eqref{spin-parity}, and $v_j$
is the string parity defined as
\begin{equation}
v_j := \begin{cases} 
v(n_j) & \text{for $ j \neq m_1 $} \\
-1 & \text{for $ j = m_1 $}
\label{string-parity}
\end{cases}.
\end{equation}
Using the definitions \eqref{def-w}, \eqref{def-yz} and \eqref{def-p}, 
one can readily verify \eqref{v-w}. 

Finally, we also introduce $\{q_j\}_{\ge 1}$ and 
$\{\wt{q}_j\}_{\ge 1}$ as
\begin{alignat}{3}
&q_j=(-1)^r\left[p_{r}-(j-m_r)p_{r+1}\right] \quad &&(m_r\le j < m_{r+1}), \nn \\
&\wt{q}_j=(-1)^r\left[p_{r}-(j-m_r)p_{r+1}\right] \quad &&(m_r< j \le m_{r+1}).
\label{def-q}
\end{alignat}

\section{Proof of the equivalence between \eqref{restriction1} and \eqref{restriction2}}\label{TS2}
%
Let us show that the condition \eqref{restriction1} is 
equivalent to \eqref{restriction2}. This is the same as showing that 
the set of integers $n$  satisfying
\begin{equation}
v(n)\sin\left(\frac{\pi}{p_0}j\right)\sin\left(\frac{\pi}{p_0}(n-j)\right)>0 \quad (1\le j\le n-1)
\label{condition1}
\end{equation}
is equivalent to $\{\wt{n}_j\}_{j=1}^{m_{\alpha}}$, where $p_0$ is a rational number
given by \eqref{cfrac}. This can be achieved by using the method similar to that in 
\cite{takahashi1972one}.

\begin{table}[tb]
\begin{tabular}{|c |l l l l |l l l l | c| }
\hline
$r$ &\multicolumn{4}{c|}{0} &\multicolumn{4}{c|}{1} & $\cdots$ \\
\hline
$j$ & 1 & 2 & $\cdots$ & $m_1$ &  
$m_1+1$ &$m_1+2$ &$\cdots$ &$m_2$ &$\cdots$ \\
\hline
$\wt{n}_j$ & 1 & 2 & $\cdots$ & $y_1$ & 
$1+y_1$ & $1+2y_1$ & $\cdots$& $y_2$ & $\cdots$  \\
\hline
\end{tabular}
\vspace*{5mm}

\begin{tabular}{|c | l l l l |l l l l |c }
\hline
$r$ & \multicolumn{4}{c|}{$\alpha-2$}  &\multicolumn{4}{c|}{$\alpha-1$}  \\
\hline
$j$  &$m_{\alpha-2}+1$& $m_{\alpha-2}+2$ & $\cdots$ & $m_{\alpha-1}$ 
&$m_{\alpha-1}+1$ & $m_{\alpha-1}+2$ &$\cdots$ & $m_{\alpha}$\\
\hline
$\wt{n}_j$  & $y_{\alpha-3}+y_{\alpha-2}$ & $y_{\alpha-3}+2y_{\alpha-2}$ 
& $\cdots$ & $y_{\alpha-1}$ 
&  $y_{\alpha-2}+y_{\alpha-1}$ & $y_{\alpha-2}+2y_{\alpha-1}$ 
& $\cdots$ & $y_{\alpha}$  \\
\hline
\end{tabular}
\caption{The modified TS-numbers $\wt{n}_j$ defined in \eqref{TS-numbers}.}
\label{TS-table}
\end{table}
Substituting \eqref{p0-yz} into \eqref{condition1}, one sees that $n$ is restricted
in the region $1\le n\le y_{\alpha}$. Otherwise, the LHS of \eqref{condition1}
must be $0$, when $j=y_\alpha$.
Let us rewrite the condition \eqref{condition1} as
\begin{equation}
\left\lfloor 
\frac{n-1}{p_0}\right\rfloor=\left\lfloor \frac{j}{p_0}\right\rfloor+
\left\lfloor \frac{n-j}{p_0}\right\rfloor \quad (1\le j\le n-1).
\label{condition2}
\end{equation}
For $p_0=\nu_1(=y_1)\in\mathbb{N}_{\ge 2}$ (i.e., $\alpha=1$), 
we easily see
that the integers $n$ satisfying \eqref{condition2} are
\begin{equation}
n\in \{1,2,\dots, \nu_1\},
\end{equation}
which are nothing but the modified TS-numbers for $\alpha=1$
(See also Table~\ref{TS-table}). 

Let us consider the case  $\alpha>1$, where
the following lemma is useful.
\begin{lemma}\label{lem}
Let $x\in\mathbb{N}$, $y\in\mathbb{R}_{>0}$ and $m\in\mathbb{N}$
be numbers such that $\lfloor x/y \rfloor=m$ and $x/y\notin\mathbb{N}$.
Then $\lfloor m y\rfloor<x\le \lfloor (m+1)y \rfloor$ holds.
\end{lemma}
\noindent
For $\alpha>1$, all the integers $n$ such that $n \le \lfloor p_0 \rfloor+1$, 
i.e., 
\begin{equation}
n\in\{1,2,\dots,\nu_1+1\},
\end{equation}
satisfy \eqref{condition2}. 
These integers are the modified TS-numbers $\wt{n}_j$ ($1 \le j \le m_1+1$) 
(see Table~\ref{TS-table}). On the other hand, the integers 
$n > \lfloor p_0 \rfloor+1$ can be found by considering \eqref{condition2} 
under the cases $j = \lfloor j_1 p_0 \rfloor$ and $j = \lfloor j_1 p_0 
\rfloor+1$, where $j_1\in\mathbb{N}_+$
satisfies 
\begin{equation}
1 \le \lfloor j_1 p_0 \rfloor + 1 \le n - 1.
\label{range}
\end{equation}
Inserting $j = \lfloor j_1 p_0 \rfloor$ into \eqref{condition2}, 
and applying  Lemma~\ref{lem}
by setting $x=n-j$, 
$y=p_0$ and $m=\lfloor (n-1)/p_0\rfloor-\lfloor j/p_0 \rfloor$, 
then using $\lfloor j/p_0 \rfloor = j_1 - 1$, we find 
\begin{equation}
\lfloor j_1 p_0 \rfloor + \lfloor (\ell_1 - j_1) p_0 \rfloor < n,
\end{equation}
where $\ell_1 \in \mathbb{N}_+$ is defined by
\begin{equation}
\ell_1:=\left\lfloor \frac{n-1}{p_0} \right\rfloor + 1.
\end{equation}
Similarly, for $j = \lfloor j_1 p_0 \rfloor + 1$, we see 
$\lfloor j/p_0 \rfloor = j_1$, and hence Lemma~\ref{lem} 
yields
\begin{equation}
n \le 1+\lfloor j_1 p_0 \rfloor + \lfloor (\ell_1 - j_1) p_0 \rfloor.
\end{equation} 
These two conditions for $n$ determine
\begin{equation}
n = 1+\lfloor j_1 p_0 \rfloor  + \lfloor (\ell_1 - j_1) p_0 \rfloor
\quad (1 \le j_1 \le \ell_1 - 1),
\end{equation}
where the range of $j_1$ comes from the condition \eqref{range}.
By substituting $p_2 = p_0 - \nu_1$ (see \eqref{def-p}) into the 
above expression, we can rewrite it as
\begin{equation}
n = \ell_2 + \ell_1 \nu_1=\ell_2+\ell_1 y_1,
\label{first}
\end{equation}
where
\begin{equation}
\ell_2 := 1 + \lfloor j_1 p_2 \rfloor + \lfloor (\ell_1 - j_1) p_2 \rfloor
\quad (1 \le j_1 \le \ell_1 - 1).
\label{l2}
\end{equation}
For $2 \le \ell_1 \le \lfloor 1/p_2 \rfloor + 1 =
 \nu_2 + 1$, since $\ell_2=1$,
\begin{equation}
n \in \{1 + 2 y_1, 1 + 3 y_1, \dots, 1 +  (\nu_2+1) y_1 = y_1+y_2\},
\end{equation}
are also the modified TS-numbers $\{\wt{n}_j\}_{j=m_1+2}^{m_2+1}$. 
For $\ell_1 > \lfloor 1/p_2\rfloor + 1$, the same procedure used to derive \eqref{first} is 
applicable. Namely, using the condition \eqref{l2} for
 $j_1 = \lfloor j_2/p_2 \rfloor$ and $j_1 = \lfloor j_2/p_2 \rfloor+1$, 
where $j_2 \in \mathbb{N}_+$ 
satisfies $1\le \lfloor j_2/p_2 \rfloor+1\le \ell_1-1$, 
and applying Lemma~\ref{lem}, we obtain
\begin{equation}
\ell_1 = 1 + \left\lfloor j_2\frac{1}{p_2} \right\rfloor +
\left\lfloor (\ell_2 - j_2)\frac{1}{p_2} \right\rfloor
\quad (1 \le j_2 \le \ell_2-1).
\end{equation}
Substituting $p_3=1-\nu_2 p_2$ into the equation above, we find
\begin{equation}
\ell_1 = \ell_3 + \ell_2 \nu_2,
\label{l1}
\end{equation}
where
\begin{equation}
\ell_3 := 1 + \left\lfloor j_2\frac{p_3}{p_2}\right\rfloor +
\left\lfloor (\ell_2 - j_2)\frac{p_3}{p_2} \right\rfloor
\quad (1 \le j_2 \le \ell_2-1).
\end{equation}
For $2 \le \ell_2 \le \lfloor p_2/p_3 \rfloor + 1 = \nu_3 + 1$,
$\ell_3=1$. Hence, using  \eqref{l1} and \eqref{first}, we find
\begin{equation}
n \in \{y_1 + 2y_2, y_1 + 3y_2, \dots, y_1 + (\nu_3+1)y_2 = y_2 + y_3\} = \left\{\wt{n}_j\right\}_{j = m_2+2}^{m_3+1}.
\end{equation}

Applying the above procedure repeatedly, at the last stage we arrive at
\begin{equation}
\ell_{\alpha-1}:=
1+\left\lfloor j_{\alpha-2}\frac{p_{\alpha-1}}{p_{\alpha-2}}\right\rfloor+
\left\lfloor (\ell_{\alpha-2}-j_{\alpha-2})\frac{p_{\alpha-1}}{p_{\alpha-2}} 
\right\rfloor \quad (1\le j_{\alpha-2} \le \ell_{\alpha-2}-1),
\end{equation}
and for $\ell_{\alpha-2}>\lfloor p_{\alpha-2}/p_{\alpha-1} \rfloor+1$, 
\begin{equation}
\ell_{\alpha-2}=\ell_\alpha+\ell_{\alpha-1}\nu_{\alpha-1},
\end{equation}
where
\begin{equation}
\ell_{\alpha}:=1+\left\lfloor j_{\alpha-1}\frac{p_{\alpha}}{p_{\alpha-1}}
\right\rfloor+
\left\lfloor (\ell_{\alpha-1}-j_{\alpha-1})\frac{p_{\alpha}}{p_{\alpha-1}} 
\right\rfloor \quad (1\le j_{\alpha-1} \le \ell_{\alpha-1}-1).
\end{equation}
For $2\le \ell_{\alpha-1}\le \lfloor p_{\alpha-1}/p_{\alpha}\rfloor=\nu_\alpha+1$,
\begin{equation}
\ell_{\alpha-2}=1+\ell_{\alpha-1}\nu_{\alpha-1}.
\end{equation}
Using the above equality with 
$
\ell_k = \ell_{k+2} +  \ell_{k+1} \nu_{k+1} \quad (0 \le k \le \alpha - 2),
$
where $\ell_0 := n$ and $\ell_\alpha:=1$, and taking into account
the definition of $y_r$ given by \eqref{def-yz}, 
we obtain
\begin{equation}
n = y_{\alpha - 2} + \ell_{\alpha - 1} y_{\alpha - 1} \quad (2 \le \ell_{\alpha - 1} 
\le \nu_\alpha + 1).
\end{equation}
Since $n$ is restricted to $n \le y_{\alpha}$, 
as previously explained, we have
\begin{equation}
n \in \left\{ y_{\alpha - 2} + 2y_{\alpha - 1}, y_{\alpha - 2} + 3y_{\alpha - 1}, \dots, y_{\alpha - 2} + \nu_{\alpha} y_{\alpha - 1} = y_\alpha \right\} = 
\left\{ \wt{n}_j \right\}_{j = m_{\alpha - 1}+2}^{m_\alpha}.
\end{equation}
Thus, we have shown that the integers fulfill the condition \eqref{condition1}
is the modified TS numbers $\{\wt{n}_j\}_{j=1}^{m_{\alpha}}$.
Consequently, the equivalence between \eqref{restriction1} and \eqref{restriction2} has been proved.

If $p_0$ is an irrational number, the above discussion can be 
extended by letting $\alpha\to\infty$. In this case, the sets 
$\{\wt{n}_j\}$ and $\{n_j\}$ coincide when duplicate elements 
are removed. Thus, if $p_0$ is an irrational number, the condition 
\eqref{restriction2} is equivalent to \eqref{S-restriction}.

%
\section{Derivation of the TBA equation}\label{NLIE-derivation}

%
We briefly outline the method for deriving the TBA equations, as given in 
\eqref{TBA-QTM}, starting from the $ Y $-systems in \eqref{y-system}. Although the procedure described in \cite{kuniba1998continued} can be applied here---since the $ Y $-systems are identical to those in the spin-$1/2 $ 
case (i.e., $ \sigma=1 $)---the analytical properties become more complex for 
spins higher than $1/2 $ (i.e., $ \sigma > 1 $). As a result, the
 driving terms in the TBA equations manifest differently compared to 
the spin-$1/2$ case.  
Let $ \mathcal{S}[x] := \{ z \in \mathbb{C} \mid \Im(z) \in [-x, x] \} $ where $ x \in \mathbb{R}_{> 0} $. As in \cite{kuniba1998continued}, the following lemma is useful for deriving the TBA equations.
\begin{lemma}\label{ANZC}
Let $ f_j(v) $ for $ j \ge 0 $ be functions satisfying
\begin{equation}
f_0(v + i v_0) f_0(v - i v_0) = \prod_{j \ge 1} f_j(v + i v_j) f_j(v - i v_j),
\end{equation}
where $ 0 \le v_j < v_0 $ for $ j > 1 $. Also assume that
$ f_j(v) $ is analytic, non-zero, and has constant asymptotics (ANZC) in the domain
$ v \in \mathcal{S}[w_j] $ where $ v_j < w_j $. Then $ f_0(v) $ can be given by 
the following NLIE:
\begin{equation}
\log f_0(v) = \sum_{j \ge 1} K_j \ast \log f_j(v) + \text{const.},
\end{equation}
where 
\begin{equation}
K_j(v) =\int_{-\infty}^{\infty} e^{i k v} \frac{\ch(v_j k)}
{2\pi\ch(v_0 k)} \, dk,
\end{equation}
and the constant term is determined by the asymptotic values on both sides.
\end{lemma}
This lemma is applicable to the $Y$-system \eqref{y-system} after some modification 
to the $Y$-functions defined in \eqref{y-function1} and \eqref{y-function2}.
Specifically, for  $m_{r-1}\le j\le m_r-2$ 
($1\le r \le \alpha$) and $j=m_r-1$ ($r_\sigma \le r \le \alpha$), we introduce 
\begin{align}
&\wt{Y}_j(v):=\frac{Y_j(v)}{\psi_j(v)},\nn  \\
&\psi_j(v):=
\begin{cases}
\left[
\tnh\frac{\pi}{4p_{r_\sigma}}(v+i(p_{r_\sigma}\pm u_N))
\tnh\frac{\pi}{4p_{r_\sigma}}(v-i(p_{r_\sigma}\pm u_N))
\right]^{\pm \frac{N}{2}(-1)^{r_{\sigma}-1}}  & \text{if $j=j_\sigma$} \\
1              & \text {otherwise}
\end{cases},
\label{psi}
\end{align}
where $j_\sigma$ and $r_\sigma$ are defined by \eqref{jsigma}, and
the signs $+$ and $-$ in the exponent and before $u_N$ correspond to 
the sign of $J$, with $+$ chosen when $J>0$ ($u_N<0$) and $-$ 
when $J<0$ ($u_N>0$). See \eqref{un} for the relationship between $J$ and $u_N$.
Additionally, 
for $j=m_r-1$ ($1\le r \le r_\sigma-1$),
we also define 
\begin{align}
&\wt{X}_{j}^{(\pm)}(v):=\frac{Y_{j}(v+ip_{r+1})Y_{j}(v-ip_{r+1})}
{(1+Y_{j+1}(v))(1+Y_{j}(v\mp ip_{r+1}))}\frac{1}{\psi_{j}(v)},  \nn \\
&\psi_{j}(v):=\begin{cases}
\left[
\frac{
\tnh\frac{\pi}{4p_r}(v+i(p_r-|\wt{q}_{j_\sigma+1}|-u_N))
\tnh\frac{\pi}{4p_r}(v-i(p_r-|\wt{q}_{j_\sigma+1}|-u_N))}
{\tnh\frac{\pi}{4p_r}(v+i(p_r-|\wt{q}_{j_\sigma+1}|+u_N))
\tnh\frac{\pi}{4p_r}(v-i(p_r-|\wt{q}_{j_\sigma+1}|+u_N))}\right]^{\frac{N}{2}(-1)^r} 
& \text{for $j_\sigma\ne m_\alpha-1$} \\
\left[
\tnh\frac{\pi}{4p_r}(v+i(p_r\pm u_N))\tnh\frac{\pi}{4p_r}(v-i(p_r\pm u_N))
\right]^{ \pm N(-1)^{r-1}} & \text{for $j_\sigma=m_\alpha-1$}
\end{cases}.
\end{align}
Again, the signs $\pm$ in the exponent and before $u_N$ in $\psi_{r,\sigma}(v)$
for $j_\sigma=m_\alpha-1$ are determined according to the
specification given in \eqref{psi}.
Since the identity
\begin{equation}
\tanh\frac{\pi}{4c}(v+ic)\tanh\frac{\pi}{4c}(v-ic)=1
\end{equation}
is valid for arbitrary $c\in\mathbb{C}$, the $Y$-functions
in the $Y$-system can be replaced by the above modified ones:
\begin{align}
&\textrm{For } m_{r-1} \le j \le m_r - 2\,\, ( 1 \le r \le \alpha), \nn \\
&\quad \wt{Y}_j(v+ip_r)\wt{Y}_j(v-ip_r) = 
\left( 1+Y_{j-1}(v)\right)^{1-2\delta_{j,m_{r-1}}}
\left(1+Y_{j+1}(v)\right)
 \left( 1+Y_{j+2}(v)^{-1}\right)^{\delta_{j,m_{\alpha}-2}}
,\nn \\
&\textrm{for } j = m_r -1\,\, (r_\sigma \le r \le \alpha -1 ),  \nn \\
&\quad \wt{Y}_j(v + ip_r + ip_{r+1})\wt{Y}_j(v + ip_r - ip_{r+1})
\wt{Y}_j(v - ip_r + ip_{r+1})\wt{Y}_j(v - ip_r - ip_{r+1}) 
\nn \\
&\qquad = \left[ \left(1+Y_{j-1}(v+ip_{r+1})\right)
\left(1+Y_{j-1}(v-ip_{r+1})\right) \right]^{1-2\delta_{j,m_{r-1}}}
\left(1+Y_{j+1}(v+ip_r)\right)\nn \\
&\qquad \quad\times \left(1+Y_{j+1}(v-ip_r)\right)
\left(1+Y_j(v + ip_r - ip_{r+1})\right)
\left(1+Y_j(v - ip_r + ip_{r+1})\right),
\nn \\
&\textrm{for } j = m_r -1\,\, (1 \le r \le r_\sigma -1 ),  \nn \\
&\quad \wt{X}^{(+)}_j(v + ip_r)\wt{X}^{(-)}_j(v - ip_r)
= \left[ \left(1+Y_{j-1}(v+ip_{r+1})\right)
\left(1+Y_{j-1}(v-ip_{r+1})\right) \right]^{1-2\delta_{j,m_{r-1}}},
\nn \\
&\quad \wt{Y}_{m_\alpha-1}(v+ip_\alpha)\wt{Y}_{m_\alpha-1}(v-ip_\alpha)
=e^{2 y_\alpha \beta h}(1+Y_{m_\alpha-2}(v)), \nn \\
&\quad \left(\wt{Y}_{m_\alpha}(v+ip_\alpha)\wt{Y}_{m_\alpha}
(v-ip_\alpha)\right)^{-1}
=e^{-2 y_\alpha \beta h}(1+Y_{m_\alpha-2}(v)),
\label{y-system-mod}
\end{align}
where $\wt{Y}_{m_\alpha}(v)^{-1}:=e^{-2y_\alpha\beta h}\wt{Y}_{m_\alpha-1}(v)$.
In the above modification, Lemma~\ref{ANZC} is applicable to \eqref{y-system-mod}
and the resultant NLIEs read
\begin{align}
\text{For}\ &m_{r-1}\le j\le m_r-2\ (1\le r\le \alpha),\nn \\
& \log Y_j(v)=\log \psi_j(v)+(1-2\delta_{j,m_{r-1}})s_r*\log(1+ Y_{j-1})(v)
    \nn \\
&\qquad \qquad \quad + s_r*\log(1+ Y_{j+1})
(1+ Y_{j+2}^{-1})^{\delta_{j,m_{\alpha}-2}}(v), \nn \\
\text{for}\ & j=m_r-1 \ (r_{\sigma}\le r\le \alpha-1) ,  \nn \\
& \log  Y_j(v)=\log \psi_j(v)+(1-2\delta_{j,m_{r-1}})s_r*\log(1+ Y_{j-1})(v)
    \nn \\ 
&\qquad \qquad \quad +d_r*\log(1+ Y_j)(v) +s_{r+1}*\log(1+ Y_{j+1})(v), \nn\\
\text{for}\ & j=m_r-1 \ (1\le r \le r_{\sigma}-1), \nn \\ 
& \log  Y_j(v)=s_{r+1}*\log \psi_{j}(v)
+(1-2\delta_{j,m_{r-1}})s_r*\log(1+ Y_{j-1})(v)    \nn \\ 
&\qquad \qquad \quad +d_r*\log(1+ Y_j)(v) +s_{r+1}*\log(1+ Y_{j+1})(v), \nn\\
& \log  Y_{m_{\alpha}-1}(v)=y_{\alpha}\beta h+ 
\log \psi_{m_\alpha-1}(v)
+s_\alpha*\log(1+ Y_{m_\alpha-2})(v), \nn \\
& \log  Y_{m_{\alpha}}(v)^{-1}=-y_{\alpha}\beta h+ 
\log \psi_{m_{\alpha}-1}(v)
+s_\alpha*\log(1+ Y_{m_\alpha-2})(v).
\label{TBA-finite-N}
\end{align}
In the above NLIEs, the Trotter limit $N\to\infty$ can be taken analytically and
the resultant equations lead to the TBA equations \eqref{TBA-QTM}.

%
\section{Free energy} \label{free-energy}
Let us express the free energy, as given by \eqref{free-largest}, 
in terms of the solutions to the TBA equations \eqref{TBA-QTM}. 
For this purpose, we first express $ \Lambda(u_N,v)=T_\sigma(v) $ in terms of 
$ Y $-functions. Using \eqref{y-function3} and \eqref{jsigma}, we obtain
\begin{align}
&\frac{T^{y_{r_\sigma-1}}_\sigma\left(\wt{v}+iw_{j_\sigma}p_0\right)
 T^{-y_{r_\sigma-1}}_\sigma\left(\wt{v}+i w_{j_\sigma}p_0\right)}
{T^{\wt{n}_{j_\sigma+1}}_{y_{r_\sigma-1}-1}\left(\wt{v}+iw_{j_\sigma}p_0\right)
T^{-\wt{n}_{j_\sigma+1}}_{y_{r_\sigma-1}-1}\left(\wt{v}+iw_{j_\sigma}p_0\right)}
=\left(1+Y_{j_\sigma}(v)\right)
\left(1+Y_{j_\sigma+1}(v)^{-1}\right)^{\delta_{j_\sigma,m_{\alpha}-1}}.
\end{align}
Here we have introduced the notations 
\begin{equation}
\wt{v}:=v+i\wt{w}_{j_\sigma+1}p_0, \quad 
T^{\pm x}(v):=T(v\pm i x)
\end{equation}
to save space.
Referring to the definitions of the sequences in Appendix~\ref{TS}
and the periodicity \eqref{periodicity}, 
we can further modify this equation to obtain
\begin{align}
&\frac{T^{p_{r_\sigma}}_\sigma(v)T^{-p_{r_\sigma}}_\sigma(v)}
{ T^{\wt{q}_{j_{\sigma+1}}}_{y_{r_\sigma-1}-1}\left(\wt{v}+
i(z_{r_{\sigma}-1}-1)p_0\right)
T^{-\wt{q}_{j_{\sigma+1}}}_{y_{r_\sigma-1}-1}
\left(\wt{v}+i(z_{r_{\sigma}-1}-1)p_0\right)}\nn \\
&\qquad \qquad \qquad \qquad \qquad \qquad \qquad \qquad  =
\left(1+Y_{j_\sigma}(v)\right)
\left(1+Y_{j_\sigma+1}(v)^{-1}\right)^{\delta_{j_\sigma,m_{\alpha}-1}}.
\label{free1}
\end{align}
From \eqref{DVF}, we observe that $ T_n(v) $ possesses trivial zeros 
stemming from the vacuum part. Crucially, these zeros obstruct the 
proper application of Lemma~\ref{ANZC}. To avoid this, we modify $ T_n(v) $ 
to remove these zeros
\begin{align}
&\wt{T}_{n}(v)=\frac{T_{n}(v)}{\varphi_n(v)}, \nn \\
&\varphi_n(v):=\prod_{k=1}^{\sigma-n}
\phi\left[v+i(\left(2k-(\sigma-n)+u\right)\right]
\phi\left[v-i(\left(2k-(\sigma-n)+u\right)\right].
\end{align}
Applying Lemma~\ref{ANZC} to \eqref{free1} leads to
\begin{align}
&\log T_\sigma(v)=s_{r_\sigma}*\log(1+Y_{j_\sigma})
(1+Y_{j_\sigma+1}^{-1})^{\delta_{j_\sigma,m_{\alpha}-1}}(v) \nn\\
& \qquad \qquad \quad
+ s_{r_\sigma}*\log \varphi_{y_{r_\sigma-1}-1}
\left(\wt{v}+i\wt{q}_{j_\sigma+1}+i(z_{r_\sigma-1}-1)p_0 \right) \nn\\
& \qquad \qquad \quad+ s_{r_\sigma}*\log \varphi_{y_{r_\sigma-1}-1}
\left(\wt{v}-i\wt{q}_{j_\sigma+1}+i(z_{r_\sigma-1}-1)p_0 \right)  \nn \\
&\qquad \qquad \quad
+s_{r_\sigma,\sigma}*
\log \wt{T}_{y_{r_\sigma-1}-1}\left(\wt{v}+i(z_{r_{\sigma}-1}-1)p_0\right), \nn \\
&s_{r,\sigma}(v):=\int_{-\infty}^{\infty}e^{ikv}
\frac{\ch(\wt{q}_{j_\sigma+1}k)}
{2\pi\ch(p_r k)}dk.
\label{free2}
\end{align}
To remove the $T$-function from the RHS, set 
$j = m_{r_\sigma - \ell} - 1$ ($\ell=1$) in \eqref{y-function3}. 
Using \eqref{TS-numbers} and \eqref{def-yz}, we obtain 
$\wt{n}_{j+1} = y_{r_\sigma-\ell}$ ($\ell=1$). Following a procedure 
similar to that used to derive \eqref{free2}, we have
\begin{align}
&\log \wt{T}_{y_{r_{\sigma}-\ell}-1}\left(\wt{v}+i(z_{r_{\sigma}-\ell}-1)p_0\right)
=s_{r_\sigma-\ell}*\log(1+Y_{m_{r_\sigma-\ell}-1})(v) \nn \\
& \qquad \qquad \qquad \qquad
+ s_{r_\sigma-\ell}*\log \varphi_{y_{r_\sigma-1-\ell}-1}
\left(\wt{v}+ip_{r_\sigma+1-\ell}+i(z_{r_\sigma-1-\ell}-1)p_0 \right) \nn\\
& \qquad \qquad \qquad \qquad
+ s_{r_\sigma-\ell}*\log \varphi_{y_{r_\sigma-1-\ell}-1}
\left(\wt{v}-ip_{r_\sigma+1-\ell}+i(z_{r_\sigma-1-\ell}-1)p_0 \right) \nn\\
& \qquad \qquad \qquad \qquad
- s_{r_\sigma-\ell}*\log \varphi_{y_{r_\sigma-\ell}-1}
\left(\wt{v}+ip_{r_\sigma-\ell}+i(z_{r_\sigma-\ell}-1)p_0 \right) \nn\\
& \qquad \qquad \qquad \qquad
- s_{r_\sigma-\ell}*\log \varphi_{y_{r_\sigma-\ell}-1}
\left(\wt{v}-ip_{r_\sigma-\ell}+i(z_{r_\sigma-\ell}-1)p_0 \right) \nn\\
&\qquad \qquad \qquad \qquad+\wt{s}_{r_\sigma-\ell}*
\log \wt{T}_{y_{r_\sigma-1-\ell}-1}
\left(\wt{v}+i(z_{r_\sigma-1-\ell}-1)p_0\right), \nn\\
&\wt{s}_{r}(v):=\int_{-\infty}^{\infty}e^{ikv}
\frac{\ch(p_{r+1} k)}
{2\pi\ch(p_{r} k)}dk,
\label{free-l}
\end{align}
where $\ell=1$.
Again, by setting $j = m_{r_\sigma - \ell} - 1$ (where $\ell=2$ and 
hence $\wt{n}_{j+1} = y_{r_\sigma-\ell}$) in \eqref{y-function3}, 
we can rewrite the $T$-function in the RHS using $Y$-functions
and another $T$-functions.
The resulting expression 
is provided by \eqref{free-l} for $\ell=2$. By repeatedly applying 
this procedure, the $T$-function appearing in the RHS of 
\eqref{free2} can be expressed using the $Y$-functions and some known functions. 
Explicitly it reads
\begin{align}
&\log T_\sigma(v)=s_{r_\sigma}*\log(1+Y_{j_\sigma})
(1+Y_{j_\sigma+1}^{-1})^{\delta_{j_\sigma,m_{\alpha}-1}}(v)+\sum_{r=1}^{r_\sigma-1}
\wt{d}_r*\log(1+Y_{m_r-1})(v) \nn \\
&\quad +s_{r_\sigma}*\log \varphi_{y_{r_\sigma-1}-1}
\left(\wt{v}+i\wt{q}_{j_\sigma+1}+i(z_{r_\sigma-1}-1)p_0 \right) \nn\\
&\quad +s_{r_\sigma}*\log \varphi_{y_{r_\sigma-1}-1}
\left(\wt{v}-i\wt{q}_{j_\sigma+1}+i(z_{r_\sigma-1}-1)p_0 \right) \nn\\
&\quad+ \sum_{r=1}^{r_\sigma-1} 
\wt{d}_{r}*\log 
\frac{
\varphi_{y_{r-1}-1}
\left(\wt{v}+ip_{r+1}+i(z_{r-1}-1)p_0 \right) 
\varphi_{y_{r-1}-1}
\left(\wt{v}-ip_{r+1}+i(z_{r-1}-1)p_0 \right) 
}
{
\varphi_{y_{r}-1}
\left(\wt{v}+ip_{r}+i(z_{r}-1)p_0 \right) 
\varphi_{y_{r}-1}
\left(\wt{v}-ip_{r}+i(z_{r}-1)p_0 \right) 
},
\nn \\
&\wt{d}_{r}(v):=\int_{-\infty}^{\infty}e^{ikv}
\frac{\ch(\wt{q}_{j_\sigma+1} k)}
{4\pi\ch(p_{r} k)\ch(p_{r+1}k)}dk.
\label{finite-N}
\end{align}

At infinite temperature $\beta\to 0$ (i.e., $u_N\to0$), 
$T_\sigma(v)$ 
has been already calculated in \eqref{u=0}. On the other hand,  $Y_j(v)$
in the RHS in \eqref{finite-N} becomes constant in this limit. Hence, one obtains
\begin{align}
\left.(1+Y_j(v))(1+Y_{j+1}^{-1}(v))^{\delta_{j,m_{\alpha}-1}}
\right|_{u_N=0}&=\lim_{\beta\to 0}\lim_{v\to\infty}(1+Y_j(v))
(1+Y_{j+1}^{-1}(v))^{\delta_{j,m_{\alpha}-1}}\nn \\
&=
\lim_{\beta\to 0}\frac{\sh^2(\beta h \wt{n}_{j+1})}{\sh^2(\beta h y_r)}\nn \\
&=\frac{\wt{n}_{j+1}^2}{y_r^2},
\end{align}
where $m_r\le j < m_{r+1}$ ($0\le r \le \alpha-1$). To derive
this, we have used \eqref{DVF} at the sector $M=N\sigma/2$.
Thus, 
\begin{equation}
s_{r_\sigma}*\log(1+Y_{j_\sigma})
(1+Y_{j_\sigma+1}^{-1})^{\delta_{j_\sigma,m_{\alpha}-1}}(v)+
\sum_{r=1}^{r_\sigma-1}
\wt{d}_r*\log(1+Y_{m_r-1})(v)=\sigma+1 \quad (u_N\to 0).
\end{equation}
Comparing both sides of \eqref{finite-N} as $u_N \to 0$, 
we conclude that the sum of the last three terms reduces 
to $\sum_{\ell=1}^\sigma \log\left[\phi(v+2i\ell)\phi(v-2i\ell)\right] $ as $u_N \to 0$.
The Trotter limit $ N \to \infty$ in \eqref{finite-N} can be taken analytically by 
expanding the last three terms up to $O(u_N)$. Using \eqref{jsigma} and 
the properties of the sequences defined in Appendix~\ref{TS}, these 
terms can be summarized as:
\begin{align}
\log T_\sigma(v)&=s_{r_\sigma}*\log(1+Y_{j_\sigma})
(1+Y_{j_\sigma+1}^{-1})^{\delta_{j_\sigma,m_{\alpha}-1}}(v)+
\sum_{r=1}^{r_\sigma-1}\wt{d}_r*\log(1+Y_{m_r-1})(v)\nn \\
&\quad +\sum_{\ell=1}^\sigma \log\left[\phi(v+2i\ell)\phi(v-2i\ell)\right] +
\beta \Phi(v),
\end{align}
where
\begin{align}
&\Phi(v)=\frac{\pi J \sin\theta}{\theta}\Biggl\{(-1)^{r_\sigma-1} s_{r_\sigma}*
\left[a(v,p_{r_\sigma})+a(v,p_{r_\sigma}+2(-1)^{r_\sigma-1}\wt{q}_{j_\sigma+1})\right]
\nn \\
&\qquad \qquad -\sum_{r=1}^{r_\sigma-1}(-1)^r\wt{d}_r
*\left[a(v,p_r+p_{r+1}+\wt{q}_{j_\sigma+1})+a(v,p_r+p_{r+1}-\wt{q}_{j_\sigma+1})\right]
\Biggr\}, 
\nn \\
&a(v,p):=\frac{\theta}{2\pi}\frac{\sin(\theta p)}{\ch(\theta v)-\cos (\theta p)}.
\end{align}
From \eqref{free-largest}, the free energy reads
\begin{align}
f=-\Phi(0)-\frac{1}{\beta}\left[ s_{r_\sigma}*\log(1+Y_{j_\sigma})
(1+Y_{j_\sigma+1}^{-1})^{\delta_{j_\sigma,m_{\alpha}-1}}(0)
+\sum_{r=1}^{r_\sigma-1}\wt{d}_r*\log(1+Y_{m_r-1})(0)\right].
\label{free-TBA} 
\end{align}

\section{Drude weight via GHD}\label{derivation-Drude}
%
Here we  summarize how to derive the Drude weight 
from the perspective of the GHD, making sure that this paper 
is self-contained. The discussion here is primarily based 
on \cite{doyon2017drude, doyon2020lecture}.

The Drude weight associated with the current densities $ j_\mu(t,x) $ and $ j_\nu(t,x) $ 
is described by the Kubo formula in the form of the dynamical
correlation functions in the equilibrium state:
\begin{equation}
\ms{D}_{\mu\nu}(\beta):=\lim_{t\to\infty}\frac{\beta}{t}\int_0^t 
ds \int dx \bra j_\mu(s,x)j_\nu(0,0)\ket^\textrm{c}
= \beta \lim_{t\to\infty}
\int dx \bra j_\mu(t,x)j_\nu(0,0)\ket^\textrm{c},
\label{def-Drude}
\end{equation}
where $ \bra \cdots \ket^\textrm{c} $ stands for the connected 
correlation function. 
(Note that 
$\ms{D}_{\mu\nu}/\beta$ is sometimes used as an alternative definition
of the Drude weight.)
These current densities satisfy the continuity equation
\begin{equation}
\del_t q_\mu(t,x)+\del_x j_\mu(t,x)=0,
\label{continuity}
\end{equation}
where $ q_\mu(t,x) $ is the charge density of the (quasi) 
local conserved charge $ Q_\mu $ ($ \mu\in\mathbb{N} $). That is,
\begin{equation}
Q_\mu=\int dx\, q_\mu(t,x), \quad \del_t Q_\mu=0.
\end{equation}

Quantum integrable many-body systems with short-range interactions, 
including the model under consideration, possess an infinite set of 
conserved quantities $\{Q_\mu\}_{\mu=0}^\infty$. In these systems, the 
maximal entropy state is, in general, characterized by the GGE, whose density matrix 
is given by
\begin{align}
&\rho_\textrm{GGE}=\frac{\exp\left(-\sum_\mu\beta^\mu Q_\mu\right)}
{Z_\textrm{GGE}}
=\frac{\exp\left[-\sum_\mu\beta^{\mu}\int dx , q_\mu(0,x)
\right]}{Z_\textrm{GGE}},
\nn \\
&Z_\textrm{GGE}:=\Tr\left[\exp\left(-\sum_\mu\beta^\mu Q_\mu\right)\right],
\end{align}
where $\beta^\mu$ represents the generalized inverse temperature.
Let us denote 
\begin{equation}
\bra  o \ket_{\ul{\beta}}:=\lim_{L\to\infty}
\Tr (o \rho_{\textrm{GGE}})
\end{equation}
as the average of any operator $o$ over the GGE.
In particular, let $ \ms{q}_\mu $ and $ \ms{j}_\mu $ be the average charge 
and current, respectively:
\begin{equation}
\ms{q}_\mu=\bra q_\mu(0,0)\ket_{\ul{\beta}}, 
\quad \ms{j}_\mu=\bra j_\mu(0,0)\ket_{\ul{\beta}}.
\label{cc-GGE}
\end{equation}
Then, the relations $\del_{\beta^\mu}\ms{q}_\nu=\del_{\beta^\nu}\ms{q}_\mu$ and
$\del_{\beta^\mu}\ms{j}_\nu=\del_{\beta^\nu}\ms{j}_\mu$ hold \cite{doyon2020lecture},
which ensure the existence of the free energy density
\begin{equation}
\ms{f}:=-\lim_{L\to\infty}\frac{1}{L}\log Z_{\textrm{GGE}}
\label{free-density}
\end{equation}
and the free energy flux $\ms{g}$ (whose explicit form
is defined later) 
such that \begin{equation}
\ms{q}_\mu=\del_{\beta^\mu}\ms{f}, \quad \ms{j}_\mu=\del_{\beta^\mu}\ms{g}.
\label{flux}
\end{equation}
Now, we introduce the equal-time charge-charge and current-charge 
correlation functions:
\begin{equation}
\ms{B}_{\mu\nu}:=(j_\mu,q_\nu)=-\del_{\beta^\mu}\del_{\beta^\nu}\ms{g}=-\del_{\beta^\nu}\ms{j}_\mu,
\quad
\ms{C}_{\mu\nu}:=(q_\mu,q_\nu)=-\del_{\beta^\mu}\del_{\beta^\nu}\ms{f}=-\del_{\beta^\nu}\ms{q}_\mu,
\label{BC}
\end{equation}
where $(o_1,o_2)$ represents the inner product defined as
\begin{equation}
(o_1,o_2):=\int dx\, \bra o_1(0,0) o_2(0,x)\ket^\textrm{c}_{\ul{\beta}}.
\end{equation}
Obviously, the matrices $\ms{B}$ and $\ms{C}$ are symmetric matrices.
Introducing the flux Jacobian,
\begin{equation}
\ms{A}_\mu^{\ \nu}:=\del_{\ms{q}_\nu}\ms{j}_\mu,
\label{flux-Jacobian}
\end{equation}
we write $\ms{B}$ in terms of $\ms{A}$ and $\ms{C}$:
\begin{equation}
\ms{B}=\ms{A} \ms{C}=\ms{C} \ms{A}^\textrm{T}.
\label{AC}
\end{equation}
We denote $\ms{D}_{\mu\nu}(\ul{\beta})$ be a generalization
of the Drude weight \eqref{def-Drude} by replacing 
the current-current correlation function over the 
canonical Gibbs ensemble with those over the GGE.
Then we express $\ms{D}_{\mu\nu}(\ul{\beta})$  using the hydrodynamic 
projection method \cite{mazur1969non,suzuki1971ergodicity, doyon2022hydrodynamic}:
\begin{equation}
\beta^{-1}\ms{D}_{\mu\nu}(\ul{\beta})=(j_\mu,\mathbb{P}j_\nu)=
(\mathbb{P}j_\mu,j_\nu)=(\mathbb{P}j_\mu,\mathbb{P}j_\nu),
\end{equation}
where the operator $\mathbb{P}$ projects 
$o$ onto the space spanned by local or quasi-local conserved charges:
\begin{equation}
\mathbb{P} o=\sum_{\mu,\nu} q_\mu \ms{C}^{\mu \nu}(q_\nu,o).
\end{equation}
The matrix $\ms{C}^{\mu\nu}$ denotes the inverse of $\ms{C}_{\mu\nu}$. Thus, 
as referenced in \cite{doyon2017drude}, the Drude weight can be rewritten as
\begin{equation}
\beta^{-1}\ms{D}_{\mu\nu}(\ul{\beta})=\sum_{\gamma,\delta} (j_\mu,q_\gamma) 
\ms{C}^{\gamma\delta}(q_\delta,j_\nu).
\end{equation}
Using the definition \eqref{BC} and the relation \eqref{AC}, 
we find that $\ms{D}_{\mu\nu}(\ul{\beta})$ is given by  the following 
matrix forms:
\begin{equation}
\beta^{-1}\ms{D}(\ul{\beta})=\ms{B}\ms{C}^{-1}\ms{B}^\textrm{T}=\ms{A}\ms{C}\ms{A}^\textrm{T}=
\ms{A}^2\ms{C}.
\label{drude-matrix}
\end{equation}

Let us now apply the argument discussed above to a generic fermionic 
quantum integrable system (which of course includes the present model), 
whose thermodynamic quantities can be described by the TBA.
In such a system, the local conserved charges $\ms{q}_\mu$ are decomposed 
in terms of the quasi-particles of type $j$ (corresponding to the $n_j$-string 
in the case of our present model):
\begin{equation}
\ms{q}_\mu=\sum_{j} \int dv\, \rho_j(v) h_{\mu;j}(v)=\sum_{j}
\int dv\, \rho^\textrm{tot}_j(v)\vartheta_j(v)  h_{\mu;j}(v).
\label{charge}
\end{equation}
Here, 
$\rho_j(v)$ is the distribution density of the 
quasi-particles of type $j$, and $\rho^\textrm{tot}(v):=
\rho_j(v) + \rho_j^\textrm{h}(v)$ denotes the density of 
states, where $\rho_j^\textrm{h}$
is the density of holes. 
The quantity $ h_{\mu;j}(v) $ is referred to as a bare 
charge. In particular, we set 
\begin{equation}
h_{1;j}(v) = \varepsilon_j(v) 
\label{bare-energy}
\end{equation}
to be the bare energy, and accordingly, 
\begin{equation}
\beta^1=\beta.
\label{beta1}
\end{equation}
The quantity 
\begin{equation}
\vartheta_j(v):=\frac{\rho_j(v)}{\rho_j^\textrm{tot}(v)}
\label{stat-factor}
\end{equation}
represents the fermionic distribution function. 
From the BAE (see \eqref{BAE} for 
instance), we obtain $\rho^\text{tot}_j(v)$ as 
the solution to the integral equation:
\begin{equation}
\xi_j\rho_j^\text{tot}(v)
=\frac{p'_j(v)}{2\pi}-\sum_k \xi_k K_{jk}\ast(\vartheta_k
\xi_k\rho^\textrm{tot}_k)(v),
\label{rho-total}
\end{equation}
where
$\xi_j$ is a sign factor, 
and $K_{jk}(v)$ is a kernel assumed to be symmetric $K_{jk}(v)=K_{kj}(v)$.
The quantity $p_j(v)$ is the bare momentum, related to the bare energy 
\eqref{bare-energy} by 
\begin{equation}
p'_j(v)=\epsilon \varepsilon_j(v)=\epsilon h_{1,j}(v),
\label{e-p}
\end{equation}
where $\epsilon\in\mathbb{R}$ is a constant depending the
definition of the overall scaling factor of the Hamiltonian. 
For instance, 
\begin{equation}
\epsilon=-\frac{\theta}{J\sin\theta}
\end{equation}
for our present model.

In the framework of the TBA, the free energy density $\ms{f}$
\eqref{free-density} is written as
\begin{equation}
\ms{f}=-\frac{1}{2\pi}\sum_j \int dv \xi_j p_j'(v)\log\left(1+\eta_j^{-1}(v)\right).
\label{free-energy-density}
\end{equation}
Correspondingly, the free energy flux
$\ms{g}$ (see \eqref{flux}) is given by \cite{castro2016emergent} as
\begin{equation}
\ms{g}=-\frac{1}{2\pi}\sum_j \int dv \xi_j \varepsilon_j'(v)
\log\left(1+\eta_j^{-1}(v)\right).
\label{free-energy-flux}
\end{equation}
The function  $\eta_j(v)$ defined by
\begin{equation}
\eta_j(v):=\frac{\rho_j^\textrm{h}(v)}{\rho_j(v)}
\label{eta}
\end{equation}
is determined by the generalized TBA equations (cf. \eqref{TBA-xxz}):
\begin{equation}
\log \eta_j(v)=\sum_{\mu}\beta^{\mu} h_{\mu;j}(v)+
\sum_{k}\xi_k K_{jk}*\log\left(1+\eta^{-1}_k\right)(v).
\label{TBA-general}
\end{equation}
By differentiating \eqref{TBA-general} with respect to 
$\beta^\mu$, we define the dressed charge $h_{\mu;j}^\dr(v)$ as
\begin{equation}
h_{\mu;j}^\dr(v) := \del_{\beta^{\mu}}\log\eta_j(v),
\label{dress-eta}
\end{equation}
which fulfills the integral equation
\begin{equation}
h_{\mu;j}^\dr(v) = h_{\mu;j}(v) - \sum_{k}\xi_k 
K_{jk}*(\vartheta_k h_{\mu;k}^\dr)(v).
\label{dress}
\end{equation}
The above dressed operation can also similarly apply for any 
function $h_j(v)$ by just replacing $h_{\mu,j}(v)$  with $h_j(v)$.
Some crucial quantities related to the Drude weight are expressed by the dressed 
functions. For example, the density of state $\rho_j^\textrm{tot}(v)$ 
\eqref{rho-total} is expressed as the dressed energy:
\begin{equation}
\rho_j^\textrm{tot}(v) = \frac{\epsilon}{2\pi} \xi_j h_{1;j}^\dr(v)
= \frac{\epsilon}{2\pi}\xi_j \varepsilon_j^\dr(v)
= \frac{1}{2\pi} \xi_j \del_v\left(p_j^\dr\right)(v).
\label{rho-dress}
\end{equation}
This can be easily followed from \eqref{rho-total} in conjunction with \eqref{e-p} 
and \eqref{dress}. In particular, the effective velocity $v_j^\eff(v)$ of 
the quasi-particle of type $j$ is given by
\begin{equation}
v_j^{\eff}(v):= \frac{\del_v[\varepsilon_j^{\dr}(v)]}
{\del_v[p_j^{\dr}(v)]} = \frac{\beta^{-1}\del_v\log\eta_j(v)}
{\epsilon \del_\beta \log \eta_j(v)}.
\label{velocity}
\end{equation}

In addition to \eqref{charge}, we derive $\ms{q}_\mu$ using 
\eqref{flux} and \eqref{free-energy-density}:
\begin{equation}
\ms{q}_{\mu} = \frac{1}{2\pi} \sum_j \int dv \, p_j'(v) \xi_j 
\vartheta_j(v) h_{\mu;j}^{\dr}(v).
\label{q-dress}
\end{equation}
Noticing the relation in \eqref{e-p} and applying the dressed function 
formalism to \eqref{dress}, we find that \eqref{q-dress} aligns with \eqref{charge}. 
Similarly, from \eqref{flux} and \eqref{free-energy-flux}, we obtain
\begin{align}
\ms{j}_{\mu} &= \frac{1}{2\pi} \sum_j \int dv \, \varepsilon_j'(v) \xi_j 
\vartheta_j(v) h_{\mu;j}^{\dr}(v) = \frac{1}{2\pi} 
\sum_j \int dv \, \del_v(\varepsilon_j^{\dr})(v) 
\xi_j  \vartheta_j(v) h_{\mu;j}(v) \nn \\
&= \sum_j \int dv \, \rho_j(v) v_j^{\eff}(v) h_{\mu,j}(v),
\label{j-dress}
\end{align}
where the dressed function formalism is applied to derive the
second equality. Eq.~\eqref{rho-dress} has been used 
to derive the third equality which provides a physically 
reasonable picture. The relation \eqref{j-dress} was originally 
proposed by \cite{castro2016emergent,bertini2016transport}, and 
has recently been confirmed by the Bethe ansatz technique 
\cite{urichuk2019spin,borsi2020current,pozsgay2020current}. 
According to the definitions given in \eqref{BC}, we  
differentiate \eqref{q-dress} and \eqref{j-dress} with
respect to $\beta^{\nu}$ to derive  $\ms{B}$ and $\ms{C}$.
Using the relation
\begin{equation}
\del_{\beta^{\mu}}\vartheta_j(v)=-(1-\vartheta_j(v))\vartheta_j(v)h_{\mu;j}^{\dr},
\end{equation}
which can be followed from  \eqref{eta} and \eqref{dress-eta},
and applying the dressed function formalism, we have \cite{doyon2017drude}
\begin{align}
&\ms{B}_{\mu\nu}=\sum_j \int dv\,  \rho_j(v)\left(1-\vartheta_j(v)\right)
\left[v_j^{\eff}(v) h_{\mu;j}^\dr(v)\right] h_{\nu;j}^\dr(v), \nn \\
&\ms{C}_{\mu\nu}=\sum_j \int dv\, \rho_j(v)\left(1-\vartheta_j(v)\right)
h_{\mu;j}^\dr(v) h_{\nu;j}^\dr(v).
\end{align}
The relation between $\ms{B}$ and $\ms{C}$ as given by \eqref{AC}
(more explicitly $B_{\mu\nu} = \sum_{\gamma} A_{\mu}^{\ \gamma} C_{\gamma\nu}$ 
in terms of the matrix elements) 
leads to
\begin{equation}
A_{\mu}^{\ \nu}h_{\nu;j}^\dr(v)=v_j^\eff h_{\mu;j}^\dr(v).
\label{A-eigen}
\end{equation}
This indicates that the dressed charges serve as the eigenfunctions of the flux Jacobian, and their corresponding eigenvalues are the effective velocities.
Then, using \eqref{drude-matrix} with \eqref{A-eigen}, 
we finally arrive at 
\begin{equation}
\ms{D}_{\mu\nu}(\ul{\beta})=\beta \sum_{\gamma,\delta}
\ms{A}_{\mu}^{\ \gamma} \ms{A}_\gamma^{\ \delta} \ms{C}_{\delta\nu}
=\beta \sum_j \int dv\, \rho_j(v)\left(1-\vartheta_j(v)\right)
\left[v^\eff_j(v)\right]^2 h_{\mu;j}^\dr(v)h_{\nu;j}^\dr(v).
\label{drude-general}
\end{equation}
All the quantities in the integrand are calculated 
using the generalized TBA equations \eqref{TBA-general}. 
After replacing them with those obtained via the standard TBA,
we also evaluate the original Drude weight $\ms{D}_{\mu\nu}(\beta)$
\eqref{def-Drude}
by \eqref{drude-general}.

\end{appendix}

\bibliography{BibFile}

\begin{thebibliography}{76}
\providecommand{\natexlab}[1]{#1}
\providecommand{\url}[1]{\texttt{#1}}
\expandafter\ifx\csname urlstyle\endcsname\relax
  \providecommand{\doi}[1]{doi: #1}\else
  \providecommand{\doi}{doi: \begingroup \urlstyle{rm}\Url}\fi

\bibitem[Rigol et~al.(2006)Rigol, Muramatsu, and Olshanii]{rigol2006hard}
Marcos Rigol, Alejandro Muramatsu, and Maxim Olshanii.
\newblock Hard-core bosons on optical superlattices: Dynamics and relaxation in
  the superfluid and insulating regimes.
\newblock \emph{Physical Review A}, 74\penalty0 (5):\penalty0 053616, 2006.

\bibitem[Rigol et~al.(2007)Rigol, Dunjko, Yurovsky, and
  Olshanii]{rigol2007relaxation}
Marcos Rigol, Vanja Dunjko, Vladimir Yurovsky, and Maxim Olshanii.
\newblock Relaxation in a completely integrable many-body quantum system: an ab
  initio study of the dynamics of the highly excited states of 1d lattice
  hard-core bosons.
\newblock \emph{Physical review letters}, 98\penalty0 (5):\penalty0 050405,
  2007.

\bibitem[Vidmar and Rigol(2016)]{vidmar2016generalized}
Lev Vidmar and Marcos Rigol.
\newblock Generalized gibbs ensemble in integrable lattice models.
\newblock \emph{Journal of Statistical Mechanics: Theory and Experiment},
  2016\penalty0 (6):\penalty0 064007, 2016.

\bibitem[Rigol and Srednicki(2012)]{rigol2012alternatives}
Marcos Rigol and Mark Srednicki.
\newblock Alternatives to eigenstate thermalization.
\newblock \emph{Physical review letters}, 108\penalty0 (11):\penalty0 110601,
  2012.

\bibitem[Castro-Alvaredo et~al.(2016)Castro-Alvaredo, Doyon, and
  Yoshimura]{castro2016emergent}
Olalla~A Castro-Alvaredo, Benjamin Doyon, and Takato Yoshimura.
\newblock Emergent hydrodynamics in integrable quantum systems out of
  equilibrium.
\newblock \emph{Physical Review X}, 6\penalty0 (4):\penalty0 041065, 2016.

\bibitem[Bertini et~al.(2016)Bertini, Collura, De~Nardis, and
  Fagotti]{bertini2016transport}
Bruno Bertini, Mario Collura, Jacopo De~Nardis, and Maurizio Fagotti.
\newblock Transport in out-of-equilibrium {XXZ} chains: Exact profiles of
  charges and currents.
\newblock \emph{Physical review letters}, 117\penalty0 (20):\penalty0 207201,
  2016.

\bibitem[Doyon(2020)]{doyon2020lecture}
Benjamin Doyon.
\newblock Lecture notes on generalised hydrodynamics.
\newblock \emph{SciPost Physics Lecture Notes}, page 018, 2020.

\bibitem[Bulchandani et~al.(2021)Bulchandani, Gopalakrishnan, and
  Ilievski]{bulchandani2021superdiffusion}
Vir~B Bulchandani, Sarang Gopalakrishnan, and Enej Ilievski.
\newblock Superdiffusion in spin chains.
\newblock \emph{Journal of Statistical Mechanics: Theory and Experiment},
  2021\penalty0 (8):\penalty0 084001, 2021.

\bibitem[De~Nardis et~al.(2022)De~Nardis, Doyon, Medenjak, and
  Panfil]{de2022correlation}
Jacopo De~Nardis, Benjamin Doyon, Marko Medenjak, and Mi{\l}osz Panfil.
\newblock Correlation functions and transport coefficients in generalised
  hydrodynamics.
\newblock \emph{Journal of Statistical Mechanics: Theory and Experiment},
  2022\penalty0 (1):\penalty0 014002, 2022.

\bibitem[Essler(2022)]{essler2022short}
Fabian~HL Essler.
\newblock A short introduction to generalized hydrodynamics.
\newblock \emph{Physica A: Statistical Mechanics and its Applications}, page
  127572, 2022.

\bibitem[Bastianello et~al.(2021)Bastianello, De~Luca, and
  Vasseur]{bastianello2021hydrodynamics}
Alvise Bastianello, Andrea De~Luca, and Romain Vasseur.
\newblock Hydrodynamics of weak integrability breaking.
\newblock \emph{Journal of Statistical Mechanics: Theory and Experiment},
  2021\penalty0 (11):\penalty0 114003, 2021.

\bibitem[Gopalakrishnan and Vasseur(2023)]{gopalakrishnan2023anomalous}
Sarang Gopalakrishnan and Romain Vasseur.
\newblock Anomalous transport from hot quasiparticles in interacting spin
  chains.
\newblock \emph{Reports on Progress in Physics}, 2023.

\bibitem[Bertini et~al.(2021)Bertini, Heidrich-Meisner, Karrasch, Prosen,
  Steinigeweg, and {\v{Z}}nidari{\v{c}}]{bertini2021finite}
Bruno Bertini, Fabian Heidrich-Meisner, Christoph Karrasch, Toma{\v{z}} Prosen,
  R~Steinigeweg, and Marko {\v{Z}}nidari{\v{c}}.
\newblock Finite-temperature transport in one-dimensional quantum lattice
  models.
\newblock \emph{Reviews of Modern Physics}, 93\penalty0 (2):\penalty0 025003,
  2021.

\bibitem[Zotos(1999)]{zotos1999finite}
X~Zotos.
\newblock Finite temperature {Drude} weight of the one-dimensional spin-1/2
  {Heisenberg} model.
\newblock \emph{Physical review letters}, 82\penalty0 (8):\penalty0 1764, 1999.

\bibitem[Prosen and Ilievski(2013)]{prosen2013families}
Toma{\v{z}} Prosen and Enej Ilievski.
\newblock Families of quasilocal conservation laws and quantum spin transport.
\newblock \emph{Physical review letters}, 111\penalty0 (5):\penalty0 057203,
  2013.

\bibitem[Urichuk et~al.(2019)Urichuk, Oez, Kl{\"u}mper, and
  Sirker]{urichuk2019spin}
Andrew Urichuk, Yahya Oez, Andreas Kl{\"u}mper, and Jesko Sirker.
\newblock The spin {Drude} weight of the {XXZ} chain and generalized
  hydrodynamics.
\newblock \emph{SciPost Physics}, 6\penalty0 (1):\penalty0 005, 2019.

\bibitem[Kl{\"u}mper and Sakai(2019)]{klumper2019spin}
Andreas Kl{\"u}mper and Kazumitsu Sakai.
\newblock The spin {Drude} weight of the spin-1/2$ {XXZ} $ chain: An analytic
  finite size study.
\newblock \emph{arXiv preprint arXiv:1904.11253}, 2019.

\bibitem[{\v{Z}}nidari{\v{c}}(2011)]{vznidarivc2011spin}
Marko {\v{Z}}nidari{\v{c}}.
\newblock Spin transport in a one-dimensional anisotropic {Heisenberg} model.
\newblock \emph{Physical Review Letters}, 106\penalty0 (22):\penalty0 220601,
  2011.

\bibitem[Ilievski et~al.(2018)Ilievski, De~Nardis, Medenjak, and
  Prosen]{ilievski2018superdiffusion}
Enej Ilievski, Jacopo De~Nardis, Marko Medenjak, and Toma{\v{z}} Prosen.
\newblock Superdiffusion in one-dimensional quantum lattice models.
\newblock \emph{Physical review letters}, 121\penalty0 (23):\penalty0 230602,
  2018.

\bibitem[Ljubotina et~al.(2017)Ljubotina, {\v{Z}}nidari{\v{c}}, and
  Prosen]{ljubotina2017spin}
Marko Ljubotina, Marko {\v{Z}}nidari{\v{c}}, and Toma{\v{z}} Prosen.
\newblock Spin diffusion from an inhomogeneous quench in an integrable system.
\newblock \emph{Nature communications}, 8\penalty0 (1):\penalty0 16117, 2017.

\bibitem[Gopalakrishnan and Vasseur(2019)]{gopalakrishnan2019kinetic}
Sarang Gopalakrishnan and Romain Vasseur.
\newblock Kinetic theory of spin diffusion and superdiffusion in {XXZ} spin
  chains.
\newblock \emph{Physical review letters}, 122\penalty0 (12):\penalty0 127202,
  2019.

\bibitem[Ljubotina et~al.(2019)Ljubotina, {\v{Z}}nidari{\v{c}}, and
  Prosen]{ljubotina2019kardar}
Marko Ljubotina, Marko {\v{Z}}nidari{\v{c}}, and Toma{\v{z}} Prosen.
\newblock {Kardar-Parisi-Zhang} physics in the quantum {Heisenberg} magnet.
\newblock \emph{Physical review letters}, 122\penalty0 (21):\penalty0 210602,
  2019.

\bibitem[Kardar et~al.(1986)Kardar, Parisi, and Zhang]{kardar1986dynamic}
Mehran Kardar, Giorgio Parisi, and Yi-Cheng Zhang.
\newblock Dynamic scaling of growing interfaces.
\newblock \emph{Physical Review Letters}, 56\penalty0 (9):\penalty0 889, 1986.

\bibitem[Krajnik et~al.(2022)Krajnik, Ilievski, and Prosen]{krajnik2022absence}
{\v{Z}}iga Krajnik, Enej Ilievski, and Toma{\v{z}} Prosen.
\newblock Absence of normal fluctuations in an integrable magnet.
\newblock \emph{Physical Review Letters}, 128\penalty0 (9):\penalty0 090604,
  2022.

\bibitem[Rosenberg et~al.(2023)Rosenberg, Andersen, Samajdar, Petukhov, Hoke,
  Abanin, Bengtsson, Drozdov, Erickson, Klimov, et~al.]{rosenberg2023dynamics}
Eliott Rosenberg, Trond Andersen, Rhine Samajdar, Andre Petukhov, Jesse Hoke,
  Dmitry Abanin, Andreas Bengtsson, Ilya Drozdov, Catherine Erickson, Paul
  Klimov, et~al.
\newblock Dynamics of magnetization at infinite temperature in a {Heisenberg}
  spin chain.
\newblock \emph{arXiv preprint arXiv:2306.09333}, 2023.

\bibitem[Kohn(1964)]{kohn1964theory}
Walter Kohn.
\newblock Theory of the insulating state.
\newblock \emph{Physical review}, 133\penalty0 (1A):\penalty0 A171, 1964.

\bibitem[Takahashi and Suzuki(1972)]{takahashi1972one}
Minoru Takahashi and Masuo Suzuki.
\newblock One-dimensional anisotropic {Heisenberg} model at finite
  temperatures.
\newblock \emph{Progress of theoretical physics}, 48\penalty0 (6):\penalty0
  2187--2209, 1972.

\bibitem[Ilievski(2023)]{ilievski2023popcorn}
Enej Ilievski.
\newblock Popcorn {Drude} weights from quantum symmetry.
\newblock \emph{Journal of Physics A: Mathematical and Theoretical},
  55\penalty0 (50):\penalty0 504005, 2023.

\bibitem[Ilievski and De~Nardis(2017)]{ilievski2017ballistic}
Enej Ilievski and Jacopo De~Nardis.
\newblock Ballistic transport in the one-dimensional {Hubbard} model: The
  hydrodynamic approach.
\newblock \emph{Physical Review B}, 96\penalty0 (8):\penalty0 081118, 2017.

\bibitem[Doyon and Spohn(2017)]{doyon2017drude}
Benjamin Doyon and Herbert Spohn.
\newblock {Drude} weight for the {Lieb-Liniger} {Bose} gas.
\newblock \emph{SciPost Physics}, 3\penalty0 (6):\penalty0 039, 2017.

\bibitem[Nagy et~al.(2023{\natexlab{a}})Nagy, Kormos, and
  Tak{\'a}cs]{nagy2023thermodynamics}
Botond~C Nagy, M{\'a}rton Kormos, and G{\'a}bor Tak{\'a}cs.
\newblock Thermodynamics and fractal {Drude} weights in the sine-{Gordon}
  model.
\newblock \emph{arXiv preprint arXiv:2305.15474}, 2023{\natexlab{a}}.

\bibitem[Nagy et~al.(2023{\natexlab{b}})Nagy, Tak{\'a}cs, and
  Kormos]{nagy2023thermodynamic2}
BC~Nagy, G~Tak{\'a}cs, and M~Kormos.
\newblock Thermodynamic {Bethe} ansatz and generalised hydrodynamics in the
  sine-{Gordon} model.
\newblock \emph{arXiv preprint arXiv:2312.03909}, 2023{\natexlab{b}}.

\bibitem[Kirillov and Reshetikhin(1987{\natexlab{a}})]{kirillov1987exact1}
Anatol~N Kirillov and N~Yu Reshetikhin.
\newblock Exact solution of the integrable {XXZ} {Heisenberg} model with
  arbitrary spin. i. the ground state and the excitation spectrum.
\newblock \emph{Journal of Physics A: Mathematical and General}, 20\penalty0
  (6):\penalty0 1565, 1987{\natexlab{a}}.

\bibitem[Zamolodchikov and Fateev(1980)]{Zamolodchikov1980}
A.~B. Zamolodchikov and V.~A. Fateev.
\newblock Model factorized {S} matrix and an integrable {Heisenberg} chain with
  spin 1.
\newblock \emph{Sov. J. Nucl. Phys.}, 32:\penalty0 298, 1980.

\bibitem[Piroli and Vernier(2016)]{piroli2016quasi}
Lorenzo Piroli and Eric Vernier.
\newblock Quasi-local conserved charges and spin transport in spin-1 integrable
  chains.
\newblock \emph{Journal of Statistical Mechanics: Theory and Experiment},
  2016\penalty0 (5):\penalty0 053106, 2016.

\bibitem[Frahm et~al.(1990)Frahm, Yu, and Fowler]{frahm1990integrable}
Holger Frahm, Nai-Chang Yu, and Michael Fowler.
\newblock The integrable {XXZ} {Heisenberg} model with arbitrary spin:
  construction of the {Hamiltonian}, the ground-state configuration and
  conformal properties.
\newblock \emph{Nuclear Physics B}, 336\penalty0 (3):\penalty0 396--434, 1990.

\bibitem[Frahm and Yu(1990)]{frahm1990finite}
H~Frahm and Nai-C Yu.
\newblock Finite-size effects in the integrable {XXZ} {Heisenberg} model with
  arbitrary spin.
\newblock \emph{Journal of Physics A: Mathematical and General}, 23\penalty0
  (11):\penalty0 2115, 1990.

\bibitem[Kuniba et~al.(1998)Kuniba, Sakai, and Suzuki]{kuniba1998continued}
Atsuo Kuniba, Kazumitsu Sakai, and Junji Suzuki.
\newblock Continued fraction tba and functional relations in {XXZ} model at
  root of unity.
\newblock \emph{Nuclear Physics B}, 525\penalty0 (3):\penalty0 597--626, 1998.

\bibitem[Kulish et~al.(1981)Kulish, Reshetikhin, and Sklyanin]{kulish1981yang}
Petr~P Kulish, N~Yu Reshetikhin, and EK~Sklyanin.
\newblock {Yang-Baxter} equation and representation theory: I.
\newblock \emph{Letters in Mathematical Physics}, 5:\penalty0 393--403, 1981.

\bibitem[Reshetikhin(2010)]{reshetikhin2010lectures}
N~Reshetikhin.
\newblock Lectures on the integrability of the six-vertex model.
\newblock \emph{Exact methods in low-dimensional statistical physics and
  quantum computing}, pages 197--266, 2010.

\bibitem[Fonseca and Balogh(2015)]{fonseca2015higher}
Tiago Fonseca and Ferenc Balogh.
\newblock The higher spin generalization of the 6-vertex model with domain wall
  boundary conditions and {Macdonald} polynomials.
\newblock \emph{Journal of Algebraic Combinatorics}, 41\penalty0 (3):\penalty0
  843--866, 2015.

\bibitem[Kulish and Reshetikhin(1981)]{kulish1981quantum}
Petr~Petrovich Kulish and Nikolai~Yur'evich Reshetikhin.
\newblock Quantum linear problem for the sine-{Gordon} equation and higher
  representations.
\newblock \emph{Zapiski Nauchnykh Seminarov POMI}, 101:\penalty0 101--110,
  1981.

\bibitem[Sogo et~al.(1983)Sogo, Akutsu, and Abe]{sogo1983new}
Kiyoshi Sogo, Yasuhiro Akutsu, and Takayuki Abe.
\newblock New factorized s-matrix and its application to exactly solvable
  q-state model. {I}.
\newblock \emph{Progress of theoretical physics}, 70\penalty0 (3):\penalty0
  730--738, 1983.

\bibitem[Nepomechie(2002)]{nepomechie2002solving}
Rafael~I Nepomechie.
\newblock Solving the open {XXZ} spin chain with nondiagonal boundary terms at
  roots of unity.
\newblock \emph{Nuclear Physics B}, 622\penalty0 (3):\penalty0 615--632, 2002.

\bibitem[Frappat et~al.(2007)Frappat, Nepomechie, and
  Ragoucy]{frappat2007complete}
Luc Frappat, Rafael~I Nepomechie, and Eric Ragoucy.
\newblock A complete bethe ansatz solution for the open spin-s xxz chain with
  general integrable boundary terms.
\newblock \emph{Journal of Statistical Mechanics: Theory and Experiment},
  2007\penalty0 (09):\penalty0 P09009, 2007.

\bibitem[Bytsko(2003)]{bytsko2003integrable}
Andrei~G Bytsko.
\newblock On integrable {Hamiltonians} for higher spin {XXZ} chain.
\newblock \emph{Journal of Mathematical Physics}, 44\penalty0 (9):\penalty0
  3698--3717, 2003.

\bibitem[Reshetikhin(1983)]{reshetikhin1983functional}
N~Yu Reshetikhin.
\newblock The functional equation method in the theory of exactly soluble
  quantum systems.
\newblock \emph{Zh. Eksp. Teor. Fiz}, 84\penalty0 (1):\penalty0 190--1201,
  1983.

\bibitem[Kirillov and Reshetikhin(1985)]{kirillov1985}
Anatol~N Kirillov and N~Yu Reshetikhin.
\newblock Classification of the string solutions of {Bethe} equations in the
  {XXZ}-model of arbitrary spin.
\newblock \emph{Zapiski Nauchnykh Seminarov LOMI}, 146:\penalty0 31--46, 1985.

\bibitem[Kirillov and Reshetikhin(1988)]{kirillov1988classification}
AN~Kirillov and N~Yu Reshetikhin.
\newblock Classification of the string solutions of {Bethe} equations in an
  {XXZ} model of arbitrary spin.
\newblock \emph{Journal of Soviet Mathematics}, 40\penalty0 (1):\penalty0
  22--35, 1988.

\bibitem[Kirillov and Reshetikhin(1987{\natexlab{b}})]{kirillov1987exact2}
AN~Kirillov and N~Yu Reshetikhin.
\newblock Exact solution of the integrable {XXZ} {Heisenberg} model with
  arbitrary spin. ii. thermodynamics of the system.
\newblock \emph{Journal of Physics A: Mathematical and General}, 20\penalty0
  (6):\penalty0 1587, 1987{\natexlab{b}}.

\bibitem[Johannesson(1988{\natexlab{a}})]{johannesson1988central}
Henrik Johannesson.
\newblock Central charge for the integrable higher-spin {XXZ} model.
\newblock \emph{Journal of Physics A: Mathematical and General}, 21\penalty0
  (11):\penalty0 L611, 1988{\natexlab{a}}.

\bibitem[Johannesson(1988{\natexlab{b}})]{johannesson1988universality}
Henrik Johannesson.
\newblock Universality classes of critical antiferromagnets.
\newblock \emph{Journal of Physics A: Mathematical and General}, 21\penalty0
  (23):\penalty0 L1157, 1988{\natexlab{b}}.

\bibitem[Suzuki(1976)]{suzuki1976relationship}
Masuo Suzuki.
\newblock Relationship between d-dimensional quantal spin systems and (d+
  1)-dimensional ising systems: Equivalence, critical exponents and systematic
  approximants of the partition function and spin correlations.
\newblock \emph{Progress of theoretical physics}, 56\penalty0 (5):\penalty0
  1454--1469, 1976.

\bibitem[Suzuki(1985)]{suzuki1985transfer}
Masuo Suzuki.
\newblock Transfer-matrix method and {Monte Carlo} simulation in quantum spin
  systems.
\newblock \emph{Physical Review B}, 31\penalty0 (5):\penalty0 2957, 1985.

\bibitem[Suzuki and Inoue(1987)]{suzuki1987st}
Masuo Suzuki and Makoto Inoue.
\newblock The {ST}-transformation approach to analytic solutions of quantum
  systems. i: General formulations and basic limit theorems.
\newblock \emph{Progress of theoretical physics}, 78\penalty0 (4):\penalty0
  787--799, 1987.

\bibitem[Koma(1987)]{koma1987thermal}
Tohru Koma.
\newblock Thermal {Bethe}-ansatz method for the one-dimensional {Heisenberg}
  model.
\newblock \emph{Progress of theoretical physics}, 78\penalty0 (6):\penalty0
  1213--1218, 1987.

\bibitem[Suzuki et~al.(1990)Suzuki, Akutsu, and Wadati]{suzuki1990new}
Junji Suzuki, Yasuhiro Akutsu, and Miki Wadati.
\newblock A new approach to quantum spin chains at finite temperature.
\newblock \emph{Journal of the Physical Society of Japan}, 59\penalty0
  (8):\penalty0 2667--2680, 1990.

\bibitem[Takahashi(1991)]{takahashi1991correlation}
Minoru Takahashi.
\newblock Correlation length and free energy of the s= 1/2 {XYZ} chain.
\newblock \emph{Physical Review B}, 43\penalty0 (7):\penalty0 5788, 1991.

\bibitem[Destri and De~Vega(1992)]{destri1992new}
C~Destri and HJ~De~Vega.
\newblock New thermodynamic {Bethe} ansatz equations without strings.
\newblock \emph{Physical review letters}, 69\penalty0 (16):\penalty0 2313,
  1992.

\bibitem[Kl{\"u}mper(1993)]{klumper1993thermodynamics}
Andreas Kl{\"u}mper.
\newblock Thermodynamics of the anisotropic spin-1/2 {Heisenberg} chain and
  related quantum chains.
\newblock \emph{Zeitschrift f{\"u}r Physik B Condensed Matter}, 91:\penalty0
  507--519, 1993.

\bibitem[Kl{\"u}mper(1992)]{klumper1992free}
A~Kl{\"u}mper.
\newblock Free energy and correlation lengths of quantum chains related to
  restricted solid-on-solid lattice models.
\newblock \emph{Annalen der Physik}, 504\penalty0 (7):\penalty0 540--553, 1992.

\bibitem[Takahashi(1999)]{takahashi1999thermodynamics}
M.~Takahashi.
\newblock \emph{Thermodynamics of One-Dimensional Solvable Models}.
\newblock Cambridge University Press, 1999.
\newblock ISBN 9780521551434.
\newblock URL \url{https://books.google.co.jp/books?id=kX1FAwEACAAJ}.

\bibitem[Essler et~al.(2005)Essler, Frahm, G{\"o}hmann, Kl{\"u}mper, and
  Korepin]{essler2005one}
Fabian~HL Essler, Holger Frahm, Frank G{\"o}hmann, Andreas Kl{\"u}mper, and
  Vladimir~E Korepin.
\newblock \emph{The one-dimensional {Hubbard} model}.
\newblock Cambridge University Press, 2005.

\bibitem[{\v{S}}amaj and Bajnok(2013)]{vsamaj2013introduction}
Ladislav {\v{S}}amaj and Zolt{\'a}n Bajnok.
\newblock \emph{Introduction to the statistical physics of integrable many-body
  systems}.
\newblock Cambridge University Press, 2013.

\bibitem[Benz et~al.(2005)Benz, Fukui, Kl{\"u}mper, and
  Scheeren]{benz2005finite}
J~Benz, T~Fukui, A~Kl{\"u}mper, and C~Scheeren.
\newblock On the finite temperature {Drude} weight of the anisotropic
  {Heisenberg} chain.
\newblock \emph{Journal of the Physical Society of Japan}, 74\penalty0
  (Suppl):\penalty0 181--190, 2005.

\bibitem[Prosen(2014)]{prosen2014quasilocal}
Toma{\v{z}} Prosen.
\newblock Quasilocal conservation laws in {XXZ} spin-1/2 chains: Open, periodic
  and twisted boundary conditions.
\newblock \emph{Nuclear Physics B}, 886:\penalty0 1177--1198, 2014.

\bibitem[Pereira et~al.(2014)Pereira, Pasquier, Sirker, and
  Affleck]{pereira2014exactly}
Rodrigo~Gon{\c{c}}alves Pereira, V~Pasquier, J~Sirker, and I~Affleck.
\newblock Exactly conserved quasilocal operators for the {XXZ} spin chain.
\newblock \emph{Journal of Statistical Mechanics: Theory and Experiment},
  2014\penalty0 (9):\penalty0 P09037, 2014.

\bibitem[Urichuk et~al.(2021)Urichuk, Sirker, and
  Kl{\"u}mper]{urichuk2021analytical}
Andrew Urichuk, Jesko Sirker, and Andreas Kl{\"u}mper.
\newblock Analytical results for the low-temperature {Drude} weight of the
  {XXZ} spin chain.
\newblock \emph{Physical Review B}, 103\penalty0 (24):\penalty0 245108, 2021.

\bibitem[De~Nardis et~al.(2019)De~Nardis, Bernard, and Doyon]{de2019diffusion}
Jacopo De~Nardis, Denis Bernard, and Benjamin Doyon.
\newblock Diffusion in generalized hydrodynamics and quasiparticle scattering.
\newblock \emph{SciPost Physics}, 6\penalty0 (4):\penalty0 049, 2019.

\bibitem[Ilievski et~al.(2021)Ilievski, De~Nardis, Gopalakrishnan, Vasseur, and
  Ware]{ilievski2021superuniversality}
Enej Ilievski, Jacopo De~Nardis, Sarang Gopalakrishnan, Romain Vasseur, and
  Brayden Ware.
\newblock Superuniversality of superdiffusion.
\newblock \emph{Physical Review X}, 11\penalty0 (3):\penalty0 031023, 2021.

\bibitem[Ae(2023)]{ae2023spin}
Shinya Ae.
\newblock Spin element of onsager matrix for spin-1/2 critical {XXZ} chain at
  infinite temperature and zero magnetic field.
\newblock \emph{arXiv preprint arXiv:2310.04790}, 2023.

\bibitem[Mazur(1969)]{mazur1969non}
P~Mazur.
\newblock Non-ergodicity of phase functions in certain systems.
\newblock \emph{Physica}, 43\penalty0 (4):\penalty0 533--545, 1969.

\bibitem[Suzuki(1971)]{suzuki1971ergodicity}
M~Suzuki.
\newblock Ergodicity, constants of motion, and bounds for susceptibilities.
\newblock \emph{Physica}, 51\penalty0 (2):\penalty0 277--291, 1971.

\bibitem[Doyon(2022)]{doyon2022hydrodynamic}
Benjamin Doyon.
\newblock Hydrodynamic projections and the emergence of linearised euler
  equations in one-dimensional isolated systems.
\newblock \emph{Communications in Mathematical Physics}, 391\penalty0
  (1):\penalty0 293--356, 2022.

\bibitem[Borsi et~al.(2020)Borsi, Pozsgay, and Pristy{\'a}k]{borsi2020current}
M{\'a}rton Borsi, Bal{\'a}zs Pozsgay, and Levente Pristy{\'a}k.
\newblock Current operators in bethe ansatz and generalized hydrodynamics: An
  exact quantum-classical correspondence.
\newblock \emph{Physical Review X}, 10\penalty0 (1):\penalty0 011054, 2020.

\bibitem[Pozsgay(2020)]{pozsgay2020current}
Bal{\'a}zs Pozsgay.
\newblock Current operators in integrable spin chains: lessons from long range
  deformations.
\newblock \emph{SciPost Physics}, 8\penalty0 (2):\penalty0 016, 2020.

\end{thebibliography}

\end{document}